\documentclass[useAMS,usenatbib]{mn2e}

\usepackage{amssymb}
\usepackage{amsmath}
\usepackage{times}
\usepackage{graphicx}
\usepackage{pgf}

\def\crab{PSR\,B0531+21}
\def\crabj{PSR\,J0534+2200}
\def\vela{PSR\,B0833-45}
\def\velaj{PSR\,J0835-4510}
\def\msh{PSR\,B1509-58}
\def\mshj{PSR\,J1513-5908}
\def\lmcpsr{PSR\,B0540-69}
\def\lmcpsra{PSR\,J0540-6919}

\def\psra{PSR\,J1846-0258} 
\def\psrb{PSR\,J1811-1925} 
\def\psrc{PSR\,J1617-5055} 
\def\psrd{PSR\,J1930+1852} 
\def\psre{PSR\,J2229+6114}
\def\psrf{PSR\,J0205+6449}
\def\psrg{PSR\,J0537-6910}
\def\psrh{PSR\,J2022+3842}
\def\psri{PSR\,J1813-1749}
\def\axj{AX\,J1838.0-0655}
\def\psraxj{PSR\,J1838-0655}
\def\igra{IGR\,J14003-6326}
\def\psrigra{PSR\,J1400-6325}
\def\igrb{IGR\,J18490-0000}
\def\psrigrb{PSR\,J1849-0001}

\def\psrcanda{PSR\,J1640-4631}
\def\psrcandb{PSR\,J1813-1246}

\newcommand{\gr}{$\gamma$}
\newcommand{\xmm}{{\it XMM-Newton}}
\newcommand{\fermi}{{\it Fermi}}
\newcommand{\integral}{{\it INTEGRAL}}
\newcommand{\cxo}{{\it Chandra}}
\newcommand{\rxte}{{\it RXTE}}
\newcommand{\asca}{{\it ASCA}}
\newcommand{\cgro}{{\it CGRO}}
\newcommand{\nustar}{{\it NuSTAR}}
\newcommand{\Suzaku}{{\it Suzaku}}
\newcommand{\sax}{{\it BeppoSAX}}
\newcommand{\rosat}{{\it ROSAT}}
\newcommand{\exosat}{{\it EXOSAT}}
\newcommand{\einstein}{{\it Einstein}}
\newcommand{\ginga}{{\it Ginga}}
\newcommand{\hst}{{\it HST}}
\newcommand{\swift}{{\it Swift}}
\newcommand{\agile}{{\it AGILE}}

\title[The soft \gr-ray pulsar population]{The soft $\gamma$-ray pulsar population: an high-energy overview}
\author[L. Kuiper and W. Hermsen]{L. Kuiper$^{1}$\thanks{E-mail:
L.M.Kuiper@sron.nl (LK); W.Hermsen@sron.nl (WH)} and W. Hermsen
$^{1,2}$\footnotemark[1]\\
$^{1}$SRON-National Institute for Space Research, Sorbonnelaan 2, 
3584 CA, Utrecht, The Netherlands\\
$^{2}$Astronomical Institute ``Anton Pannekoek", University of 
Amsterdam, Science Park 904, 1098 XH Amsterdam, The Netherlands}
\begin{document}

\date{Accepted ..... Received 2014 December 11; in original form 2014 December 11}

\pagerange{\pageref{firstpage}--\pageref{lastpage}} \pubyear{2014}

\maketitle

\label{firstpage}

\begin{abstract}
At high-energy \gr-rays ($>100$ MeV) the Large Area Telescope (LAT) on the \fermi\ satellite  already detected 
more than 145 rotation-powered pulsars (RPPs), while the number of pulsars seen at soft \gr-rays 
(20 keV - 30 MeV) remained small. 
We present a catalogue of 18 non-recycled RPPs from which presently non-thermal pulsed emission has been securely 
detected at soft  \gr-rays above 20 keV, and characterize their pulse profiles and energy spectra. 
For 14 of them we report new results, (re)analysing mainly data from \rxte, \integral,
\xmm\ and \cxo. The soft \gr-pulsars are all fast rotators and on average $\sim 9.3\times$ younger and  
$\sim 43 \times$ more energetic than the \fermi\ LAT sample.
The majority (11 members) exhibits broad, structured single pulse profiles, and only 6 have double (or even multiple, 
Vela) pulses. Fifteen soft \gr-ray pulsar show hard power-law spectra in the hard X-ray band and reach maximum 
luminosities typically in the MeV range. For only 7 of the 18 soft \gr-ray pulsars pulsed emission has also been 
detected by the LAT, but 12 have a pulsar wind nebula (PWN) detected at TeV energies. For six pulsars with PWNe, 
we present also the spectra of the total emissions at hard X-rays, and for \igrb, associated with HESS J1849-000 and \psrigrb, 
we used our \cxo\ data to resolve and characterize the contributions from the point-source and PWN.
Finally, we also discuss a sample of 15 pulsars which are candidates for future detection of pulsed soft \gr-rays, 
given their characteristics at other wavelengths.
\end{abstract}

\begin{keywords}
radiation mechanisms: non-thermal -- 
stars: neutron --
X-rays: general --
gamma-rays: general --
pulsars: individual: \psrf, \crab, \psrg, \lmcpsr, \vela, \igra, \msh, \psrc, \psrcanda, \psrb, \psrcandb, \psri, \axj, \psra, \igrb, \psrd, \psrh, \psre, 
PSR J1119-6127, PSR J1124-5916, PSR J1357-6429, PSR J1833-1034, PSR J1420-6048, PSR J1418-6058, PSR J1023-5746, PSR J1838-0537, PSR J1826-1256, PSR J1301-6305, 
PSR J1341-6220, PSR J1614-5048, PSR J1803-2137, PSR J1856+0245, IGR J11014-6103

\end{keywords}

\section{Introduction}
Rotation-powered pulsars (RPPs) form a well-established galactic \gr-ray source population known to emit steadily from radio frequencies up to 
high-energy \gr-rays. Despite decades of theoretical modeling the origin and nature of the non-thermal high-energy ($\ga 10$ keV) emission are still being debated. 
The most developed models assume as production site of the high-energy emission either a location in the pulsar magnetosphere near (or starting near) 
the magnetic pole (polar cap models \citep{harding2005}; slot gap models \citep{muslimov2004, dyks2003})
or in vacuum gaps along the last closed field lines in the outer magnetosphere (outer gap models \citep{cheng1986, romani1996, cheng2000}), 
or even in the stripe of the pulsar wind outside the light cylinder radius \citep{petri2012a}.
Modeling its emission characteristics (e.g. phase-resolved spectra, pulse morphology and polarization) drives ever-more sophisticated electrodynamic calculations \citep[e.g.][]{wang2011,kalapotharakos2012a,li2012,petri2012b,hirotani2015}. Simulations have been performed for the 
different geometrical models to compute beaming patterns and light curves as a function of magnetic inclination angle and viewing direction, 
resulting in atlasses of gamma-ray profiles and radio lags \citep[e.g.][]{watters2009,romani2010,bai2010a,bai2010b,kalapotharakos2012b}.
These recent theoretical developents were driven by the major observational progress achieved by the Large Area Telescope (LAT; 20 MeV - 300 GeV ) 
aboard the \fermi\ Gamma-ray Space Telescope, launched June 11, 2008. After more than four years of operations \fermi\ LAT had detected and
characterized the pulsed emission
\citep[see e.g.][]{abdo2010a,ray2011a,abdo2013} of about 145 pulsars (status May 8, 2013), a significant fraction of these is \gr-ray only i.e. 
Geminga-like with no obvious counterparts at other (less energetic) wavelengths regimes. This number can be compared with only seven pulsars
securely detected above 100 MeV with EGRET on the preceding Compton Gamma-Ray Observatory (\cgro; April 5, 1991 -- June 4, 2000).

The observational progress at softer \gr-rays proceeded much slower. After the \cgro\ mission (till $\sim$ 2005), at soft \gr-rays above 20 keV up to $\sim 30$ MeV the pulsed signals from only four pulsars had been detected, hampered mainly by the intrinsic source weaknesses in this energy band with the relatively poorer sensitivities of the operational instruments. These were the Crab (\crab) and the Vela (\vela) pulsars, detected up to GeV energies, \msh\ in MSH 15-52, detected up to $\sim 10$ MeV, and the twin of the Crab pulsar in the Large Magellanic Cloud (LMC) \lmcpsr\ detected up to $\sim 50$ keV (see for references section 5).

Significant progress could be achieved in the hard X-ray/soft \gr-ray band when the high-resolution imaging (\cxo) and large sensitive area X-ray observatories (\rxte, \xmm,
\Suzaku\ and recently \nustar) became available, allowing the detections of weak energetic point-sources at soft/medium X-rays (0.1-10 keV) in young supernova 
remnants (SNRs), often discovered at radio-frequencies, or in location-error boxes of unidentified \cgro\ EGRET/\fermi\ LAT ($\ga 100$ MeV), \integral\ ISGRI (20-300 keV) or H.E.S.S./VERITAS/MAGIC ($\ga 30/100$ GeV) sources. In the X-ray band below $\sim$ 10 keV, the instruments aboard these observatories were sufficiently
sensitive to allow for the measurements of pulsar ephemerides, when there were no radio ephemerides available (many are radio-weak or even radio-quiet). For energies above 20 keV, the operational instruments were relatively insensitive, and could not detect independently the pulsed signals. But, once these timing solutions were determined at lower energies, phase folding of the event arrival times measured with \rxte\ PCA \& HEXTE and \integral\ ISGRI in the hard X-ray band could be performed to search for and reveal the pulsed signals above 20 keV.

Currently, there are 18 pulsars for which pulsed emission has been detected in the soft \gr-ray band above 20 keV up to $\sim 50-150$ keV.
In this paper we present the catalogue of soft \gr-ray pulsars, and for each entry an overview with the latest observational results.  
For 14 of them we present in this work new results, mainly from analyses of archival \rxte\ PCA \& HEXTE and \integral\ ISGRI data for energies 
above 20 keV, but also from analyses of \xmm, \cxo\ and \asca\ data below 10 keV, or \cgro\ COMPTEL and BATSE data at MeV energies. 
For one pulsar and its nebula, the \integral\ source 
\igrb/\psrigrb\, we present the results from our follow-up \cxo\ observations of this source together with new results from analyses 
of archival \xmm, \rxte\ and \integral\ data.

The structure of the paper is as follows. Section 2 describes the \fermi\-LAT high-energy \gr-ray pulsar population of which the hard non-thermal 
power-law spectra are expected to extrapolate into the soft gamma-ray band. We will therefore compare the average characteristics of the 
pulsars in the two catalogues in the discussion. Section 3 presents briefly the properties of the instruments used in this work, and section 4 briefly 
the timing-analysis methods. In section 5, a status review is given for each of the 18 pulsars with new results for 14 of them. Section 6 presents the observational 
status for very young ($\la 15$ kyr) and energetic ($L_{sd} \ga 10^{36}$ erg/cm$^{2}$ s) pulsars so far not detected at hard X-rays, but that are good candidates 
for future detections of pulsed soft \gr-rays, given their characteristics at other wavelengths.
Finally, in section 7 the results are summarized and discussed, including the above-mentioned comparison between the \fermi-LAT and soft \gr-ray pulsar 
populations. 

\section{The \fermi-LAT high-energy ($>100$ MeV) gamma-ray pulsar population}

The launch of \fermi\ heralded a new era in high-energy gamma-ray pulsar research.
Analyzing the first three years of observations, \citet{abdo2013} present the second \fermi\ LAT pulsar catalogue, listing
117 gamma-ray pulsars (an additional $28$ were discovered by the LAT collaboration and 
other groups during the preparation of the catalogue). The 117 pulsars are categorized in three groups: 
radio-loud pulsars (number 42, obtained through pulse-phase folding using radio-ephemerides), radio-quiet pulsars 
(number 35, discovered at \gr-energies through blind searches) and millisecond (i.e. recycled) pulsars (number 40). All have spin-down 
powers $L_{sd}$ above $3\times 10^{33}$ erg/s, which seems to be a lower-limit for detectability as \gr-ray ($>100$ MeV) pulsar.

The recycled millisecond \gr-ray pulsars are {\it not} addressed in this work, although some of the most energetic 
($L_{sd} > 1 \times 10^{35}$ erg/s) ones like PSR J0218+4232, PSR B1937+21 and PSR B1821-24, 
have been detected earlier as hard X-ray pulsars with {\it pulsed} emission detected up to 20--30 keV  \citep{kuiper2003a,kuiper2004}.

Of the $77\, (\,= 42 + 35)$ young/middle-aged ($\tau_{c} \la 4 \times 10^{6}$ y) non-recycled pulsars, 58 ($\sim 75 \%$) show 
at high-energy \gr-rays pulse profiles with two strong, well separated, caustic peaks, with often significant bridge emission. 
The radio phase lag $\delta$ between the \gr-ray leading pulse and the radio main pulse (often fiducial point) and the \gr-ray peak 
separation $\Delta$ show a strong anti-correlation, being a predicted general property of outer-magnetosphere models with caustic pulses 
\citep[see e.g. Fig. 3 of][]{romani1995}. 
On average the \gr-ray peak separation is in the 0.4--0.5 range (about half a rotation), while the radio lag distribution reaches 
its maximum in the 0.15 - 0.25 range. 

In the spectral domain the pulsed emission of the young/middle-aged \gr-ray pulsars can well be represented by a power law model 
in combination with a simple exponential cutoff , i.e. \(F_{\gamma}= k \cdot (E_{\gamma}/E_0)^{\Gamma}\cdot \exp(-(E_{\gamma}/E_c)^{\beta})\) 
with $\beta \equiv 1$ and $E_{\gamma}$, the photon energy, $E_0$, the pivot energy minimizing the co-variance between the parameters, 
$\Gamma$, the photon index, $E_c$, the cutoff energy and $k$ the normalization. 

Outer-gap and slot-gap models predict the production of \gr-rays at high altitudes in the magnetosphere with such a simple exponential cut-off
($\beta\sim 1$), while in polar cap models the production occurs near the magnetic poles with a sharp hyper-exponential turn over ($\beta\sim 2$)
of the spectrum at high energies. The \fermi\ spectral results thus disfavours the polar-cap models (see e.g. the detailed discussions of the 
fits to the Vela pulsar spectrum \citep{abdo2009c} and the Crab pulsar spectrum \citep{abdo2010c}).

The cutoff energies $E_c$ are on average in the 1--3 GeV range, while the photon index distribution peaks in the -1.8 -- -1.3 range. 

\citet{abdo2013} also considered the properties of the LAT detected pulsars in the soft X-ray band (0.3--10 keV). Though the X-ray coverage 
is rather uneven, with the newly discovered LAT pulsars having often only snapshots observations with \swift\ XRT, while deep observations with 
\cxo\ and/or \xmm\ exist for the well-know gamma-ray pulsars, all LAT pulsars do have X-ray coverage. Of the 77 young/middle-aged LAT 
pulsars 49 (30 radio-loud; 19 radio-quiet) have an X-ray counterpart with a non-thermal (i.e. flagged with a power law) spectral component 
obtained through model fitting in terms of a power-law in combination with one (or even two) blackbody(ies). 
The X-ray counterpart association was established either through the detection of X-ray pulsations or positional coincidence with accurately 
determined radio- or LAT timing positions.

For only 21\footnote{Table 15 of \citet{abdo2013} labeled 13 LAT pulsars with ``P'' (Pulsed X-ray emission detected), omitting by mistake 
this label for \msh\ (PSR J1513-5908, \citet{seward1982}) and PSR J2021+3651 \citep{hessels2004}, while pulsed X-ray emission has been 
detected for 6 LAT pulsars; PSR J1459-6053 \citep{pancrazi2012}, PSR J2021+4026 \citep{lin2013}, PSR J1741-2054 \citep{marelli2014}, 
PSR J1813-1246 \citep{marelli2014b}, PSR J1836+5925 \citep{arumugasamy2014b,lin2014} and \lmcpsr\ \citep{martin2014} after the publication 
of the catalogue.} of the young/middle-aged LAT pulsars the {\it pulsed\/} fingerprints are detected in the canonical X-ray band ($< 10$ keV).

\section{Instruments}

In this section we discuss briefly the general characteristics of the instruments aboard \rxte, \integral, \xmm\ and \cxo, that 
we extensively used in this work to obtain new results on the (hard) X-ray/soft \gr-ray properties of the pulsar sample presented in this paper.
For the Vela pulsar (\vela) and \msh\ we obtained new pulse profiles analysing archival \cgro\ COMPTEL 0.75--30 MeV and \cgro\ BATSE 317--1102 keV data,
respectively, from observations covering the full \cgro\ mission lifetime. For \psre\ archival \asca\ GIS data were analysed to obtain new pulse 
profiles and a new spectrum for the pulsed emission.
For the latter instruments references are given in those sections where data from these instruments are analysed by us.

\subsection{Rossi X-ray Timing Explorer (\rxte)}
In this work extensive use has been made of data from observations
of rotation powered pulsars with the two {\it non-imaging} X-ray instruments aboard \rxte, the Proportional Counter Array 
(PCA; 2-60 keV) and the High Energy X-ray Timing Experiment (HEXTE; 15-250 keV). 

\subsubsection{\rxte\ PCA}
\label{pca_char}
The PCA \citep{jahoda96} consisted of five collimated Xenon proportional 
counter units (PCUs) with a total effective area of $\sim 6500$ cm$^2$ over a $\sim 1\degr$ 
(FWHM) field of view. Each PCU had a front Propane anti-coincidence layer and three Xenon 
layers which provided the basic scientific data, and was sensitive to 
photons with energies in the range 2-60 keV. The energy resolution was about 18\% at 6 keV.
All PCA data used in this study have been collected from observations in {\tt GoodXenon} mode 
allowing high-time resolution ($0.9\mu$s) analyses in 256 spectral bins.
Since the launch of \rxte\ on Dec. 30, 1995 the PCA has experienced high voltage breakdowns for 
all constituting PCUs at irregular times. To reduce further breakdowns during its mission, which ended on Jan. 5, 2012, 
not all PCUs were simultaneously operating. The most stable PCU was PCU-2, which was
on for almost all of the time. On average one (50\%) or two (40\%) PCUs was/were operational during
a typical observation.

\subsubsection{\rxte\ HEXTE}
\label{hexte_char}
The HEXTE instrument \citep{rothschild98} consisted of two independent detector 
clusters A\& B, each containing four Na(Tl)/ CsI(Na) scintillation
detectors. The HEXTE detectors were mechanically collimated to a $\sim 1\degr$ (FWHM) 
field of view and covered the 15-250 keV energy range with an energy resolution of 
$\sim$ 15\% at 60 keV. The collecting area was 1400 cm$^2$ taking into account the 
loss of the spectral capabilities of one of the detectors. The maximum time 
resolution of the tagged events was $7.6\mu$s. In its default operation mode the 
field of view of each cluster was switched on and off source to provide instantaneous 
background measurements. However, also HEXTE suffered from aging, and since July 13, 2006
HEXTE cluster-A operated in staring mode at an on-source position, while
on March 29, 2010 cluster-B was commanded to stare at an off-source position.
Due to the co-alignment of HEXTE and the PCA, celestial targets have been observed simultaneously.

\subsection{\integral}
\label{instr_integral}
The \integral\ spacecraft \citep{winkler03}, launched 17 October 2002, carries two main 
$\gamma$-ray instruments: a high-angular-resolution imager IBIS \citep{ubertini03} and
a high-energy-resolution spectrometer SPI \citep{vedrenne03}. The payload is further
equiped with an X-ray monitoring instrument, the Joint European Monitor for X-rays 
\citep[JEM-X; ][]{lund2003} providing complementary observations in the X-ray band. 
These three high-energy instruments make use of coded aperture masks enabling 
image reconstruction in the hard X-ray/soft $\gamma$-ray band.

In this work, guided by sensitivity considerations, we only used data recorded by the \integral\ 
Soft Gamma-Ray Imager ISGRI \citep{lebrun03}, the upper detector system of IBIS, sensitive 
to photons with energies in the range $\sim$15 keV -- 1 MeV (effectively about 300 keV).
 
With an angular resolution of about $12\arcmin$ and a source location accuracy of better than 
$1\arcmin$ (for a $>10\sigma$ source) ISGRI is able to locate and separate high-energy sources 
in crowded fields within its $19\degr \times 19\degr$ field of view (50\% partially coded) with 
an unprecedented sensitivity ($\sim$ 960 cm$^2$ at 50 keV).
Its energy resolution of about 7\% at 100 keV is amply sufficient to determine the (continuum) spectral 
properties of hard X-ray sources in the $\sim$ 20 - 300 keV energy band.

The timing accuracy of the ISGRI time stamps recorded on board is about $61\mu$s. The time 
alignment between \integral\ and \rxte\ is better than $\sim 50\mu$s, verified using data 
from simultaneous \rxte\ and \integral\ observations of the accretion-powered millisecond pulsars 
IGR J00291+5934 and IGR J17511-3057 \citep[][respectively]{falanga05,falanga11}.

\subsection{\xmm}

\xmm\ launched Dec. 10, 1999, carries three different Wolter type-1 multi-shell grazing incidence 
mirrors focusing X-rays to different X-ray instruments. One of the telescopes focusses the X-rays directly
on an EPIC pn type CCD camera, while the other two telescopes have reflection grating arrays (RGA) in their light 
paths dispersing by reflection about 40\% of the incoming X-ray radiation to a linear strip of CCDs 
(RGS detector) at their secondary foci, while about 44\% of the incoming X-ray radiation passes to the prime foci 
onto MOS type CCD cameras. With the EPIC cameras very sensitive imaging observations can be performed over a 
$30\arcmin$ (diameter) field of view across the 0.15-12 keV band with a spectral resolution ($E/\Delta E$) of 20--50 and angular
resolution of $\sim 6\farcs6$ (FWHM).

In this work we only used data collected by the EPIC pn instrument \citep{struder2001}, consisting of 12 back-illuminated CCDs,
because the operation modes of this CCD camera are well suited for accurate studies of timing phenomena 
occuring at milli second scales. In particular, the Small window mode offers a fast read-out/frame time of 
about 5.67 ms, amply sufficient for most of the soft \gr-pulsars studied in this work, while 
imaging information is kept only for the central CCD number 4 in a frame of $63 \times 64$ pixels
of $4\farcs1 \times 4\farcs1$ size each offering thus a $4\farcm3 \times 4\farcm37$ field of view.

For one pulsar in our sample, \psrg, we analyzed the EPIC pn data collected in timing mode, in which the time stamps are registered
at a $\sim 30\mu$s resolution, while one-dimensional imaging information is conserved.
The other pulsars for which we analyzed \xmm\ EPIC pn data were  \psrf, \vela, \psrcandb, \psri\, and \igrb, and in these cases 
the data were taken in Small window mode.

\subsection{\cxo}

The \cxo\ X-ray Observatory CXO \citep{weisskopf2000} was launched on July 23, 1999, and combines a sub-arcsecond resolution X-ray telescope with advanced imaging and spectroscopic instruments in the focal plane of the High Resolution Mirror Assembly (HRMA),
which is composed of a nested set of four Wolter type-1 grazing-incidence X-ray mirror pairs with an outer diameter of 1.2 m and focal length of 10 m. Within the optical path (between the HRMA and focal plane instruments) two different types of grating assemblies, the High-Energy Transmission Grating (HETG) - comprised of a High Energy Grating (HEG) and Medium Energy Grating 
(MEG) on a single structure - and the Low Energy Transmission Grating (LETG), can be placed mechanically on command.

The Science Instrument Module (SIM) at the focal plane houses two instruments, the Advanced CCD Imaging Spectrometer (ACIS) and High Resolution Camera (HRC). 

The ACIS instrument \citep{garmire2003} contains 10 planar, $1024 \times 1024$ pixels (pixel scale is $0\farcs4920\pm 0\farcs0001$) CCDs of which 4 are arranged in a $2 \times 2$ array ($16\farcm9 \times 16\farcm9$ field of view) for imaging purposes (ACIS-I) and the other 6 in a $1 \times 6$ array ($8\farcm3 \times 50\farcm6$ field of view) serving either as grating readout or as imaging detector (ACIS-S; containing two back-illuminated CCDs). 

The energy resolving power ($E/\Delta E$) is moderate and depends on the CCD type (front or back-illuminated), is energy and time (Charge Transfer Inefficiency) dependent, and is in the range $\sim$10--60 for the FI CCDs and $\sim$4--45 for the BI CCDs. The BI type CCDs ($\sim$ 0.18-10 keV) have higher effective area at lower energies than the FI type CCDs ($\sim$0.48-10 keV). The effective area reaches a maximum of about 600 cm$^2$ at 1.4 keV for the FI CCDs. The nominal frame time in Timed Exposure (TE) mode is 3.2 s with a 40.96 ms read-out time. This relatively long integration time can cause pile-up (more than one event registered in a pixel during a frame time) effects even for moderate count rates.

Although in TE mode other frame times are possible using subarray read-outs, the default TE mode is not suitable for pulsar timing studies at millisecond time scales, but it is very useful to obtain sub-arcsecond (energy resolved) images of the pulsar environment. The other operation mode of the ACIS is Continuous Clocking (CC). In this mode the integration time is 2.85 ms, and this is reached at the expense of spatial information in one direction. 

The other focal plane instrument is the HRC \citep{murray2000}, which is comprised of two multichannel plate imaging detectors: the HRC-I offering wide-field imaging ($30\arcmin \times 30\arcmin$ field of view with 0\farcs13175/pixel), and the HRC-S serving mainly as readout for the LETG. The HRC detectors do not/hardly have energy resolving power. Due to a wiring problem the time resolution of the HRC-I/S
detector is governed by the actual count rate and is much worse than the anticipated resolution of about 16 $\mu$s. However, the central segment of the HRC-S (6\arcmin $\times$ 30\arcmin field of view) can be operated in a special mode, designed for accurate timing studies, in which the original 16 $\mu$s time resolution can be recovered.

In this work we present (PI-obtained) \cxo\ data for \igrb\ from two 25 ks observations, one with the HRC-S in timing mode and one with the ACIS-S. Moreover, we have estimated the genuine pulsed fraction of \psrc\ using \cxo\ HRC-S timing mode data. 

\section{Analysis methods}

In the analysis of the high-energy data from the non-imaging \rxte\ instruments PCA and HEXTE, the coded-mask imager ISGRI aboard \integral\ and 
the X-ray telescopes aboard \cxo\ and \xmm\ some common analysis procedures exist. To obtain pulsed flux measurements accurate 
timing models (ephemerides) are required describing every revolution of the neutron star accurately. These so-called phase coherent ephemerides 
are mainly based, in this work, on Time-of-Arrival (ToA) determinations of a pulse template using \rxte\ PCA data, and are listed in 
Table \ref{eph_table} for a limited set of soft \gr-ray pulsars, for which \rxte\ PCA monitoring observations exist. 
The method to obtain accurate timing models is extensively described in Sect. 4.1 of \citet{kuiper2009}.

Irrespective of the high-energy instrument with which the data have been collected a typical timing analysis consists of 1) data selection 
using e.g. the HEASARC browse interface for \rxte, \xmm\ and \cxo\ observations fulfilling the source search criteria or the browse 
facility at the ISDC for \integral\ data, 2) screening the data by creating good-time intervals (GTI), 3) barycentering the event time tags 
using the instantaneous spacecraft ephemeris (position and velocity) information, the JPL solar system ephemeris information (DE200/DE405) 
and an accurate source position to convert the times from Terrestial Time scale (TT or TDT, which differs from Coordinated Universal Time 
(UTC) by a number of leap seconds and a fixed offset of 32.184 s) into Barycentric Dynamical Time (TDB) scale, a time standard for 
Solar system ephemerides. 

Subsequently, the TDB time tags are converted to pulse 
phases \citep[see formula (1) in Sect. 3.4 of][]{kuiper2010} adopting appropriate timing models (phase folding). Next, the events are 
sorted on pulse phase and energy to obtain 2d event distributions, storing thus pulse profiles for any energy band covering the passband 
of the involved instrument. The detection of a pulsed signal is expressed in a number of $\sigma$'s applying the $Z_n^2$ test described
in \citet{buccheri1983}.
Typical timing analyses of \rxte\ PCA, HEXTE, \xmm\ and \integral\ ISGRI data are described in Sect. 3.1, Sect. 3.7 and Sect. 3.8 of 
\citet{kuiper2010} and Sect. 4.3 of \citet{kuiper2009}, respectively. A typical timing analysis for \cxo\ HRC data is given in Sect. 4 
of \citet{kuiper2002} and Sect. \ref{sectigrj1849} of this work. 

From the pulse-phase distributions per energy band we derived the number of pulsed excess counts by fitting with (a) pulse profile template(s) or
a truncated Fourier series. The pulsed fluxes were obtained by applying forward folding spectral fitting procedures taking into account the 
instrument spectral response, effective area, exposure time and absorption in the interstellar medium. 
This is explained for example in Sect. 5 of \citet{kuiper2009} for PCA, HEXTE and ISGRI data. 
For \xmm\ \citep[see e.g. Sect. 4 of][]{kuiper2010} and \cxo\ equivalent procedures exist.

\section{The soft \gr-ray pulsar population}

\subsection{\psrf}

\psrf\ is a young ($\tau \sim 5.4$ kyr) 65 ms pulsar located near the center of super nova remnant/pulsar wind nebula 3C58 with a spin-down luminosity of $2.7 \times 10^{37}$ erg/s. Its pulsation was discovered in the X-ray band using \cxo\ data and confirmed using \rxte\ PCA data \citep{murray2002}, and its weak pulsed signal was later detected at radio frequencies.

At TeV-energies ($> 300$ GeV) VERITAS did not (yet) detect emission from the sky region including 3C58, yielding a $99\%$ flux upper-limit of about $2\%$ of the Crab nebula \citep{humensky2009}. However, the upgraded stereoscopic MAGIC telescope detected 3C58 as unresolved source 
at a $5.7\sigma$ level with an integral flux above 1 TeV of about 0.65\% of the Crab \citep{aleksic2014b}.   

Detailed timing studies in the X-ray and radio-bands have been performed by both \citet{livingstone2009} and \citet{kuiper2010}. The X-ray pulse profile shows two sharp peaks separated in phase by $0.488\pm 0.002$ with the main peak lagging the single radio pulse by 0.089 in phase. The pulsed X-ray/soft $\gamma$-ray spectrum over the energy band 0.56-267.5 keV (based on \xmm\ EPIC-pn, \rxte\ PCA and HEXTE measurements) for the sum of the two pulses is very hard with photon index $-1.03\pm 0.02$ \citep{kuiper2010}. No difference in spectral shape is found for the two pulses in the X-ray/soft $\gamma$-ray band. High-energy $\gamma$-ray pulsations ($>100$ MeV) from
\psrf\ have been detected using \fermi\ LAT data \citep{abdo2009a}. The  $\gamma$-ray lightcurve ($>100$ MeV) shows also two pulses, aligned with the X-ray pulses, but now the strongest $\gamma$-ray pulse coincides with the weakest X-ray pulse, indicating
different spectral behaviours of the two pulses when considering the full 0.5 keV to 10 GeV band. For more detailed information about the high-energy characteristics of the pulsed emission we refer to \citet{kuiper2010} and \citet{abdo2009a}.

\subsection{The Crab pulsar, \crab\ / \crabj}

The 33 ms pulsar, \crab\ or \crabj, in the Crab nebula (3C 144; NGC 1952; CTB 18) was discovered in 1968 by \citet{staelin1968}. 
Since its detection at radio-frequencies the pulsed fingerprint of \crab\ has been detected across the full electro-magnetic spectrum from radio, optical/infra-red, 
soft X-rays (0.1-2 keV), hard X-rays (2-20 keV), soft \gr-rays (20-500 keV), medium-energy \gr-rays \citep[0.5 MeV - 30 MeV, see e.g.][for a high-energy picture from 
soft X-rays up to high-energy \gr-rays]{kuiper2001}, high-energy \gr-rays \citep[30 MeV - 100 GeV, see e.g.][]{abdo2010c} and recently even at TeV energies 
\citep[$>$ 0.1 TeV, see e.g.][]{aliu2008,aleksic2011,aleksic2014a,veritas2011}.

An intriguing feature of its pulsed emission is that the double-peaked pulse-profile with a peak separation of about 0.4 (in phase) keeps its identity and preserves its alignment
approximately (the X-ray/\gr-ray main pulse preceeds the radio main pulse by about $300\mu$s) across the electro-magnetic spectrum \citep[see e.g. Fig. 6 of][]{kuiper2003}, though the relative intensities of the two pulses vary.

The total emission of the Crab (nebula and pulsar) has long time served as flux/spectral calibration target for various instruments because of its detectability across the `full' electro-magnetic spectrum and its stability over time. The latter property has been abandoned recently by the discovery of considerable variations in the hard X-ray/soft \gr-ray band \citep{wilson2011} and flaring activity at high-energy \gr-rays \citep{tavani2011,abdo2011}. Contrarily, the pulsed emission of \crab\ as seen by \rxte\ PCA (3.2-35 keV) seems rather stable and decreases steadily consistent with the pulsar spin-down \citep{wilson2011}. 

The high-energy spectrum of the pulsed emission of \crab\ across 8 decades in energy from soft X-rays up to high-energy \gr-rays is shown in the multi-source spectral compilation depicted in 
Fig. \ref{spectralcompilation} as red-coloured datapoints. Data from \cgro\ OSSE, COMPTEL and EGRET, \sax\ LECS, MECS and PDS and the 
balloon-borne GRIS instruments are included \citep[see][for more details]{kuiper2001}, supplemented by \rxte\ PCA (5-50 keV, {\tt E\_250us\_128M\_14\_1s} event mode; this work) and HEXTE ($\sim$ 13-235 keV; this work) pulsed flux measurements. The HEXTE data have been collected during observation run 40805 covering 
the time periods March 17--31, 1999 and December 18--19, 1999, for an effective deadtime corrected ON source time of 22.7 ks and 23.8 ks for cluster A and B, respectively. 
Also included in Fig. \ref{spectralcompilation} is the model spectrum of the pulsed emission (100 MeV -- 10 GeV) as obtained by \cite{abdo2010c} using \fermi\ LAT data. 
Maximum pulsed flux is reached near 100 keV, while beyond 30 MeV a flattening occurs, followed beyond $\sim$ 10 GeV by a power-law like decay, 
detected up to about 400 GeV by the VERITAS TeV telescope \citep{veritas2011}.

The (unabsorbed; using a N$_{\hbox{\scriptsize H}}$ of $3.61 \times 10^{21}$ cm$^{-2}$) 2--10 keV, 20--100 keV, 1--10 MeV and 0.1--1 GeV pulsed fluxes for the Crab pulsar are $(1.948\pm 0.005)\cdot 10^{-9}$, $(2.766\pm 0.008)\cdot 10^{-9}$, $(1.36\pm 0.05)\cdot 10^{-9}$ and $(7.87\pm 0.28)\cdot 10^{-10}$ erg/cm$^2$s, respectively, derived from \rxte\ PCA/HEXTE data (2--10 and 20--100 keV, this work), \cgro\ COMPTEL \citep[see e.g.][from the pulsed flux measurements reported in that work we derived the 1--10 MeV pulsed flux value]{kuiper2001} and \fermi\ LAT \citep[flux value derived from the spectral parameters and their $1\sigma$ statistical errors as given in Sect. 4.3 of][]{abdo2010c}.

\begin{table*}
\caption{Phase-coherent ephemerides, derived in this work, for the subset of soft $\gamma$-ray pulsars for which \rxte\ PCA monitoring data exist.}
\label{eph_table}
\begin{center}

\begin{tabular}{lccclllc}
\hline
Pulsar &  Start &  End  &   t$_0$, Epoch   & \multicolumn{1}{c}{$\nu$}   & \multicolumn{1}{c}{$\dot\nu$}               & \multicolumn{1}{c}{$\ddot\nu$}                  &  Validity range\\
 \#   &  [MJD] & [MJD] &     [MJD,TDB]    & \multicolumn{1}{c}{[Hz]}     & \multicolumn{1}{c}{$\times 10^{-11}$ Hz/s}  & \multicolumn{1}{c}{$\times 10^{-22}$ Hz/s$^2$}  &  \multicolumn{1}{c}{(days)}   \\
\hline
\psrg     & 51197  & 51263 & 51237.0      &  62.0402460583(64)    & -19.9221(2)                 &  99(40)               & 67\\
\psrg     & 51294  & 51423 & 51366.0      &  62.0380675984(19)    & -19.92497(3)                &  84(5)                &130\\
\psrg     & 51422  & 51547 & 51480.0      &  62.0361053238(9)     & -19.92024(2)                &  46(6)                &126\\
\psrg     & 51576  & 51706 & 51628.0      &  62.0335856000(14)    & -19.92746(3)                &  72(3)                &131\\
\psrg     & 51715  & 51786 & 51761.0      &  62.0313152172(24)    & -19.9279(1)                 & 248(16)               & 72\\
\psrg     & 51795  & 51818 & 51795.0      &  62.03072978(2)       & -19.920(2)                  &   0.0                 & 24\\
\psrg     & 51833  & 51875 & 51833.0      &  62.0300843300(32)    & -19.9294(2)                 &   0.0                 & 43\\
\psrg     & 51886  & 51955 & 51886.0      &  62.02918045(1)       & -19.9357(8)                 & 276(30)               & 70\\
\psrg     & 51964  & 51996 & 51964.0      &  62.0278657375(87)    & -19.9354(7)                 &   0.0                 & 33\\
\psrg     & 51997  & 52145 & 52049.0      &  62.0264019706(14)    & -19.92818(3)                &  63(2)                &149\\
\psrg     & 52132  & 52166 & 52132.0      &  62.0249729876(73)    & -19.9153(6)                 &   0.0                 & 35\\
\psrg     & 52175  & 52230 & 52175.0      &  62.0242443562(98)    & -19.9364(9)                 & 167(38)               & 56\\
\psrg\dag & 52252  & 52368 & 52252.0      &  62.0229447197(59)    & -19.9342(2)                 &  68(5)                &117\\
\psrg     & 52389  & 52446 & 52389.0      &  62.02059598(1)       & -19.936(1)                  & 134(56)               & 58\\
\psrg     & 52459  & 52540 & 52459.0      &  62.019403934(12)     & -19.9359(7)                 & 103(22)               & 82\\
\psrg     & 52551  & 52648 & 52551.0      &  62.0178456346(56)    & -19.9374(3)                 &  94(7)                & 98\\
\psrg     & 52647  & 52718 & 52647.0      &  62.016192268(14)     & -19.9327(9)                 & 148(29)               & 72\\
\psrg     & 52745  & 52792 & 52745.0      &  62.0145138255(43)    & -19.9360(3)                 &   0.0                 & 48\\
\psrg     & 52822  & 52884 & 52822.0      &  62.013203356(14)     & -19.9401(9)                 & 141(36)               & 63\\
\psrg     & 52889  & 53008 & 52889.0      &  62.0120638182(43)    & -19.9409(2)                 & 132(3)                &120\\
\psrg     & 53019  & 53122 & 53019.0      &  62.0098457965(56)    & -19.9405(3)                 &  95(7)                &104\\
\psrg     & 53128  & 53143 & 53128.0      &  62.007969319(34)     & -19.939(6)                  &   0.0                 & 16\\
\psrg     & 53147  & 53285 & 53147.0      &  62.0076662548(52)    & -19.9426(2)                 &  95(3)                &139\\
\psrg     & 53290  & 53366 & 53290.0      &  62.005227527(14)     & -19.9488(8)                 & 240(29)               & 77\\
\psrg     & 53365  & 53444 & 53365.0      &  62.003935314(18)     & -19.9404(9)                 & 192(30)               & 80\\
\psrg     & 53446  & 53549 & 53446.0      &  62.002556305(10)     & -19.9471(4)                 & 165(10)               &104\\
\psrg     & 53551  & 53682 & 53609.0      &  61.9997679030(28)    & -19.94013(3)                & 106(5)                &132\\
\psrg     & 53710  & 53860 & 53792.0      &  61.9966405925(17)    & -19.93914(2)                &  83(3)                &151\\
\psrg     & 53862  & 53947 & 53862.0      &  61.995449381(11)     & -19.9510(5)                 & 230(17)               & 86\\
\psrg     & 53952  & 53996 & 53952.0      &  61.993899826(27)     & -19.945(3)                  & 531(154)              & 45\\
\vspace{-2mm}\\
\psrb     & 52341  & 52541 & 52341.0      &  15.4612089999(14)    & -1.056106(19)               & 0.0                   & 201\\
\psrb     & 52595  & 52904 & 52595.0      &  15.4609772117(10)    & -1.056411(9)                & 0.0                   & 310\\
\psrb     & 52904  & 53333 & 52904.0      &  15.4606951476(22)    & -1.056670(25)               & -2.03(15)             & 430\\
\vspace{-2mm}\\
\axj      & 54513  & 55002 & 54513.0      &  14.1847581888(11)    & -0.992954(11)               & 1.95(6)               & 488\\
\axj      & 55038  & 55537 & 55136.0      &  14.1842449934(1)     & -0.999054(6)                & 3.8(1.2)              & 497\\
\vspace{-2mm}\\
\igrb     & 55525  & 55545 & 55526.0      &  25.961259794(16)     & -0.961(2)                   & 0.0                   & 19\\
\vspace{-2mm}\\
\psrd     & 52529  & 52633 & 52529.0      &  7.30613850(9)        & -4.008(2)                   & 0.0                   & 104\\
\vspace{-2mm}\\
\psrh     & 55223  & 55232 & 55223.0      &  20.58651709(2)       & -3.649(6)                   & 0.0                   & 10\\
\hline
\multicolumn{8}{l}{\dag This entry has been used for \xmm\ EPIC-pn observation 0113020201 in timing mode}\\
\end{tabular}
\end{center}
\end{table*}

\subsection{\psrg}

\citet{marshall1998a,marshall1998b} reported the discovery of a young ($\tau \sim 4.9$ kyr) fast 16 ms pulsar in LMC supernova remnant N157B (NGC 2060) 
using \rxte\ PCA data taken from 2 observation runs in 1996 centered on the nearby SN 1987A and archival \asca\ GIS observations to pin down its location 
and derive its spin-down. So far, \psrg\ is the fastest and most energetic ($\dot{E} \sim 4.9 \times 10^{39}$ erg/s) non-recycled spin-down 
powered pulsar. The pulsed signal was also found in archival \sax\ MECS \citep[2--10 keV;][]{cusumano1998a,cusumano1998b} and \rosat\ HRI data 
\citep[0.1--2.4 keV;][]{wang1998}, and in follow-up \cxo\ HRC-S data \citep{wang2001}.

On January 19, 1999 an intensive \rxte\ monitoring (observation id. 40139) of \psrg\ commenced, which ended with cessation of the \rxte\ mission in early 
January 2012, aimed at following its timing behaviour with time. Analyzing the first 2.6 years of this monitoring campaign resulted in a glitch rate 
of 2.3 per year, the highest seen for any pulsar \citep{marshall2004}. An update on this work was provided by \citet{middleditch2006} analyzing 7.6 years 
of \rxte\ monitoring data, and finding 23 glitches at a rate of 3.3 per year. The latter authors found a correlation between the time interval between 
two glitches and the amplitude of the first glitch (of the interval), yielding a prediction of these intervals with an accuracy of a few days.

\begin{figure}
  \begin{center}
     \includegraphics[width=6cm,height=12cm,angle=0,bb=170 95 420 695]{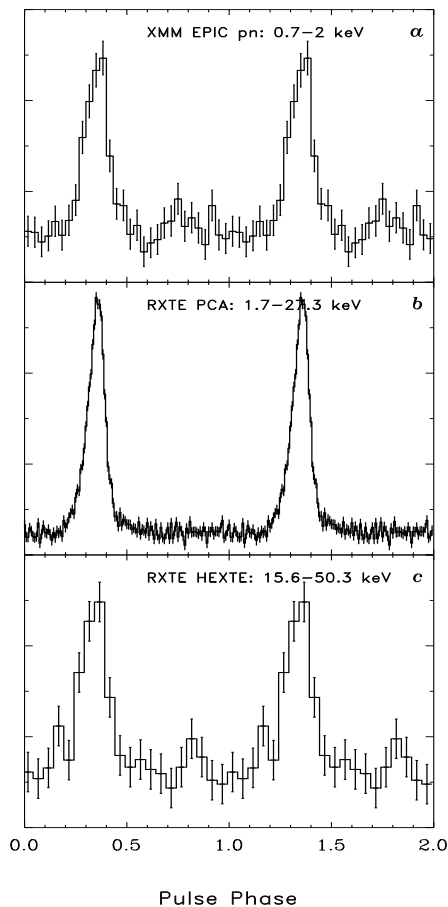}
     \caption{\label{psrj0537_prof}{\bf \psrg;} pulse profiles for \xmm\ EPIC pn (0.7-2 keV; panel a), \rxte\ PCA (1.7-27.3 keV; panel b) and 
     \rxte\ HEXTE (15.6-50.3 keV; panel c). Pulsed emission has been significantly detected up to $\sim 50$ keV.}
  \end{center}
\end{figure}

While its timing characteristics have been studied intensively, the high-energy spectral characteristics of \psrg\ are poorly determined. 
Therefore, we invoked the huge \rxte\ data archive on \psrg\ to establish its high-energy spectral characteristics. 
For this purpose we used a (slightly larger) dataset than \citet{middleditch2006} and revisited their analyses to obtain the alignment for each 
of the 24 timing segments enabling us to co-add the pulse profiles for each of the 24 timing segments directly \citep[Note: this alignment information 
is not given in][]{middleditch2006}\footnote{Analyzing the full \rxte\ database on \psrg\ is outside the scope of this paper, but work on this, 
in particular to derive the characteristics for all the glitches across the full \rxte\ data period, is in progress.}.

We started with the derivation of ToA's for all observations between Jan. 19, 1999 and Oct. 5, 2006 (time span MJD 51197 -- 53996; 7.7 years). 
From these ToA's we determined in a consistent way phase coherent timing models, which are in good agreement with those obtained by \citet{middleditch2006} 
and are listed in Table \ref{eph_table}. 
Across this period we found 24 glitches of which 23 have been reported on by \citet{middleditch2006}, resulting in 24 timing segments for which phase 
coherent solutions (all with proper alignment) have been derived. 

Subsequent pulse-phase folding of PCA events (from all detector layers), screened according to standard criteria and removing break-down periods 
of the detector units, yielded high-statistics 2 dimensional event distributions sorted on pulse-phase and energy (PHA). In Fig. \ref{psrj0537_prof} 
panel b the 120 bin 1.7--27.3 keV profile is shown, analyzing PCA data from time segments 1--18, which provides us already with very high pulsed-signal statistics. 
Similar procedures for HEXTE yielded, now analyzing all 24 segments given the poor pulsed-signal statistics contrary to the PCA case, 
a $\sim 10\sigma$ pulsed signal for the 15.6-50.3 keV band (see Fig. \ref{psrj0537_prof} panel c), and even a 
$5.2\sigma$ signal for the 31.0-50.3 keV band, establishing \psrg\ as a soft \gr-ray pulsar. Indications for pulsed emission were found for 
energies above 50 keV. 

In order to determine the timing characteristics of \psrg\ below $\sim 2$ keV we also analyzed EPIC pn timing mode data from a 35.9 ks \xmm\ 
observation (obs. id. 0113020201) performed on November 19, 2001 (data period MJD 52232.949 - 52233.366). Default screening and optimizing on the 
RAWX event parameter by allowing only events with RAWX in the range $[26,42]$ yielded highly significant pulse profiles after phase folding using 
the \psrg\ ephemeris entry with epoch MJD 52252 (see Table \ref{eph_table})\footnote{\citet{middleditch2006} reported a glitch (\#8 in their 
table 4) occuring at MJD $52241.6\pm 7.8$, however, the \xmm\ data can still be correctly described by the post-glitch ephemeris entry and 
{\it not} with the pre-glitch ephemeris, indicating that the glitch must have occured somewhere between MJD 52230 (the end time of the pre-glitch 
ephemeris) and MJD 52232.949, the start time of the \xmm\ observation.}. The resulting 0.7-2 keV \xmm\ EPIC pn pulse profile is shown in 
Fig.\ref{psrj0537_prof} panel a, and the deviation from uniformity is about $16.4\sigma$ applying a $Z_5^2$-test.

In order to derive the pulsed spectrum of \psrg\ we fitted a model composed of a (flat) background component and a component describing the shape 
of the single pulse, as derived from the high-statistics \rxte\ PCA profile, to the pulse profiles of any of the energy bands in the band width of 
\xmm\ EPIC pn, \rxte\ PCA and HEXTE. The resulting pulsed excess counts are converted to pulsed fluxes in a forward folding procedure assuming 
an underlying (photon) power-law model and adopting interstellar absorption under a column of 
N$_{\hbox{\scriptsize H}}$ of $7 \times 10^{21}$ cm$^{-2}$ \citep[see e.g. Table 4 of][for X-ray spectroscopic information of some sources in 
the 30 Doradus region, including \psrg\ and SNR N157B, using \cxo\ ACIS data]{townsley2006}. 

\begin{figure}
  \begin{center}
     \includegraphics[width=8cm,bb=40 150 545 658]{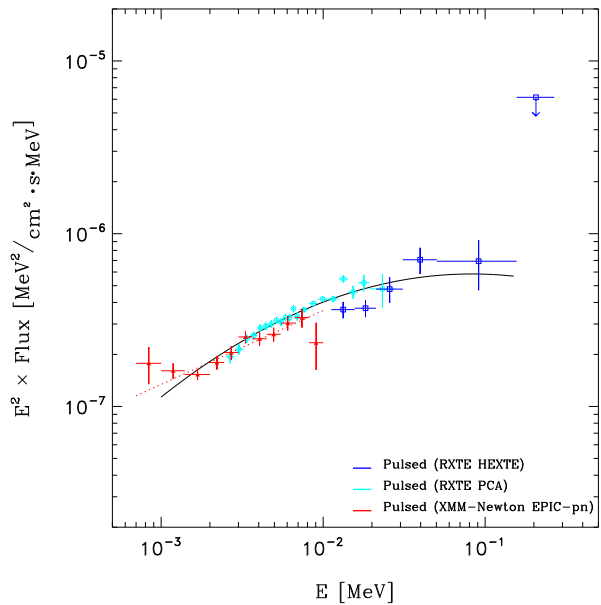}
     \caption{\label{psr0537_spc}{\bf \psrg;} high-energy (0.7-250 keV) unabsorbed pulsed spectrum as derived from
                                measurements by \xmm\ EPIC-pn (0.7-10 keV; red), \rxte\ PCA 
                                (2.5-30 keV; aqua) and \rxte\ HEXTE (15-250 keV; blue). The best fit model 
                                (``curved'' power-law) to all these pulsed 
                                flux measurements is superposed as solid black line. The red dotted 
                                line represents the best power-law fit model to solely EPIC-pn flux measurements.}
  \end{center}
\end{figure}

In Fig. \ref{psr0537_spc} the pulsed flux measurements across the 0.7-250 keV band are shown for \xmm\ EPIC pn (red data points), \rxte\ 
PCA (aqua) and \rxte\ HEXTE (blue) along with fits to solely EPIC-pn data (power-law model; red dotted line; 0.7-10 keV) and to all data 
(``curved'' power-law; black solid line; 0.7-250 keV). The later model provides a $4.3\sigma$ improvement over a simple power-law model 
fit considering the full dataset, and thus proves the necessity of a gradually breaking spectrum. 
The unabsorbed pulsed flux in the 2--10 keV band for the ``curved'' PL-model is $(7.42\pm 0.26)\times 10^{-13}$ erg/cm$^2$s, about 10\% 
higher than the 2--10 keV flux of $(6.72\pm 0.38)\times 10^{-13}$ erg/cm$^2$s derived from the PL-fit using solely EPIC-pn data (the photon 
index was $-1.57\pm 0.03$ in this case). 
Both values are, however, consistent with the value of $(6.7\pm 0.6)\times 10^{-13}$ erg/cm$^2$s given by \citet{marshall1998b}. 
Comparing the unabsorbed pulsed flux in the 2--8 keV band of $(6.04\pm 0.24)\times 10^{-13}$ erg/cm$^2$s (this work) with the unabsorbed flux 
value of $3.09\times 10^{-12}$ erg/cm$^2$s as given by \citet{townsley2006} for the total emission of \psrg\ indicates that about 18\% of the 
emission is pulsed in this band, similar to what has been found for the Crab pulsar/nebula system. 

Noteworthy is the recent detection of \psrg\ by \integral\  ISGRI in the 20--60 keV band, at a flux of $0.44\pm 0.08$ mCrab (total emission), 
resulting from a very deep survey of the LMC region \citep{grebenev2013}.

On the other hand no pulsed radio and high-energy \gr-ray emissions have been detected so far from \psrg\ \citep[see][for the radio and 
high-energy \gr-ray band, respectively]{crawford1998,abdo2013}. In the TeV domain, however, the H.E.S.S. collaboration reported recently 
on the detection, at a $14 \sigma$ level, of a point source positionally consistent with N157B/\psrg\, making this the first extragalactic 
PWN association \citep{abramowski2012}. 

\begin{figure}
  \begin{center}
     \includegraphics[width=8.cm,height=12.cm,bb=130 170 445 630,clip=]{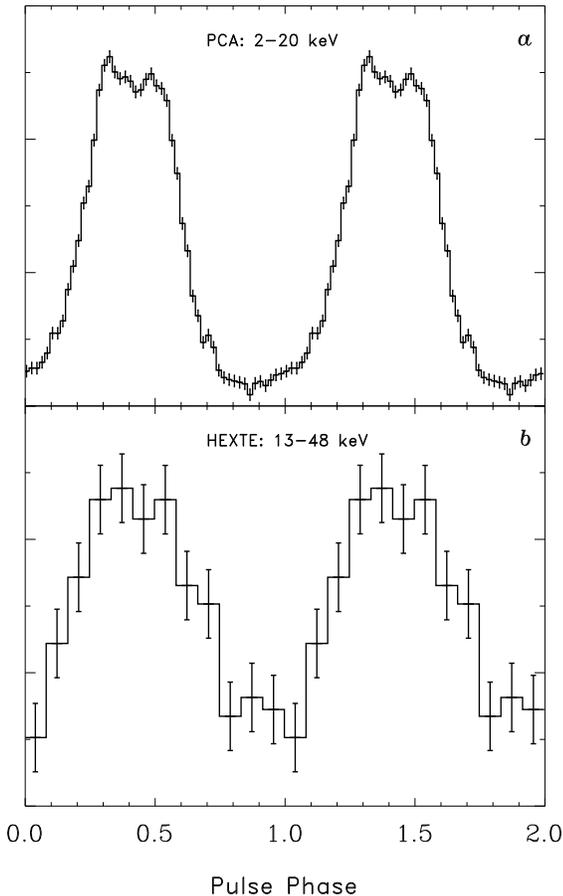}
     \caption{\label{lmcpsr0540}{\bf \lmcpsr;} pulse profile (broad single-pulse with a dip in the middle) for 
                                 the 2--20 keV band as measured by \rxte\ PCA (upper panel) combining about 700 ks of effective exposure time 
                                 \citep[see][]{deplaa2003}. In the lower panel the \rxte\ HEXTE lightcurve of \lmcpsr\ is shown for the 
                                 $\sim$ 13-48 keV band ($Z_1^2 \simeq 8.3\sigma$) using about 250 ks of dead-time and off-axis corrected 
                                 exposure time for clusters A and B.}
  \end{center}
\end{figure}

\subsection{\lmcpsr\ / \lmcpsra}

\lmcpsr\ was discovered at X-rays by \citet{seward1984} using \einstein\ X-ray observatory data, soon followed by a detection in the optical band 
\citep{middleditch1985} and a weak radio detection \citep{manchester1993} about a decade later. It is a young Crab-like pulsar located in the Large 
Magellanic cloud, and it is one of the few known spin-down powered pulsars resident outside our own galaxy \citep[see ][for an update on the spin-down 
powered pulsar content of the LMC, now 23 members, and SMC, 5 members]{ridley2013}.

Since its discovery \lmcpsr\ has been a primary target for many different X-ray instruments e.g. \ginga\ \citep{nagase1990,deeter1999}, 
\rosat\ \citep{finley1993}, \asca\ \citep{hirayama2002}, \sax\ \citep{mineo1999,cusumano2003}, \cxo\ \citep{gotthelf2000a,kaaret2001}
and \rxte\ \citep{deplaa2003,livingstone2005,campana2008}. 

\citet{deplaa2003} reported for the first time the detection of pulsed emission from \lmcpsr\ in the hard X-ray band up to $\sim 50$ keV using \rxte\ PCA and HEXTE data. 
Figure \ref{lmcpsr0540} shows the pulse profiles of \lmcpsr\ for the $\sim$ 2--20 and 13--48 keV bands, as measured by \rxte\ PCA (panel a) and HEXTE (panel b), 
respectively, using the same data as in \citet{deplaa2003}. Moreover, these authors found that the pulsed spectrum softens towards higher energies. An update of 
the findings by \citet{deplaa2003} was presented by \citet{campana2008} using about 3 times more \rxte\ PCA and 5 times more HEXTE exposure time. 
In general, these authors reached similar conclusions, namely that the pulsed spectrum softens towards higher energies, but obtained a difference in flux normalization 
of about 30\% and a somewhat different spectral shape. The first discrepancy can be traced back almost completely to improper response functions for the PCA units 
(i.e the sensitive areas) used by \citet{deplaa2003} at the time of the production of their paper, thereby overshooting the normalization by more than 20\%, while 
the shape discrepancy can be traced back by the different Hydrogen column densities used by the different authors, i.e. $4.6 \times 10^{21}$ cm$^{-2}$ by 
\citet{deplaa2003} \citep[see e.g.][]{kaaret2001}, while \citet{campana2008} derived/used a slightly lower value of $3.7 \times 10^{21}$ cm$^{-2}$. 
We prefer the higher valued number, which finds support in the work of \citet{serafimovich2004}. 

The softening of the pulsed spectrum explains also the non-detections by \cgro\ BATSE $\ga 20$ keV and \cgro\ OSSE (in the 80-210 keV band) 
\citep[see][respectively]{wilson1993b,hertz1995}. The unabsorbed pulsed fluxes of \lmcpsr\ for the 2--10 and 20--100 keV bands are 
$(6.7^{+0.4}_{-0.7})\cdot 10^{-12}$ and $(6.1^{+0.6}_{-2.1})\cdot 10^{-12}$ erg/cm$^2$s, respectively \citep[see][]{campana2008}. 
The former flux value was derived from the absorbed flux value of $(6.5^{+0.4}_{-0.7})\cdot 10^{-12}$ erg/cm$^2$s given by 
\citet{campana2008} adopting their estimated Hydrogen column density N$_{\hbox{\scriptsize H}}$ of $3.7 \times 10^{21}$ cm$^{-2}$.

A comparison of the spectral energy distributions of \lmcpsr\ and its PWN for the $\sim$ 1--100 keV band using \swift\ XRT (WT mode), 
\rxte\ PCA and HEXTE for the pulsed emission, and \swift\ XRT (PC mode) and \integral\ ISGRI for the total (=pulsar plus PWN) emission is 
given in \citet{campana2008}, and shows that about 25\% of the total emission is pulsed. Noteworthy is also the work presented by 
\citet{grebenev2013} on a deep hard X-ray survey of the LMC using \integral\ ISGRI (4.8 Ms effective exposure; \lmcpsr\ (plus PWN) 
was detected at a $22.0\sigma$ significance level in the 20-60 keV band with a 20-60 keV flux of $1.68\pm0.08$ mCrab) and \integral\ 
JEM-X (1.8 Ms effective exposure; 3-20 keV flux of $1.28\pm0.07$ mCrab ($\sim 18 \sigma$)).

At GeV energies recently the pulsed fingerprint of \lmcpsr\ has been detected at a $6.8\sigma$ level using a \rxte\ based ephemeris
covering a time span of $\sim$3.5 years \citep{martin2014}. The GeV pulse profile exhibits a single broad pulse and the GeV spectrum of the pulsed emission is soft 
and properly described by a cutoff power-law model connecting smoothly to the soft \gr-ray part of the spectrum. 

So far, (pulsed) emission has not been detected from \lmcpsr\ at TeV energies \citep[][]{komin2012}.

\subsection{The Vela pulsar, \vela\ / \velaj}

\begin{figure*}
  \begin{center}
     \includegraphics[width=12cm,height=16cm,angle=90,bb=55 135 525 705,clip=]{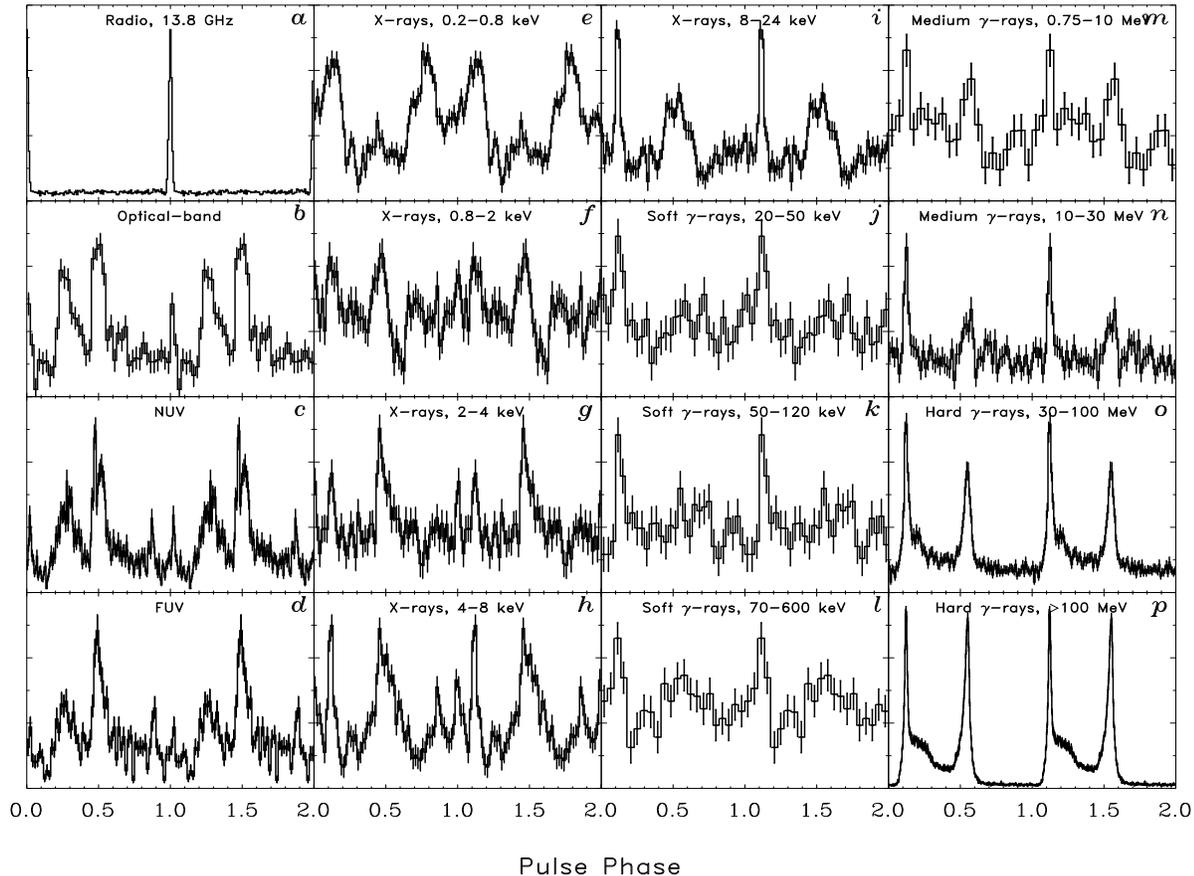}
     \caption{\label{vela_prof}{\bf \vela;} multi-wavelength radio-phase aligned pulse-profile collage from radio-frequencies up to high-energy \gr-rays.
     The left 4 panels ({\it a--d}) show the 'low' energy profiles: {\it a}) radio profile at 13.8 GHz, {\it b}) optical profile 
     \citep{gouiffes1998} and {\it c--d}) NUV (1400-3270 \.{A}) \& FUV (1140-1730 \.{A}) profiles \citep{romani2005}. Panels {\it e--i} show the 
     soft-hard X-ray pulse-profiles as obtained from XMM EPIC pn (panels {\it e--f}) and \rxte\ PCA (panels {\it g--i}) data. In panels 
     {\it j--l} the soft \gr-ray profiles are displayed obtained from multi-year \integral\ ISGRI (panels {\it j--k}) and \cgro\ OSSE 
     \citep[panel {\it l},][]{strickman1996} data. Finally, the right hand panels {\it m--p} show the medium- and high-energy \gr-ray pulse profiles 
     as obtained by analyzing multi-year \cgro\ COMPTEL (panels {\it m--n}, this work), \cgro\ EGRET (panel {\it o}) and \fermi\ LAT (panel {\it p}) data.
     Notice the dramatic morphology changes as function of energy from single peaked (radio) to double peaked with bridge emission at
     high-energy \gr-rays, with at intermediate energies very complex shapes with a plethora of pulses at optical, X-rays and soft/medium energy \gr-rays.
     }
  \end{center}
\end{figure*}

Targeting the Vela Supernova remnant at radio-frequencies \citet{large1968} discovered short period pulsations with $P \sim 89$ ms from a 
young (11 kyr) and energetic ($L_{sd} \sim 6.9\times 10^{36}$ erg/s) neutron star, known since as \vela. Since then the Vela pulsar has been 
monitored intensively at radio-frequencies, and it turns out to be a frequent glitcher, e.g. nine major glitches occured in the time period 
1969 -- 1994. 
Detailed single-pulse studies showed that the radio pulse (single; see e.g. Fig. \ref{vela_prof} panel {\it a}) could be described as a 
composite of four (Gaussian) components \citep{krishnamohan1983} covering a phase extent of $\sim 0.055$. Morphology changes 
of the radio pulse as a function of frequency (1.4-24 GHz) can be described by merely intensity variations of the constituting four Gaussian 
pulses, while maintaining their widths and positions \citep{keith2011}. Further single pulse studies \citep{johnston2001} reveal the presence 
of giant micropulses in the leading edge of the radio pulse and on the trailing edge a large Gaussian component.

In the optical domain pulsations from \vela\ were detected by \citet{wallace1977} from a sky region including the candidate proposed by 
\citet{lasker1976}. The statistics of the (double pulsed) optical profile has increased significantly since the first detection 
\citep[see e.g][]{manchester1980} with a most accurate characterisation of the profile given by \cite{gouiffes1998} using fast-photometry 
observations performed at the ESO 3.6 m telescope in 1993 and 1994. The latter observations revealed a third narrow pulse coinciding with the 
radio pulse (see e.g. Fig. \ref{vela_prof} panel {\it b}), the presence of which was confirmed at smaller wavelengths by \hst\ STIS NUV and 
FUV observations \citep[see e.g. Fig. \ref{vela_prof} panels {\it c} and {\it d},][]{romani2005}. The latter observations revealed even a 
narrow fourth pulse preceding the radio-pulse. 

Multi-band optical photometry with ESO NTT \citep{nasuti1997}, ESO VLT \citep[ISAAC, {\it J$_s$} and {\it H} bands,][]{shibanov2003}, 
\hst\ STIS \citep[NUV and FUV,][]{romani2005,kargaltsev2007} and Gemini GeMS \citep[{\it K$_s$}-band,][]{zyuzin2013} along with spectroscopy 
observations \citep[VLT {\it FORS2}; 4000- 11000 \.{A},][]{mignani2007} yielded a detailed spectral picture from near-IR to FUV of the Vela 
pulsar revealing a very flat spectrum with index $\alpha = -0.01 \pm 0.01$ \citep{zyuzin2013} across this range confirming, inline with its 
high pulsed fraction of about 80\%, the non-thermal magnetospheric nature of the near-IR to FUV emission. 

Optical astrometry with (initially) the ESO 3.5 m NTT \citep{nasuti1997}, \hst\ WFC/WFPC2 \citep{deluca2000,caraveo2001} made accurate proper 
motion and even parallax measurements of the Vela pulsar possible, yielding a distance of $294^{+76}_{-50}$ pc, consistent with the more 
accurate distance estimate of $287^{+19}_{-17}$ pc derived somewhat later from VLBI (radio) observations \citep{dodson2003}. 

In the soft X-ray band (0.1-4 keV) the first detailed imaging studies of the Vela pulsar environment have been performed with the IPC 
($\sim 1\arcmin$ spatial resolution) and HRI ($\sim 2\arcsec$ resolution) instruments aboard the {\it\/ Einstein} satellite 
(Nov. 12, 1978 - April 1981). Structures at four different spatial scales could be discerned from the images \citep{harnden1985}, 
in particular a) a pointlike object coincident with the pulsar, b) a relatively bright $\sim 4\arcmin$ diffuse nebula about the pulsar, 
c) diffuse hard emission at $\sim 1\degr$ scale between the pulsar and the radio object Vela X and finally d) thermal emission from the 
entire $\sim 5\degr$ Vela SNR. No evidence for X-ray pulsations could be found from the pointlike object associated with \vela, indicating 
a rather low pulsed fraction of $\la 9\%$ \citep[see also,][using \exosat\ 0.03-2.4 keV imaging observations]{ogelman1989}. 

The first unambiguous detection of pulsed X-ray emission from \vela\ was reported by \citet{ogelman1993} using \rosat\ PSPC (0.1-2.4 keV) data. 
This was confirmed by \rosat\ HRI monitoring observations of \vela\ yielding a pulsed fraction of about 12\% \citep{seward2000}. 
The soft X-ray (0.1-2.4 keV) pulse shape turned out to be composed of one broad and two narrower pulses. 

In the meantime \cxo\ and \xmm\ observations considerably improved the overall soft X-ray picture of the Vela pulsar and its PWN. 
The subarcsecond angular resolution of \cxo\ made it possible to separate the pulsar from its bright PWN, and to determine the genuine underlying 
pulsar spectrum. The featureless phase-averaged pulsar spectrum \citep[see][using a combination of HRC-S/LETG and ACIS-S/HETG in CC-mode data]{pavlov2001a} 
could best be described in terms of a two component model composed of a Hydrogen NS atmosphere model (to describe the soft part) and a power-law model 
with best fit photon-index of $-1.5 \pm 0.3$, nicely connecting to both the optical part and high-energy part of the Vela pulsar spectrum. 

Monitoring observations of the Vela PWN with the ACIS-S show a highly dynamical system with a complex variability and morphological changes of 
structures (e.g. inner/outer and counter jets; inner/outer arcs; shell) in the PWN  \citep{pavlov2001b,pavlov2003,durant2013}. 

In the timing domain \cxo\ observations with the HRC-I instrument showed the tripled peaked (energy integrated) soft X-ray profile in detail and 
aligned according to a contemporaneous radio ephemeris \citep{helfand2001}. These authors derived a pulsed fraction of $7.1 \pm 1.1$ across the 
0.1-10 keV band (gravity point near 1 keV) not diluted by PWN emission.

The Vela pulsar/PWN region was also observed (early in the mission on Dec. 1--2, 2000\footnote{Since April 27, 2006 there is a yearly monitoring 
of the Vela pulsar region}) by the (large collecting area) \xmm\ satellite with the EPIC pn operating in Small Window mode at a temporal resolution 
of 5.7 ms, amply sufficient to study the energy dependency of the pulse morphology \citep[see e.g.][for phase-resolved spectroscopy; however, the moderate 
angular resolution of about $6\farcs 6$ FWHM and $10\arcsec$ extraction radius used, imply a severe (not accounted for) PWN nebula contribution at 
every phase slice making their published (phase-resolved) spectral results unreliable]{manzali2007}.

We  also analysed the \xmm\ Vela observation of Dec. 2, 2000 (obs. id. 0111080201; 58.6 ks pn time) to study the energy dependency of the 
pulse profile. The radio-aligned profiles for the 0.2-0.8 and 0.8-2 keV ranges are shown in Fig. \ref{vela_prof} panels {\it e--f}
(the 2--8 keV band pn profile is statistically equivalent to the \rxte\ PCA 2--8 profile and therefore is not shown here). Drastic changes 
can be discerned between the two - soft and intermediate (transition to the non-thermal regime) - bands.

At medium energy X-rays (\rxte\ PCA; $\sim 2-30$ keV) the morphology of the lightcurve drastically changed with two (narrow) peaks roughly aligned in 
phase with the high-energy \gr-pulses \citep[][93 ks obs. performed during Jan 12--23, 1997]{strickman1999}. Much deeper \rxte\ PCA observations 
(274 ks exposure collected during April--May and July-August 1998), yielding much-improved statistics, showed the presence of several (narrow) 
emission components \citep{harding2002}. In this work we reanalyzed {\it all} available \rxte\ PCA data on \vela, totaling 360 ks screened exposure 
time (for PCA unit-2). The resulting lightcurves for the 2-4, 4-8 and 8-24 keV bands are shown in Fig. \ref{vela_prof} panels {\it g--i}. 
Notice the presence of at least 5 (narrow) pulse components, one coincident with the radio and narrow optical/NUV/FUV pulse at phase zero, rendering 
the Vela pulsar as the most complex pulsar at X-ray energies. 

Preceded by non-detections \citep[][{\it HEAO 1}; 15 keV - 11 MeV]{knight1982} and controversial (short duration) balloon flight results \citep[e.g.][]{harnden1972,tumer1984,sacco1990}, the first unambiguous detection ($4.6\sigma$ in the 70-600 keV band) of the Vela pulsar at soft \gr-rays was 
made by the OSSE instrument aboard \cgro\ \citep{strickman1996}. The pulse profile (see Fig. \ref{vela_prof} panel {\it l}; 70-600 keV) showed two 
peaks - a narrow and a more structured broad one - coincident with the pulses seen at high-energy \gr-rays.

The ISGRI instrument ($\sim$ 20-300 keV) aboard the \integral\ satellite offered a new look at the Vela pulsar at soft \gr-rays. In this work a 
timing analysis of 2244 \integral\ ISGRI observations of the Vela region, spread over the period June 12, 2003 -- Oct. 13, 2010 
(\integral\ Revolution period 81--976), has been performed to shed light on the pulse morphology in the so far poorly explored soft \gr-ray band. 
The combined ISGRI dataset represents an effective on-axis exposure time of 6.12 Ms.
The pulse profiles of \vela, generated through folding upon proper radio- or \fermi\ LAT based timing models, are shown in Fig. \ref{vela_prof} 
panels {\it j--k} for the 20-50 and 50-120 keV bands, yielding non-uniformity significances of $5\sigma$ and $4.9\sigma$, respectively, 
adopting a $Z_{11}^2$ test. The 20--300 keV pulse profile represents even a $7.9\sigma$ significance, with one prominent narrow first pulse 
and a structured second broader pulse.

At medium \gr-ray energies ($\sim$ 1--30 MeV), using data from the COMPTEL telescope aboard \cgro\ \citep{schonfelder1993}, secure detections of the 
Vela pulsar have been reported by \citet{hermsen1993,hermsen1994} and \citet{kuiper1998}. In the latter work phase-resolved spectra were 
shown combining COMPTEL MeV data collected from the first four observation cycli of \cgro. In this work we present some new, unpublished so far, COMPTEL Vela pulsar 
results based on observations performed over the full \cgro\ mission lifetime, spanning nine observation cycli and covering the period 
May 10, 1991--March 21, 2000 (23 viewing periods, typically lasting 2 weeks, with \vela\ within $30\degr$ from the pointing axis). 

Applying event selections using optimized parameter windows on Time of Flight (TOF), Pulse Shape Discrimination (PSD) and 
$\phi_{\hbox{\scriptsize arm}}$ \citep[see e.g. sections 2 \& 3 of][and references therein]{kuiper2001} and sorting the barycentered timetags 
on pulse phase and energy resulted in pulse-phase distributions as shown in Fig. \ref{vela_prof} panels m--n, for the 0.75-10 and 10-30 MeV 
bands, respectively. The $Z_9^2$-test significances represent a $6.2\sigma$ and $12.7\sigma$ deviation from uniformity for both ranges, respectively. 
Clearly visible are emission peaks coincident with the two high-energy \gr-ray pulses, P1 and P2, with in between bridge emission (most notably 
in the 0.75-10 MeV band). Indications for emission structures in the off-pulse region exist for energies above 3 MeV.

\vela\ was the {\it first} pulsar recognized in high-energy \gr-ray data from the SAS-2 satellite \citep{thompson1975,thompson1977}, 
followed by more detailed studies with COS-B \citep{bennett1977,kanbach1980,grenier1988} and later with \cgro\ EGRET \citep{kanbach1994,fierro1998}. 
The \gr-ray pulse profile shows two pulses with clear bridge emission with the main pulse trailing the radio-pulse $\sim 0.12$ in phase 
(see e.g. Fig. \ref{vela_prof} panel {\it p} for a $>100$ MeV \fermi\ LAT profile). \fermi\ LAT made high-statistics (phase-resolved) studies 
at high-energy \gr-rays possible, revealing a third component in the bridge region, which moves towards higher phase with increasing energy. 
The two \gr-ray pulses become narrower the higher the energy \citep{abdo2009c,abdo2010d} with an even vanishing first (main) pulse component. 
This has been confirmed now by the upgraded TeV telescope systeem H.E.S.S. (phase II, including a fifth 28m telescope at the center location;
energy threshold about 30 GeV), yielding a $\sim 11.0\sigma$ pulsed signal (H-test) at a mean energy of 40 GeV \citep{stegmann2014}.

Considering the full pulse-profile collage shown in Fig. \ref{vela_prof} it is clear that \vela\ shows extremely complex emission patterns, 
which change drastically as a function of energy. Therefore, it will be a major theoretical challenge to understand the multi-beam emission 
structures of \vela\ and so to unveil the topology and plasma filling of its magnetosphere.

In Fig. \ref{spectralcompilation} the total pulsed flux of \vela\ across the $\sim$ 2 keV -- 10 GeV band (thermal part excluded) is shown 
in blue color as solid curves and data symbols. Published data and spectral fit results from \cgro\ OSSE \citep{strickman1996} and Fermi 
LAT \citep{abdo2009c} are shown as well as newly derived pulsed fluxes for the $\sim$ 2-- 30 keV (\rxte\ PCA), 20--175 keV (\integral\ ISGRI) 
and 0.75-30 MeV (\cgro\ COMPTEL) bands. 

Given the complexity of the Vela lightcurve at X-rays (see e.g. Fig. \ref{vela_prof} panel {\it h}) it is rather difficult to define the 
unpulsed or DC level. For the \rxte\ data we have used the phase stretch 0.68-0.78 to determine this level. A spectral analysis adopting a 
power-law model with N$_{\hbox{\scriptsize H}}$ fixed to  $3.3 \times 10^{20}$ cm$^{-2}$ \citep[see e.g.][Table 1]{pavlov2001a} yielded a 
photon index of $-1.06 \pm 0.05$ across the 2.2--28.1 keV band. 
Pulsed fluxes (unabsorbed) are: 2--10 keV $(8.0\pm 1.0)\cdot 10^{-13}$ erg/cm$^2$s (\rxte\ PCA); 20--100 keV $(1.1\pm 0.4)\cdot 10^{-11}$ erg/cm$^2$s
(\integral\ ISGRI\footnote{Because of the complexity of the lightcurve we fitted the 20-100 keV profile in terms of a constant and 7 harmonics 
to estimate the unpulsed level, and so the number of pulsed excess counts. Varying the number of harmonics, 5, 7, 9, 11, yielded consistent 
results.}); 1--10 MeV  $(6.2\pm 0.8)\cdot 10^{-10}$ erg/cm$^2$s (\cgro\ COMPTEL; systematic errors are of the order of 30\% because of uncertainties in the unpulsed level); 0.1--1 GeV  $(3.92\pm 0.49)\cdot 10^{-9}$ erg/cm$^2$s (\fermi\ LAT).

The Vela pulsar is often considered as the canonical high-energy \gr-ray pulsar reaching maximum luminosity at GeV energies.
Note, however, that the detection at hard X-rays/soft \gr-rays is just possible because the pulsar is relatively nearby. Would \vela\ be 
placed at 2 kpc, as the Crab pulsar, then the pulsed hard X-ray/soft \gr-ray emission would reach flux levels below the detection limits 
of current instrumentation.
 
\subsection{\psrigra\ / \igra}

\igra\ was discovered by \citet{keek2006} in a deep 1.9 Ms \integral\ ISGRI mosaic image of the Circinus region for the 20--60 keV band 
using \integral\ observations performed between Febr. 28, 2003 and Mar. 3, 2005. A 20--60 keV flux of $0.97 \pm 0.15$ mCrab and 
a positional error of $3\farcm6$ at 90\% confidence were derived. 

A $\sim 5$ ks follow up observation with \cxo\ on June 29, 2008 revealed the presence of an extended X-ray source, nearly circular 
with a $1\farcm 5$ radius, within the ISGRI error region \citep{tomsick2009}. Its morphology with at the center region a PWN surrounding 
a so-far undetected pulsar made a SNR identification plausible. \citet{tomsick2009} derived spectral characteristics for the composite 
emission region within $1\farcm5$ from the central source by fitting an absorbed power-law model over the 0.3-10 keV range, and obtained 
an absorbing Hydrogen column N$_{\hbox{\scriptsize H}}$ of $(3.1 \pm 0.3) \times 10^{22}$ cm$^{-2}$, a photon index of $-1.83 \pm 0.13$ 
and an unabsorbed 2--10 keV flux of $1.9 \times 10^{-11}$ erg/cm$^2$s ($\sim 1$ mCrab).

\begin{figure}
  \begin{center}
     \includegraphics[width=6cm,height=10cm,angle=0,bb=170 175 420 630]{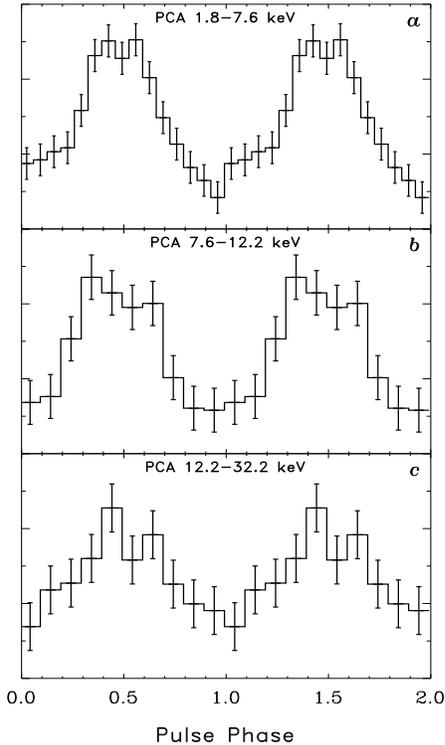}
     \caption{\label{igrj14003_prof}{\bf \igra;} pulse profiles for \rxte\ PCA in different energy bands. The significance in the  
     12.2-32.2 keV band is $3.7\sigma$ (panel c).}
  \end{center}
\end{figure}

\citet{renaud2010} reanalyzed the \cxo\ data and decomposed the emission into 3 components: 1) a point-source, extracted for a 
circular region of $1\farcs 5$ centered on source with background taken from an annulus with radii $1\farcs5$ and $2\arcsec$, 2) 
a PWN-region extracted from an annulus with radii $1\farcs5$ and $39\arcsec$ and 3) the SNR-region extracted from an annulus with 
radii $0\farcm 98$ and $1\farcm27$. Assuming absorbed power-law models they found the following photon-indices and unabsorbed 
2--10 keV fluxes for an absorbing column of $(2.09 \pm 0.12) \times 10^{22}$ cm$^{-2}$: $-1.22 \pm 0.15, -1.83 \pm 0.09$ and 
$-2.56 \pm  0.18$, and $(1.95 \pm 0.5)\times 10^{-12}, (1.51 \pm 0.2)\times 10^{-11}$ and $(1.1 \pm 0.2)\times 10^{-12}$ erg/cm$^2$s 
for the 3 components, respectively. The PWN is thus by far the most dominant source, and about 8 times brighter than the hard point-source. 

The detection of pulsed X-ray emission of \igra\ was also reported by \citet{renaud2010} using 2 sets of \rxte\ PCA observations of 44 ks 
each, taken one year apart on Sept. 29, 2008 and Sept. 30, 2009. They found an energetic ($\dot{E}=5.1\times 10^{37}$ erg/s) pulsar 
spinning with a 31.2 ms pulse period and an estimated age of 12.7 kyr. This period was later confirmed at radio-frequencies with the 
Parkes telescope, resulting furthermore in a (very) high DM value of $563 \pm 4$ cm$^{-3}$ pc and flux densities of 250 and 110 $\mu$Jy 
at 1.374 and 3.1 GHz, respectively. 

\citet{renaud2010} also derived the pulsed spectrum of \igra, through an on-off method, for the 2--10 keV band using only the top 
detection layer of each involved PCU. They found an absorbed 2--10 keV pulsed flux of $3.0\times 10^{-13}$ erg/cm$^2$s and a ({\it soft}) 
photon index of $-2.0_{-0.3}^{+0.5}$. 

We reanalyzed the PCA data including now all PCU detector layers to be more sensitive to the hard X-ray photons with energies in excess 
of 10 keV. Pulse profiles for three PCA bands are shown in Fig. \ref{igrj14003_prof}. Significant pulsed emission at $\sim 3.7\sigma$ is 
found for the 12.2-- 32.2 keV band. Pulsed excess counts have been derived through pulse template fitting using as extractor the 3 harmonics 
fit to the lightcurve of the 4--87 PHA band. These excess counts have been converted to pulsed fluxes adopting an absorbing Hydrogen column 
of $2.09 \times 10^{22}$ cm$^{-2}$ \citep{renaud2010} and assuming an underlying (absorbed) power-law model for the photon flux. 
We obtained a photon index of $-1.95 \pm 0.04$, consistent with \citet{renaud2010}, and an unabsorbed 2--10 keV flux of
$(1.3 \pm 0.1)\times 10^{-12}$ erg/cm$^2$s, about $\sim 4$ times larger than found by \citet{renaud2010}. Our pulsed flux represents $\sim 67\%$ 
of the total 2--10 keV flux, while \citet{renaud2010} found a pulsed fraction of only $\sim 18\%$.
The pulsed flux measurements of \igra\ for $\sim 2-32$ keV band are shown in Fig. \ref{spectralcompilation} (left panel) as yellow data points.

We also analyzed HEXTE data. As expected, we could not find evidence for pulsed emission for energies in excess of $\sim 15$ keV.

In the TeV domain no point-like source has been detected above $5\sigma$ in the latest map of the H.E.S.S. Galactic Plane Survey 
\citep{chaves2008,chaves2009}. Also at the GeV energies neither a point-like source is found in \fermi\ LAT data \citep{nolan2012} 
at the location of \igra\ nor pulsed emission has been detected \citep{abdo2013}.  


\subsection{\msh\ / \mshj}

\msh\ was discovered at X-rays as a 150 ms pulsar in \einstein\ HRI and IPC (0.2-4 keV) data by \citet{seward1982}, and later
confirmed at radio-wavelengths by \citet{manchester1982}. Its characteristic age of only 1570 years and spin-down luminosity of $1.8\times 10^{37}$ erg/s, 
as inferred from its timing parameters, qualifies \msh\ as a very young and energetic spin-down powered pulsar. 
Its derived surface polar magnetic field strength of $3.1\times 10^{13}$ G is close to the quantum critical field strength of $4.413\times 10^{13}$ G, above 
which exotic quantum electrodynamic effects play an important role in the high-energy pulsar emission mechanisms.

\msh\ is associated with supernova remnant MSH 15-52 (G320.4-1.2), and various detailed studies using X-ray data have been performed since the early 
eighties to disentangle the complex morphology of MSH 15-52 with at the north western rim an excess near H$\alpha$ nebula RCW 89 and near the centre 
a clump containing a diffuse synchrotron nebula surrounding the pulsar (its PWN). Using the superb X-ray imaging quality of \cxo\ \citet{gaensler2002} 
showed that, while the overall PWN shows a clear symmetry axis, the PWN also reveals several new components. Extended TeV emission ($>280$ GeV) from 
MSH 15-52 was detected at a $25\sigma$ confidence level by H.E.S.S. \citep{aharonian2005b}, and later at higher TeV \gr-rays ($>810$ GeV) by CANGAROO-III 
at a $7\sigma$ level \citep{nakamori2008}. In the GeV band extended \gr-ray emission has been detected by the \fermi\ LAT \citep{abdo2010b}.

Pulsed emission in the medium/hard X-ray band was detected by \citet{kawai1991} using \ginga\ (2-60 keV) data. These authors also found that, using 
contemporaneous X-ray (\ginga) and radio (Parkes) data, the radio pulse leads the asymmetric X-ray pulse by $0.25 \pm 0.02$ in phase. Soon after the 
launch of the Compton Gamma-Ray Observatory (\cgro) on April 16, 1991, pulsed hard X-ray/soft \gr-ray emission above $\sim 60$ keV was detected by BATSE 
using data collected in the folded-on-board data mode \citep{wilson1993a,wilson1993b}, confirmed later by \citet{ulmer1993} and \citet{matz1994} using 
\cgro\ OSSE (0.05-10 MeV) data. The radio- soft \gr-ray pulse lag found by \citet{ulmer1993} combining BATSE and OSSE data was $0.32 \pm 0.02$, somewhat 
larger than the lag found at lower energies. In the spectral domain \citet{matz1994} derived for the pulsed spectrum from 50 keV to 5 MeV, fitting a 
power-law model, a photon index of $-1.68 \pm 0.09$, significantly softer than the value of $-1.30 \pm 0.05$ obtained 
for the 2-60 keV band by \citet{kawai1991} analyzing \ginga\ data, strongly indicating a softening towards higher energies.

Pulsed emission was also detected in the medium energy \gr-ray band by \citet{kuiper1999} using \cgro\ COMPTEL (0.75-30 MeV) data. 
They found a $5.4\sigma$ pulsed signal, a broad asymmetric pulse, for the 0.75-30 MeV band analyzing COMPTEL data collected over 6 \cgro\ observation cycli. 
The pulse reaches its maximum at phase $0.38 \pm 0.03$ with respect to the radio pulse. The apparent shift of the pulse maxima from soft X-rays  to medium 
energy \gr-rays can be explained by different spectral behaviours of two Gaussian shaped pulses comprising the broad asymmetric pulse with one pulse peaking 
at $0.250\pm 0.008$ and the other at $0.386\pm 0.012$ as derived from high-statistics \rxte\ PCA data. The first (narrower) pulse seems to vanish towards higher 
energies \citep{kuiper1999}. The spectrum of the pulsed emission clearly breaks above $\sim 10$ MeV, though there were weak indications for pulsed emission 
beyond 10 MeV in both COMPTEL and \cgro\ EGRET data (also in the spatial domain).

At high-energy \gr-rays ($>100$ MeV) significant progress has been made analyzing \fermi\ LAT data \citep{abdo2010b} and \agile\ data \citep{pellizzoni2009,pilia2010} 
by detecting pulsed emission from \msh\ well above 100 MeV. In \citet{denhartog2014} the \fermi\ LAT results are presented analyzing $\sim 5$ years of data 
detecting pulsed emission from \msh\ at a $18.2\sigma$ level across the $\sim$ 30 - 1000 MeV band, enabling detailed studies of the morphology of the 
lightcurve and the breaking part of the spectrum at these energies. 

To achieve pulse profile morphology comparisons across the $\sim$ 2 keV - 1 GeV range at a high statistics level we analyzed and processed
multi-year data from \rxte\ PCA (2-30 keV; April 25, 2002 - Oct. 9, 2003), \rxte\ HEXTE (15-250 keV; Feb. 10, 1999 - Jan. 1, 2012), \cgro\ BATSE 
(22 - 4000 keV; Oct. 31, 1991 - May 9, 2000; MJD 48560--51673; \cgro\ Cycli I-IX), \cgro\ COMPTEL 
(0.75-30 MeV; Oct. 17, 1991 - Nov. 3, 1997; MJD 48546.6--50755.6; \cgro\ Cycli I-VI) and \fermi\ LAT (30-1000 MeV; Aug. 4 18:00:00 - Jan. 1, 2012; MJD 54682.75--55927). 

\begin{figure}
  \begin{center}
     \includegraphics[width=8cm,height=12cm,bb=125 130 465 665]{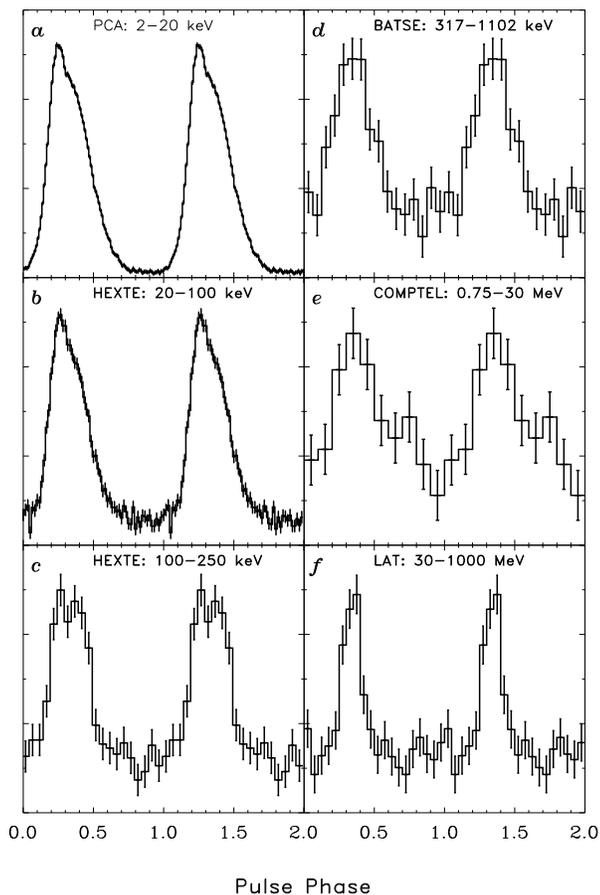}
     \caption{\label{msh_prof} {\bf \msh\ (the prototype soft \gr-pulsar);} pulse profiles from 2 keV up to 1 GeV using multi-year \rxte\ PCA \& HEXTE data (panels {\it a--c}), 
     \cgro\ BATSE (Cycli I-IX; panel {\it d}) \& COMPTEL (Cycli I-VI; panel {\it e}) and \fermi\ LAT (covering 3.4 years; panel {\it f}).}
  \end{center}
\end{figure}

\rxte\ PCA data (in {\it goodXenon} mode) were collected during observation run 70701 and part of 80803. The time intervals are fully 
covered by the validity interval of an ephemeris from the ATNF archive ensuring proper timing information. Subsequent screening adopting 
default criteria rendered for PCU-2 an effective exposure of 74.53 ks. This yielded ample statistics to study its timing characteristics 
in great detail, because \msh\ is a relatively strong pulsar at medium/hard X-ray energies. To determine the pulsed spectrum across the 
15-250 keV band we used \rxte\ HEXTE data collected over a much longer time period than used for the PCA. We combined all available HEXTE 
data on \msh\ taken during observation cycli P40704 -- 96803 (Feb. 10, 1999 -- Jan. 1, 2012) to obtain the highest possible statistics. 
For more information on these \rxte\ PCA and HEXTE data we refer to the section on \rxte\ in \citet{denhartog2014}. The resulting pulse profiles 
for the PCA (2-20 keV) and HEXTE (20-100 and 100-250 keV) are shown in panels a--c of Fig. \ref{msh_prof}. 
The $Z_2^2$-test significance of the 100-250 keV profile is about $17.1\sigma$.

To study the pulse profile of \msh\ at soft/medium energy \gr-rays we analyzed BATSE \citep[see e.g.][and references therein]{fishman1993} LAD folded-on-board data types collected during 
the {\it full} duration of the \cgro\ mission yielding a net screened exposure of about 10.6 Ms (integrated sum for all 8 LADs) for the 
time period Oct. 31, 1991 - May 9, 2000. The processing of these data is analoguous to that in \citet{kuiper2001} for the Crab pulsar 
(see Section 5.3 of that paper). The resulting 317-1102 keV BATSE profile of \msh\ is shown in panel d of Fig. \ref{msh_prof}. 
The deviation from a statistically uniform distribution is about $10.4\sigma$ adopting a $Z_2^2$-test. Panel e of Fig. \ref{msh_prof} 
shows the COMPTEL 0.75-30 MeV profile as obtained by \citet{kuiper1999}. 

Finally, panel f of Fig. \ref{msh_prof} shows the \fermi\ LAT 30-1000 MeV lightcurve using data from a $5\degr$ aperture around \msh\ 
collected during Aug. 4, 2008 18:00:00 and Jan. 1, 2012 (about 3.4 years) obtained after folding the SSB corrected arrival times with 
\rxte\ PCA based timing models. The $Z_2^2$-test significance of the 30-1000 MeV profile is about $10.2\sigma$, and a single Gaussian can 
describe the measured distribution accurately. Pulse maximum occurs at phase $0.342 \pm 0.007$ and the width (Gaussian $\sigma$) is 
$0.064 \pm 0.007$. This indicates that the first pulse indeed vanishes, while the second pulse becomes somewhat narrower, when the energy increases.

\begin{figure}
  \begin{center}
     \includegraphics[width=8.2cm,height=8.cm,bb=50 155 560 660,clip=]{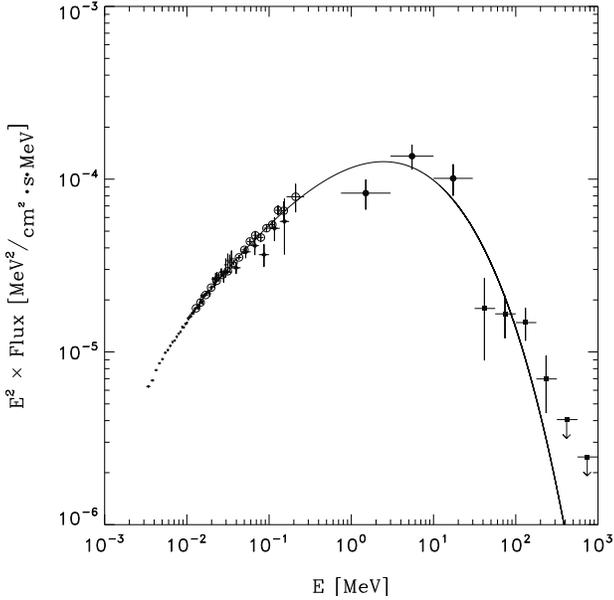}
     \caption{\label{msh_hepulspc}{\bf \msh;} unabsorbed high-energy spectrum of the pulsed emission across the $\sim$ 2 keV - 1 GeV range, 
     combining \rxte\ PCA, HEXTE, \integral\ ISGRI, \cgro\ COMPTEL and Fermi LAT measurements (see text). Maximum luminosity is reached at about 
     2.5 MeV making \msh\ an archetypal soft \gr-ray pulsar.}
  \end{center}
\end{figure}

Equiped with the multi-year \rxte\ PCA/HEXTE data covering at high statistics the $\sim$ 2 -- 250 keV band, the multi-year \fermi\ LAT 
(30-1000 MeV) data \citep{denhartog2014}, supplemented by the (archival) COMPTEL (0.75-30 MeV) data \citep{kuiper1999} and \integral\ 
ISGRI (20-175 keV; 550 ks of dedicated \msh\ observation during \integral\ AO2; this work), we attempted a broad-band spectral fit of 
the {\it pulsed} emission across the $\sim$ 2 keV - 1 GeV range adopting a photon flux model (a kind of power-law with a (modified) 
exponential cutoff) $F_{\gamma}$ with the following $E_{\gamma}$ dependency: \[F_{\gamma}= k \cdot (E_{\gamma}/E_0)^{\Gamma}\cdot \exp(-(E_{\gamma}/E_c)^{\beta})\]
Interstellar absorption effects, only applicable to \rxte\ PCA data, are modeled out adopting an absorbing Hydrogen column 
N$_{\hbox{\scriptsize H}}$ of $9.5 \times 10^{21}$ cm$^{-2}$ \citep{gaensler2002}.
For our spectral dataset the normalization energy $E_0$, which minimizes the correlation between the 4 fit parameters, was $0.100518$ MeV, 
while the best fit ($\chi^2_{r,59-4}=1.35$ i.e. acceptable) yielded the following fit parameters, 
$k=(1.574\pm0.012)\cdot 10^{-2}$ ph/cm$^2$s MeV; $\Gamma=-1.233\pm 0.005$; $E_c=0.078\pm 0.003$ MeV and $\beta=0.286\pm 0.005$. 
The broad-band $\sim$ 2 keV - 1 GeV pulsed spectrum of \msh\ is shown in Fig. \ref{msh_hepulspc} along with the best fit model. 
Maximum luminosity per energy decade is reached at $\sim 2.5$ MeV and therefore \msh\ can be considered as the ``canonical'' soft \gr-ray pulsar. 

Pulsed fluxes (unabsorbed), obtained from the best fit model, in the 2--10 keV, 20--100 keV, 1--10 MeV and 0.1--1 GeV are, 
$(2.52\pm 0.05)\cdot 10^{-11}$, $(9.59\pm 0.13)\cdot 10^{-11}$, $(4.4\pm 0.3)\cdot 10^{-10}$ and $(1.2\pm 0.4)\cdot 10^{-11}$ erg/cm$^2$s, respectively.

\begin{figure}
  \begin{center}
     \includegraphics[width=8cm,height=12cm,angle=0,bb=150 165 442 627]{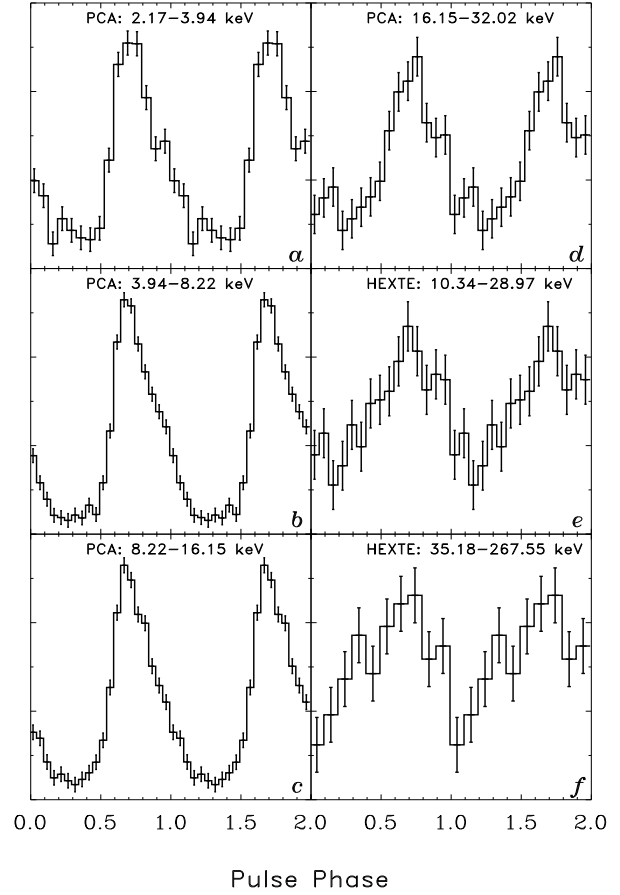}
     \caption{\label{psrj1617_prof}{\bf \psrc;} pulse profiles in various energy bands for \rxte\ PCA (panels a--d) and \rxte\ HEXTE 
                                  (panels e--f). Pulsed emission has been detected up to $\sim 64$ keV.}
  \end{center}
\end{figure}

\begin{figure}
  \begin{center}
     \includegraphics[width=8cm,bb=50 150 545 655]{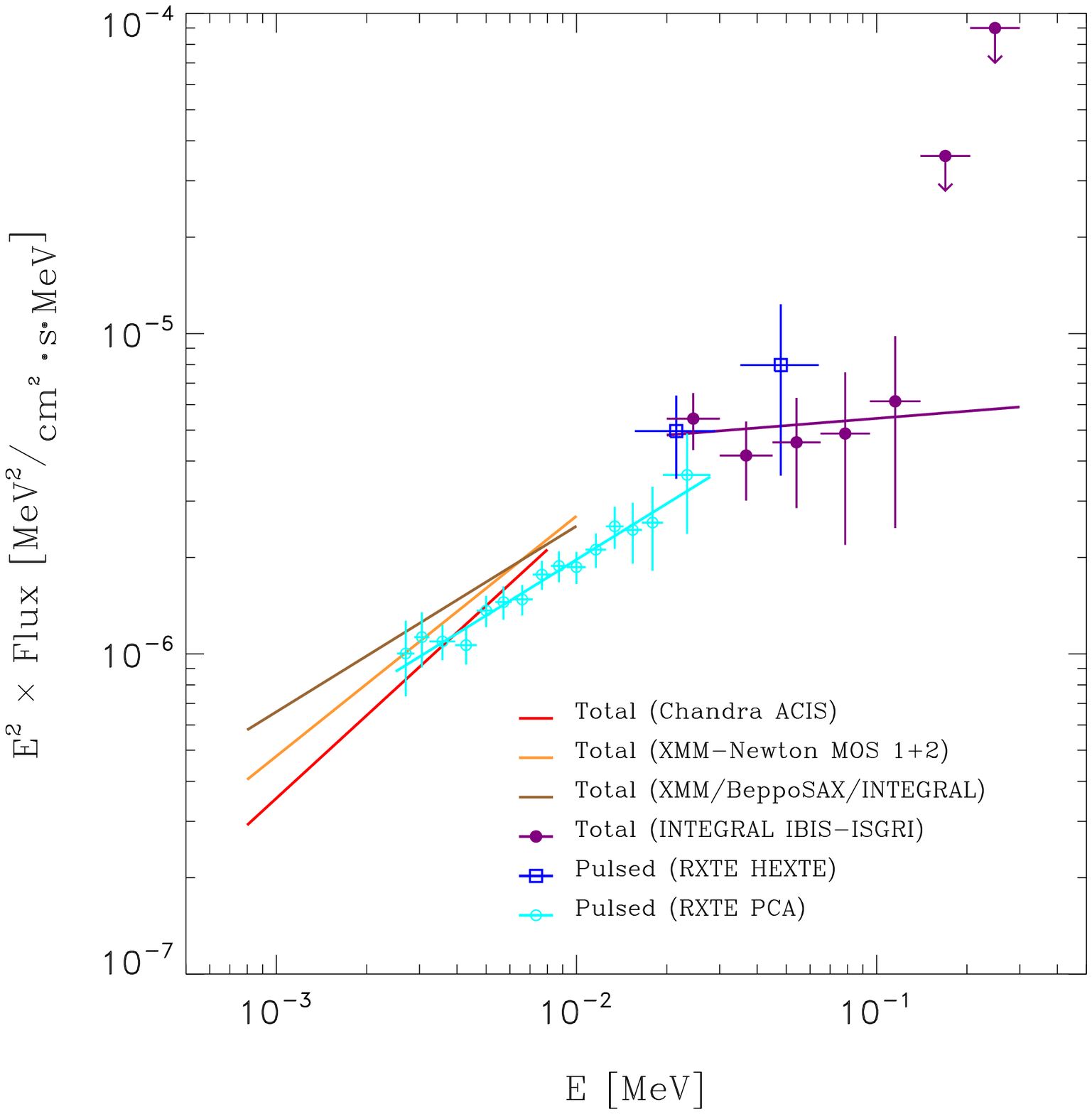}
     \caption{\label{psr1617_spc}{\bf \psrc;} high-energy (0.8-300 keV) total and pulsed spectra as derived from
                                 measurements by \cxo\ ACIS (total pulsar 0.8-8 keV; solid red; \citet{kargaltsev2009}), \xmm\ MOS 
                                (total, including compact PWN, 0.8-10 keV; solid orange; \citet{becker2002}), joint fit \xmm/\sax/ISGRI 
                                (total, including compact PWN, 0.8-10 keV; solid brown; \citet{landi2007}),
                                \rxte\ PCA (pulsed 2.5-30 keV; aqua; this work), \rxte\ HEXTE (pulsed 15-150 keV; blue; this work) 
                                and \integral\ ISGRI (total 20-300 keV; purple; this work).}
  \end{center}
\end{figure}

\subsection{\psrc}
\citet{torii1998} discovered an energetic 69 ms pulsar, \psrc, $\sim 7^{\prime}$ from SN-remnant RCW 103, analysing X-ray data 
obtained with the \asca\ GIS instrument. It is noteworthy that \citet{aoki1992} already detected pulsed emission from the RCW 103 region 
with the \ginga\ large area counters in data taken in March 1989.

Weak ($\sim 0.5$ mJy at 1.4 GHz) radio pulsations were detected by \citet{kaspi1998} using the Parkes radio telescope. Combining more
pulse period measurements from X-ray observations a characteristic age of about 8 kyr could be derived as well as strong indications
for a giant glitch \citep{torii2000} that occured between August 1993 and September 1997.

\rxte\ had the sky region containing \psrc\ several times in its field of view: on Jan. 2, 1998 and during March 5--6, 1999 in
observation run 30210 with prime target 1E 161348-5055 (the point source in the middle of RCW 103), on Jan. 22, 2001 during run 50428 
with prime target RCW 103, and finally in a dedicated observation run, 80090, performed during September 20--21, 2003 for about 
80 ks. We combined \rxte\ PCA (total screened PCU-2 exposure 146.53 ks) and HEXTE (dead-time and off-axis corrected cluster-0/1 exposures 
of 47.7 and 49.3 ks, respectively) data from these 3 observation runs and significantly detected pulsed X-ray emission  
up to $\sim 64$ keV. Pulse profiles in differential energy bands, all single peaked with a sharper rise than fall, are 
shown in Fig. \ref{psrj1617_prof}. The HEXTE profile for energies in excess of 35.3 keV (Fig. \ref{psrj1617_prof}f) has a Z$_1^2$ 
significance of $3.8\sigma$. 

We estimated the pulsed excess counts through pulse-profile model fitting with a truncated Fourier series (fundamental plus 2 harmonics) 
and converted these to photon flux values, adopting an absorbed power-law model with a fixed column density N$_{\hbox{\scriptsize H}}$ 
of $3.2 \times 10^{22}$ cm$^{-2}$ \citep{becker2002}. This yielded for the pulsed emission in the 2.5-30 keV PCA band a hard photon 
index of $-1.42\pm 0.02$ and a 2--10 keV unabsorbed flux of $(3.30\pm 0.16)\times 10^{-12}$ erg/cm$^2$s. We used all PCA detection 
layers to be more sensitive to hard X-ray photons. The reconstructed \rxte\ PCA pulsed flux values are shown in Fig. \ref{psr1617_spc} 
as aqua colored points along with the best fit. HEXTE pulsed flux measurements are also shown (blue squares).

Unfortunately, neither X-ray monitoring- nor regular radio observations (too weak, requiring long exposure times) have been performed 
on this pulsar, that would have enabled us to construct phase coherent timing models over long time stretches. Therefore, we could not 
do a timing analysis for the soft \gr-band using the low-countrate \integral\ ISGRI data. However, \citet{landi2007} demonstrated 
in a spatial analysis in  ISGRI maps the presence of a hard X-ray source coincident with \psrc\ with a significance in the 18-60 
keV band of $\sim 7.4\sigma$, combining data collected since the beginning of the mission (November 2002) up to April 2006. 
Our own spectral analysis of ISGRI data (Revs. 46-411; March 2, 2003 - Feb. 24,
2006) yielded for the {\em total} emission in the 20-300 keV band a photon index of $-1.93\pm 0.28$ and a 20-100 keV flux of 
$(1.25\pm 0.18)\times 10^{-11}$ erg/cm$^2$s, consistent with the values obtained by \citet{landi2007}. The ISGRI total flux 
measurements in the 20--300 keV band are indicated in Fig. \ref{psr1617_spc} as purple datapoints along with the best power-law fit.

The total emission spectrum of \psrc\ below 10 keV was already derived by \citet{torii2000,becker2002,landi2007,kargaltsev2009} 
using X-ray instruments on different spacecrafts -- \asca, \xmm, \sax\ and \cxo. The most recent determination by 
\citet{kargaltsev2009}, using \cxo\ ACIS-I data, resolves the emission around \psrc\ in a point-like component, associated with the pulsar, 
an inner compact pulsar wind nebula component of $\sim 1\arcsec$ size within $0\farcs5$ and $1\farcs5$ from the pulsar and an 
outer nebula extending up to $\sim 1\arcmin$ from the pulsar. Fitting a power-law to the spectral data extracted from a circular 
aperture of $0\farcs5$ centered on \psrc, \citet{kargaltsev2009} derived a power-law index of $-1.14\pm 0.06$ and a 0.5--8 keV 
(unabsorbed) flux of $(3.6\pm 0.1)\times 10^{-12}$ erg/cm$^2$s for the total (=pulsed+pulsar DC) emission of the pulsar. 
This converts to a 2--10 keV (unabsorbed) flux of $\sim 3.58 \times 10^{-12}$ erg/cm$^2$s. The flux of the inner compact PWN is 
about 10--15 \% of the total pulsar flux, and its combination is consistent with the total emission spectra derived using the X-ray 
instruments with much less spatial resolution: \citet{landi2007} find a 2--10 keV flux of $4.2\times 10^{-12}$ erg/cm$^2$s using 
\sax\ MECS, \xmm\ MOS 1+2 and \integral\ ISGRI in combination, while the 2--10 keV flux, estimated from the best fit values 
given by \citet{becker2002}, amounts  $4.0\times 10^{-12}$ erg/cm$^2$s. 
The best fit models derived by \citet{becker2002,landi2007,kargaltsev2009} are shown as solid orange, brown and red, respectively, 
lines in Fig. \ref{psr1617_spc}. 

\begin{figure*}
  \begin{center}
     \includegraphics[width=12cm,height=16cm,angle=90]{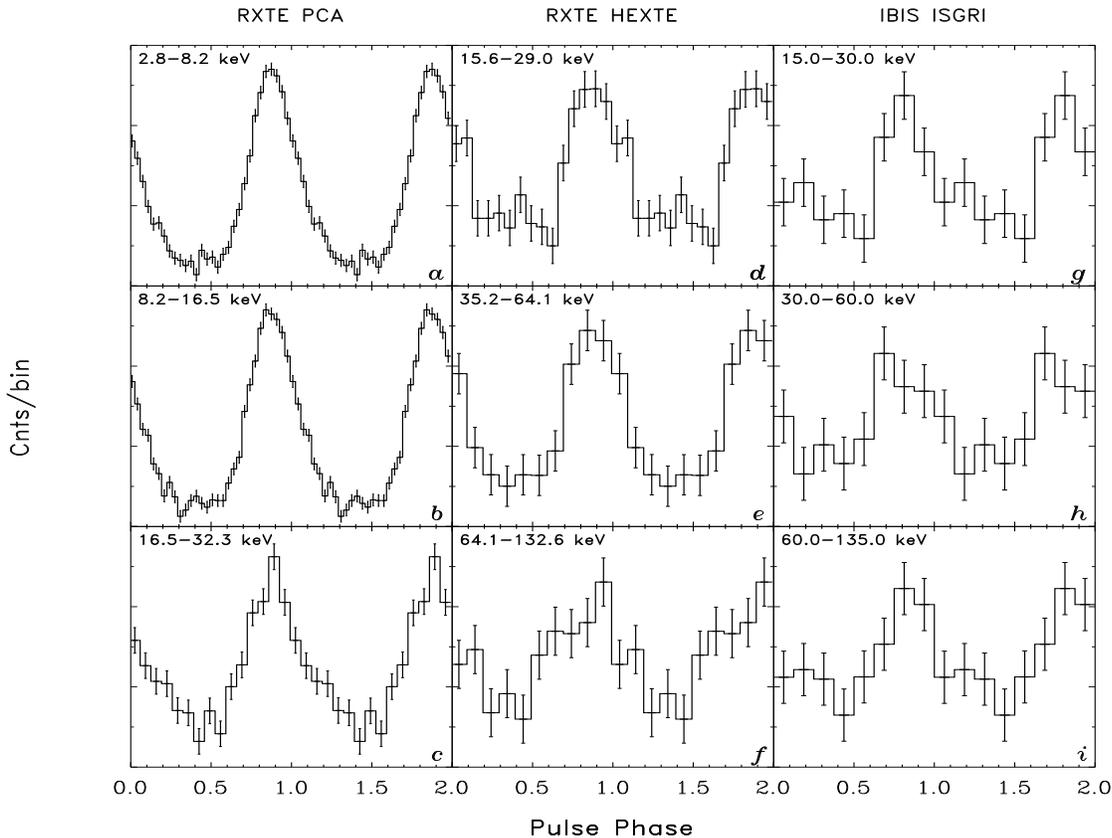}
     \caption{\label{psrg11_prof}{\bf \psrb;} pulse profiles in various energy bands for \rxte\ PCA (panels a--c), \rxte\ HEXTE (panels d--f)
               and \integral\ ISGRI (panels g--i). Pulsed emission has been detected up to $\sim 135$ keV.}
  \end{center}
\end{figure*}

Comparing the pulsed (\rxte\ PCA; this work) and total 2--10 keV flux values (\cxo; MOS 1+2; \sax) yields a (high) pulsed 
fraction in the range 79-92 \%. This value is consistent with the value, derived in this work, using data from merely \cxo\ 
HRC-S in timing mode. During a 78.2 ks exposure of 1E 161348-5055 on July 2, 2007 (obs.id. 7619) with the HRC-S \psrc\ was 
observed $7\farcm3$ off-axis. Using a $8\arcsec$ extraction radius, because of the strongly degraded PSF, and using an annulus 
centered on the pulsar with inner and outer radii of 12\arcsec and 16\arcsec, respectively, to determine the local background, 
we derived a genuine pulsed fraction (0.2-10 keV) of $78 \pm 2\% \pm 10\%$ taking into account the 10-15 \% contribution of the 
inner PWN to the total emission from the $8\arcsec$ circular (source) extraction region, which could not be resolved in this observation. 
The first error is related to the uncertainty in the inner PWN contribution and the second one to the uncertainty in the number 
of pulsed counts, derived through the method outlined by \citet{swanepoel1996}.

Also, above 20 keV there is very little room for pulsar DC and/or PWN emission.
At GeV energies no pulsed high-energy \gr-ray emission has been detected so far for \psrc\ \citep{abdo2013}.

At TeV energies ($> 200$ GeV) a bright source, HESS J1616-508, has been detected in the neighbourhood of \psrc\ \citep{aharonian2006}.  
However, due to (severe) misalignment a convincing identification at lower energies is lacking \citep[see e.g.][]{matsumoto2007,kargaltsev2009}.

\subsection{\psrcanda\ in SNR G338.3-0.0/HESS J1640-465}

HESS J1640-465 was discovered during the Galactic plane survey with H.E.S.S. performed during May-July 2004 \citep{aharonian2005,aharonian2006}. The source is
marginally extended at TeV energies and its location is consistent with the 8\arcmin diameter broken-shell SNR G338.3-0.0, which lies near the boundary of a bright 
H {\scriptsize{II}} region. At the center of SNR G338.3-0.0 an X-ray source, AX J1640.7-4632, was detected during the \asca\ Galactic plane survey \citep{sugizaki2001}.
 
HESS J1640-465 was observed by \xmm\ for 21.8 ks on August 20, 2005 with the EPIC-pn and MOS camera's operating in full-frame mode \citep{funk2007b}. 
This observation was strongly affected by soft proton flares reducing the effective exposure to only 7.3 ks. Three sources were detected in this observation of which 
XMM J164045.4-463131 is coincident with AX J1640.7-4632, and is extended in nature with a compact core and a faint tail, resembling morphologically a typical PWN. 
A spectral analysis of XMM J164045.4-463131, fitting an absorbed power-law model, yielded a rather strong absorbing Hydrogen column N$_{\hbox{\scriptsize H}}$ 
of $(6.1_{-0.6}^{+2.1})\times 10^{22}$ cm$^{-2}$ and a photon index of $-1.74\pm 0.12$. No shell-like X-ray emission was apparent in the XMM-observation.

The extended nature of XMM J164045.4-463131 was confirmed by \citet{lemiere2009}, who analysed a \cxo\ ACIS observation of HESS J1640-465 taken in May 2007 with an effective 
exposure time of 26.4 ks. Employing the sub-arcsecond scale spatial resolution of \cxo\ \citet{lemiere2009} performed a spatially resolved spectral analysis of the near field of
XMM J164045.4-463131 and found that the putative pulsar spectrum is very hard with photon index $-1.1\pm 0.4$ heavily absorbed through a Hydrogen column 
N$_{\hbox{\scriptsize H}}$ of $1.4\times 10^{23}$ cm$^{-2}$.  The unabsorbed 2--10 keV flux of the putative pulsar was $1.5\times 10^{-13}$ erg/cm$^2$s. 
The compact PWN has a softer photon index of $-2.5\pm 0.3$ and unabsorbed 2--10 keV flux of $4.2\times 10^{-13}$ erg/cm$^2$s, while the extended PWN is somewhat 
steeper with index $-2.7\pm 0.5$ and has an unabsorbed 2--10 keV flux of $4.6\times 10^{-13}$ erg/cm$^2$s.

Observations with \nustar\ (3-79 keV) of HESS J1640-465 on June 22, 2013 and Sept. 29, 2013 for 48.6 and 89.9 ks, respectively, revealed eventually the pulsed nature of 
XMM J164045.4-463131 \citep{gotthelf2014}. An energetic, $\dot{E} \simeq 4.4\times 10^{36}$ erg/s, 206 ms pulsar was found with a characteristic age of 3.35 kyr. 
The pulsar shows a single pulse - relatively sharp compared to a sinusoid - profile in the 3-25 keV band \citep[see][and also Fig. \ref{he_psr_morph} panel q]{gotthelf2014}, 
and thus can be considered as a soft \gr-ray pulsar. The pulsed fraction can be as high as $\sim 82\%$ taking into account both the 'normal' background and PWN contamination. 
The (total) pulsar spectrum combining \cxo\ and \nustar\ data, adopting an absorbed power-law model, has a hard photon index of $-1.3_{-0.5}^{+0.9}$ and an absorbed 2--10 keV flux of
$(1.8\pm 0.4)\times 10^{-13}$ erg/cm$^2$s, which translates to an unabsorbed 2--10 keV flux of $(3.63\pm 0.81)\times 10^{-13}$ erg/cm$^2$s taking into account the very high 
Hydrogen column density N$_{\hbox{\scriptsize H}}$ of $(1.8\pm 0.6)\times 10^{23}$ cm$^{-2}$. 

\citet{gotthelf2014} also searched for pulsed high-energy \gr-ray emission ($>100$ MeV) using \fermi\ LAT data collected over a more than 5 years time period, but found no 
significant pulsed signal. In the spatial domain, however, the \fermi\ LAT data \citep[first reported on by][]{slane2010} do show two distinct sources above 100 MeV in the 
region containing HESS J1640-465 \citep{lemoine2014}. In light of the new/updated findings of H.E.S.S. TeV measurements on HESS J1640-465 using all data collected between 
May 2004 and Sept. 2011 \citep{hesscol2014} and on the recently discovered (hard) nearby HESS J1641-463 (coincident with radio SNR G338.5+0.1), unnoticed using default 
detection techniques, but detected using much higher energy thresholds \citep{hesscol2014b}, \citet{lemoine2014} revisited the \fermi\ LAT data and reported a power-law 
spectrum with index $-1.99 \pm 0.04 \pm 0.07$ for HESS J1640-465, which smoothly connects to the TeV data points, and a soft power-law spectrum with index 
$-2.47 \pm 0.05 \pm 0.06$ for HESS J1641-463, requiring a significant hardening towards TeV energies. The new GeV/TeV data suggest that a 
significant part of the high-energy \gr-ray emission from HESS J1640-465 originates from proton acceleration in the supernova remnant shell \citep{hesscol2014}.

\citet{castelletti2011} presented a high-resolution imaging study of the G338.3-0.0 supernova remnant at radio frequencies acquired by the Giant Metrewave Radio
Telescope GMRT (235, 610 and 1280 MHz) and ATCA (1290 and 2300 MHz), and searched for a radio pulsar at 610 and 1280 MHz in a beam covering the XMM J164045.4-463131 
location (coincident with \psrcanda). This search yielded no
radio pulsar counterparts and resulted in flux density upper limits of 2 and 1 mJy for 610 and 1280 MHz, respectively, at $8\sigma$ confidence level.


\subsection{\psrb}
\label{psrj1811_section}
\psrb\ was discovered by \citet{torii1997} during \asca\ (0.5-10 keV) observations of G11.2-0.3, known to be a composite supernova remnant, proposed to be associated with 
the historical supernova of A.D. 386 \citep{vasisht1996}.
They found a 65 ms pulsar exhibiting a hard X-ray spectrum. Subsequent \asca\ and \sax\ observations \citep{torii1999}
allowed the measurement of the spin-down and yielded a characteristic age of $2.4\times 10^4$ yr.
No pulsed radio emission has so far been detected from the sky region containing the pulsar \citep{crawford1998}.
A high precision \cxo\ X-ray imaging observation \citep{kaspi2001} revealed at the very center of the SN-remnant
the pulsar counterpart at R.A. $18^{\hbox{\scriptsize h}} 11^{\hbox{\scriptsize m}} 29\fs22$, Decl. $-19\degr 25\arcmin 27 \farcs 6$ (Epoch J2000), embedded in a diffuse 
nebula (PWN). This provided strong evidence that the system is much younger 
than the characteristic age suggests. Assuming conventional spin-down with a constant magnetic field and breaking 
index {\it n\/} (from the assumed spin-down law: $\dot\nu \propto \nu^n$) both ages can only be reconciled if the 
pulsar was initially born with a spin period of $\sim 62$ ms unless the breaking index is unusally large.

\begin{figure}
  \begin{center}
     \includegraphics[width=8cm]{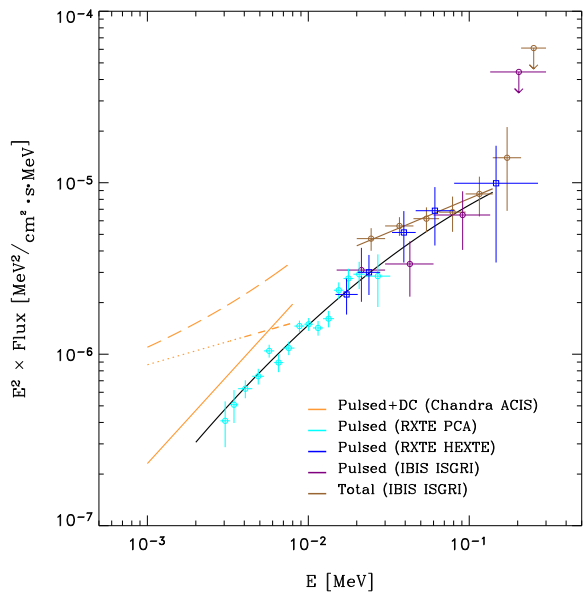}
     \caption{\label{psrg11_spc}{\bf \psrb;} high-energy (1-300 keV; unabsorbed) total and pulsed spectra as derived from measurements
      by \cxo\ ACIS (1-7 keV; pulsar Pulsed plus DC (=total; orange solid); PWN (orange dotted); PWN plus pulsar (orange long dashed)), \rxte\ PCA (aqua) and HEXTE (blue; both pulsed) and \integral\ ISGRI (total (brown) / pulsed (purple)). The ISGRI total flux measurements still contain contributions from the PWN. The spectral measurements are generally consistent with reaching a pulsed fraction of 100\% near 50 keV.}
  \end{center}
\end{figure}

Subsequent \rxte\ PCA X-ray monitoring over 992 days (March 8, 2002 -- November 24, 2004; MJD 52341 -- 53333) 
allowed us to generate phase coherent timing models (see Table \ref{eph_table}) and using these to detect the pulsed signal up to 
$\sim 100$ keV in \rxte\ HEXTE data \citep[see also][]{roberts2004}. A second \rxte\ PCA monitoring campaign was initiated on August 25, 2007
and ended on October 31, 2011 near the cessation of the \rxte\ mission. The data from this latter campaign are not included in this study.

We performed a timing analysis of \integral\  ISGRI data for the time span covering \integral\ revolutions 46-249 (February 28, 2003 -- October 28, 2004; MJD 52698 -- 53306), for which we derived accurate ephemerides (see Table \ref{eph_table}). We found
a $6.5\sigma$ signal for the 15-135 keV energy band showing a morphology consistent with that derived from 
\rxte\ PCA and HEXTE data. Pulse profiles in differential energy bands for \rxte\ PCA \& HEXTE and \integral\ ISGRI are shown in Fig. \ref{psrg11_prof}. The significances (adopting $Z_1^2$-statistics) for a deviation from a uniform distribution are: $35.6\sigma, 
35.9\sigma$ and $15.9\sigma$ for the \rxte\ PCA ranges 2.8-8.2, 8.2-16.5 and 16.5-32.3 keV (panels a--c), respectively; $11.2\sigma, 
8.6\sigma$ and $4.4\sigma$ for the \rxte\ HEXTE ranges 15.6-29.0, 35.2-64.1 and 64.1-132.6 keV (panels d--f), respectively, and finally,
$3.7\sigma, 3.6\sigma$ and $3.4\sigma$ for the \integral\ ISGRI ranges 15.0-30.0, 30.0-60.0 and 60.0-135.0 keV (panels g--i), respectively.
We derived the pulsed excess counts in differential energy bands with pulse-profile fitting, which subsequently have been converted to
photon fluxes taking into account the energy responses of the involved instruments. These (unabsorbed) pulsed flux measurements are shown 
in Fig. \ref{psrg11_spc} as aqua (PCA), blue (HEXTE) and purple (ISGRI) data points, along with the best-fit ``curved" power-law model (black) resulting from a combined fit. Pulsed fluxes derived from this fit for the 1--10 and 20--100 keV energy bands are $(2.21\pm 0.47)\times 10^{-12}$ and $(1.22\pm 0.21)\times 10^{-11}$ erg/cm$^2$s, respectively.

\citet{roberts2003} presented detailed results from spatially resolved spectral analysis for the point-source (=pulsar) in G11.2-0.3, its PWN 
and other structures in its near environment using \cxo\ ACIS data. The power-law model fits to the spectral data of the pulsar (=total emission), PWN and their combination are shown in Fig. \ref{psrg11_spc} for the 1--8 keV band. The listed absorbed (private communication M. Roberts) energy flux of the pulsar in the 1--10 keV band, $(2.82\pm 0.12)\times 10^{-12}$ erg/cm$^2$s has been converted to its unabsorbed value of $(3.48\pm 0.15)\times 10^{-12}$ erg/cm$^2$s using the estimated column density N$_{\hbox{\scriptsize H}}$ of $2.22 \times 10^{22}$ cm$^{-2}$ . Using our value for the pulsed flux, the pulsed fraction in the 1--10 keV band becomes $0.64\pm 0.14$.

In the imaging domain we detected soft \gr-ray emission up to $\sim 150$ keV using archival data 
\citep[\integral\ revolutions 46-495; February 28, 2003 -- October 5, 2006, 11.35 Ms GTI exposure; this work; see also e.g.][]{bird2007,dean2008}. 
The total (= pulsed plus unpulsed from \psrb\ and DC from the PWN and SN-remnant) emission spectrum of this source for energies 
above 20 keV can be described by a power-law model with a (hard) index of $-1.61\pm 0.15$ and a 20-100 keV flux of $(1.54\pm 0.12)\times 10^{-11}$ erg/cm$^2$s (see Fig. \ref{psrg11_spc}; brown data points). This is consistent with the index and flux derived by \citet{dean2008}, who used slightly less \integral\ exposure on the source. The pulsed fraction in the 20--100 keV band is $0.79\pm 0.15$, and
above $\sim 50$ keV the pulsed PCA/HEXTE/ISGRI spectrum is consistent with the total ISGRI spectrum, thus 
the pulsed fraction becoming consistent with 100\% for energies beyond. Therefore, the spectrum of the PWN (and SN-remnant) above $\sim 50$ keV has to bend down. 

At GeV energies and in the TeV domain no (pulsed) source with a plausible association has been detected in the vicinity of \psrb\  \citep{aharonian2007,nolan2012,abdo2013}.

\subsection{\psrcandb}

\psrcandb\ was detected in blind frequency searches using \fermi\ LAT data by \citet{abdo2009d}. With a pulse period of 48.1 ms and characteristic age of 43 kyr 
it is the second most energetic ($L_{sd}=6.3\times 10^{36}$ erg/s) (radio-quiet) blind-search pulsar. It showed a double peaked \gr-ray ($>100$ MeV) pulse profile with a peak 
separation of $0.485(3)$ and with bridge emission between the peaks. Two timing glitches have been detected in the \gr-ray data \citep{ray2011b,marelli2014b}. 
A plausible X-ray counterpart to \psrcandb\ was found in a short \swift\ XRT observation \citep{abdo2009d,ray2011b}. 

In a deep 108.9 ks \xmm\ observation performed on March 10, 2013 with the EPIC-pn operated in small window mode highly significant pulsations were found 
with a lightcurve consisting of two peaks separated $\sim 0.5$ in phase lagging the \gr-ray one by $\sim 0.25$ in phase \citep{marelli2014b}. 
The pulsed emission is very hard with a photon index of -0.85(3) and the pulsed fraction is very high, $96\pm 3\%$. We re-analysed the \xmm\ EPIC-pn 
data and confirmed the findings of \citet{marelli2014b}. The unabsorbed 2--10 keV pulsed flux is $(0.96\pm 0.015)\times 10^{-12}$ erg/cm$^2$s. The 8-12 keV 
EPIC-pn pulse profile of \psrcandb\ is shown in Fig. \ref{he_psr_morph} panel r. We verified that very significant ($12.1 \sigma$) pulsed emission is detected above 10 keV. 


\begin{figure}
  \begin{center}
     \includegraphics[width=8.25cm,bb=60 165 530 630,clip=]{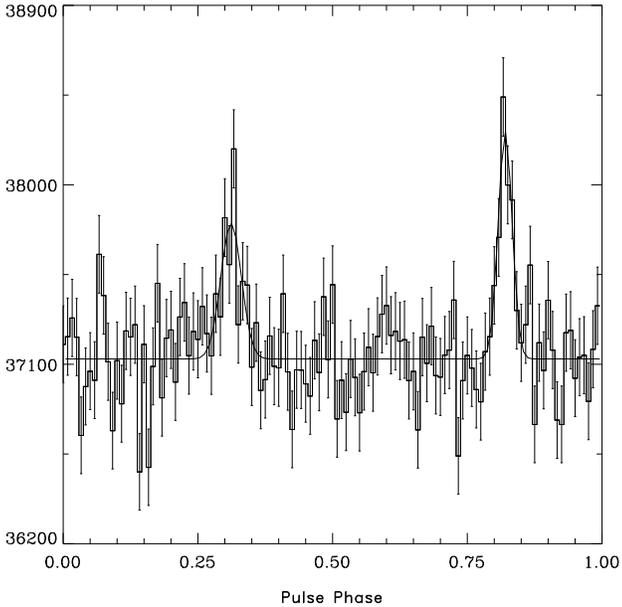}
     \caption{\label{psr1813pca}{\bf \psrcandb;} \rxte\ PCA 120 bins pulse profile for the 1.9-29.3 keV band. 
     A double Gaussian plus background fit is superposed, yielding a $10.5 \sigma$ improvement with respect to a flat background 
     description.}
  \end{center}
\end{figure}

In the on-line version of the \integral\ IBIS 9-year Galactic hard X-ray survey 
\citep[][see also {\it http://hea.iki.rssi.ru/integral/nine-years-galactic-survey}]{krivonos2012} the 35-80 keV map of the Galactic bulge region 
(6.78 Ms exposure time) shows a $3.1\sigma$ excess at the location of \psrcandb\ representing a (total) flux of $0.43\pm 0.14$ mCrab, which 
corresponds to $(1.05\pm0.34)\times 10^{-6}$ ph/cm$^2$s keV.

The hardness of the pulsed spectrum in the \xmm\ bandpass and the detection of a point-source in the 35-80 keV \integral\ ISGRI band makes it 
very plausible that \psrcandb\ shows pulsed emission at detectable levels in the hard X-ray/soft \gr-ray band. 
In the X-ray archives we found a 26.8 ks \rxte\ PCA observation (obs. id. 20090; taken in {\it E\_125us\_64M\_0\_1s} event mode with a $\sim 122.07 \mu$s 
time resolution, and all PCU's operational) performed on Nov. 7, 1997 targeting at GRO J1814-12, which has \psrcandb\ in the field of view 
at an off-axis angle of $23\farcm7$. Also, the much stronger LMXB 4U 1812-12 was in the field of view at an off-axis angle of $24\farcm3$ yielding 
increased (non-Gaussian) background levels in timing searches of \psrcandb\ in PCA data. In spite of the more than 11 years time lapse between the 
detection as (blind search) \gr-ray pulsar and the \rxte\ observation a restricted period search ($Z_8^2$ test, which turned out to be optimal for the 
EPIC-pn data, for each trial frequency) around the predicted pulse period adopting the timing parameters as given in supplementary material provided 
by \citet{abdo2013} yielded a highly significant pulsed signal, $8.5\sigma$ single trial, at 20.8047193(8) Hz (epoch MJD 50759) for the 1.9-29.3 keV 
band. Taken into account the number of scan steps of $\sim 830$ independent Fourier steps the detected signal remains still very significant. 

The 1.9-29.3 keV PCA pulse profile with superposed the best fit double Gaussian model is shown in Fig. \ref{psr1813pca}. The phase separation between 
the peaks (peak 2 is the most dominant pulse at phase $\sim 0.82$ in Fig. \ref{psr1813pca}) $\Delta\Phi_{2-1}$ is $0.509\pm 0.004$, consistent with
the \xmm\ value of $0.5043 \pm 0.0009$ \citep[see][]{marelli2014b}.
The X-ray pulses, however, are much narrower, $\sigma_1=0.019\pm 0.005$ and $\sigma_2=0.013\pm 0.002$, than observed by XMM EPIC-pn with $\sigma_1=0.0330\pm 0.0008$ and $\sigma_2=0.0305\pm 0.0005$, because of the much better time resolution of the PCA observation with repect to the \xmm\ observation, 
i.e. $\sim 122.07\ \mu$s versus $\sim 5.7$ ms.

For the differential PCA energy bands, 3.9-7.9, 7.9-15.1 and 15.1-33.5 keV, we obtained the following $Z_8^2$-test significances, $5.3\sigma, 4.3\sigma$ and $3.3\sigma$, 
respectively, proving \psrcandb\ to be a soft \gr-ray pulsar. Unfortunately, the statistics are too poor to derive reliable pulsed flux estimates for these PCA energy bands. 

The spectrum of the pulsed emission of \psrcandb\ is shown in the right panel of Fig. \ref{spectralcompilation} (red-red orange). The X-ray data points 
are derived in this work from the EPIC-pn data (template fitting per energy band, followed by a spectral fit adopting a power-law photon flux model), 
while the high-energy \gr-ray spectral points and fit are adopted from \cite{abdo2013}. This figure also includes the 35-80 keV total flux measurement of \integral\ ISGRI. The latter flux is a good estimate of the pulsed flux, because the pulsed fraction is consistent with 100\%.

No TeV counterpart has been detected so far for \psrcandb.

\subsection{\psri: The pulsar associated with HESS J1813-178/SNR G12.82-0.02/IGR J18135-1751}
\label{sectpsri}
\citet{aharonian2005} announced the detection of eight unknown VHE sources during scans of the inner galactic plane with the H.E.S.S. telescope from May to July 2004. One of these new sources is the compact TeV source HESS J1813-178 embedded in a highly obscured region, which lacked at the time of discovery plausible counterparts at other wavelength and was classified as a ``dark'' particle accelerator. Soon after its detection as TeV source a soft \gr-ray counterpart, IGR J18135-1751 \citep{ubertini2005}, of HESS J1813-178 was discovered. Moreover, radio observations revealed the presence of non-thermal radio emission from a young shell-type SNR G12.82-0.02 with a diameter of $\sim 2\farcm5$ \citep{brogan2005}. These properties do conjecture that we are dealing with a PWN powered by a so far undetected energetic pulsar. 

\begin{figure}
  \begin{center}
     \includegraphics[width=8.5cm]{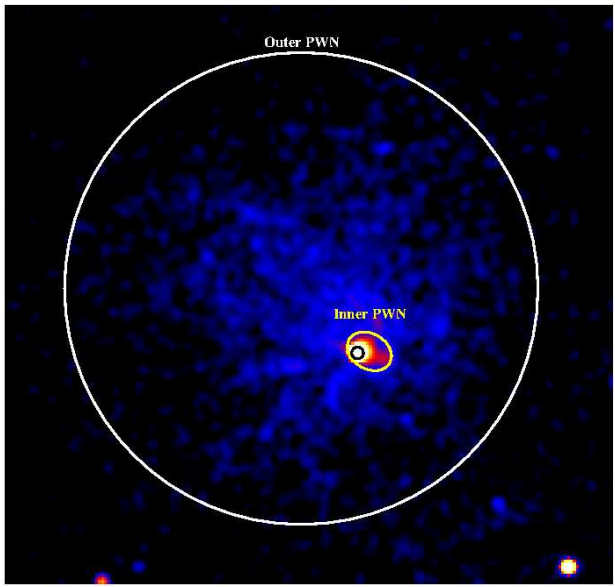}
     \caption{\label{psrj1813cxoacis} \cxo\ ACIS-I image (30 ks) of G12.82-0.02 showing clearly the point-source, the (putative) pulsar \psri\ (black $2\arcsec$ circle), 
     the inner PWN confined within a $6\arcsec \times 8\arcsec$ (yellow) elliptical region and the outer PWN confined to a $80\arcsec$ circular region offset from the point source (large white circle).}
  \end{center}
\end{figure}

Follow-up X-ray observations with \cxo\ \citep{helfand2007} and \xmm\ \citep{funk2007} indeed revealed the putative pulsar and its nebula, although pulsations still had to be detected, because the operation modes of the involved X-ray instruments had insufficient time resolution to detect pulsations at time scales of 10--100 milli-second. 

A 30 ks \cxo\ ACIS-I image of the near environment of G12.82-0.02 is presented in Fig. \ref{psrj1813cxoacis} and shows the extraction regions used in \citet{helfand2007} to derive the spectra of the (putative) pulsar (circular aperture of $2\arcsec$ in radius centered on the point source), the inner pulsar wind nebula ($6\arcsec \times 8\arcsec$ elliptical region excluding the point source region) and the outer pulsar wind nebula encompassing the bulk of the nebula extent (a $80\arcsec$ radius circular region offset from the point source excluding the point source and the inner nebula). Fitting absorbed power-law models to the various components \citep{helfand2007} yielded for the outer PWN a Hydrogen column density N$_{\hbox{\scriptsize H}}$ of $(9.8 \pm 1.2)\times 10^{22}$ cm$^{-2}$ (90\% confidence), consistent with the \xmm\ findings \citep{funk2007}, indicating highly absorbed emission. Both the pulsar and the outer PWN show hard power-law indices of about $-1.3$, while the inner nebula has even a 
harder index of $-0.4$ (see e.g. the spectral compilation in Fig. \ref{psrj1813_spc} for the fits of the inner PWN, total pulsar emission and its combination as dotted, solid and dashed orange lines, respectively), a clear manifestion of Synchrotron cooling effects. 
The total flux in the nebula is about 4.3 times larger than that measured for the point-source.

\begin{figure}
  \begin{center}
     \includegraphics[width=8.cm,height=12.cm,bb=130 170 445 630,clip=]{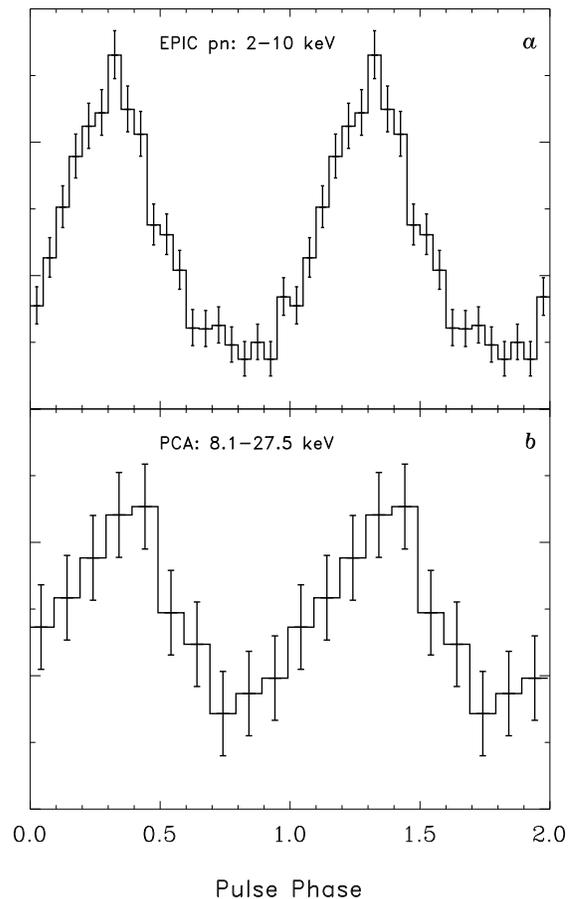}
     \caption{\label{psrj1813pulseprofiles}{\bf \psri;} the upper panel shows the pulse profile for the 2--10 keV band as measured by \xmm\ 
                                            EPIC pn during a 98 ks observation performed in small-window mode. In the lower panel the \rxte\ PCA 
                                            lightcurve of \psri\ is shown for the 8.1-27.5 keV band ($Z_1^2=4.8\sigma$) obtained from dedicated 
                                            HESS J1813-178 observations performed about 500 days before the discovery of its pulsed nature. }
  \end{center}
\end{figure}

\begin{figure}
  \begin{center}
     \includegraphics[width=7.5cm,height=8.5cm,bb=65 163 535 660,clip=]{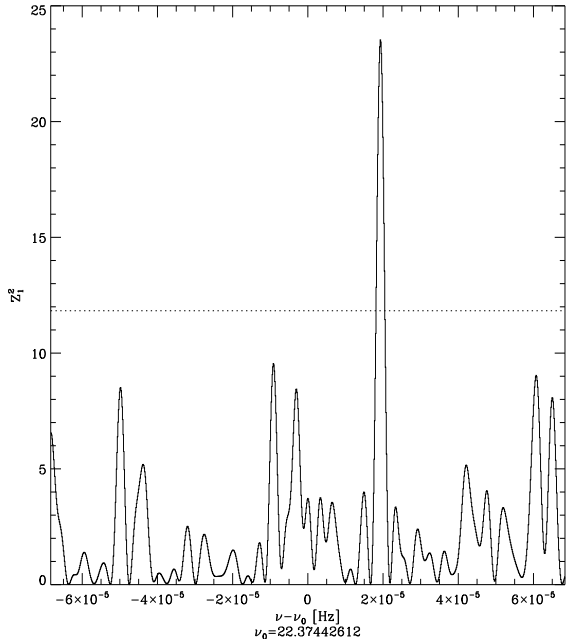}
     \caption{\label{psrj1813pcaperiodogram}{\bf \psri;} periodogram showing $Z_1^2$ versus trial frequency in a 50 IFS window around the predicted frequency $\nu_0$ using events 
     from PCU-2 with PHA in the range 5--50 and employing all detection layers. A $4.6\sigma$ single trial peak maximum is found 6.8 IFS shifted from the prediction, resulting in an overall detection significance of $3.7\sigma$. The dotted line indicates the $3\sigma$ confidence level for a single trial.}
  \end{center}
\end{figure}

Targeting at the central point source in G12.82-0.02 \citet{gotthelf2009} finally discovered 44.7 ms pulsations analysing a 98 ks \xmm\ observation performed on March 27, 2009, with the EPIC pn operating in small-window mode providing a time resolution sufficient to detect timing signals up to $\sim 125$ Hz. Follow-up observations with \cxo\ (ACIS-S3 Continuous Clocking; Feb. 12, 2012) and \xmm\ EPIC pn (small-window mode; March 13, 2011) made it possible to derive a reliable period derivative estimate of \psri, yielding a characteristic age of $5.6$ kyr and a spin-down power of $5.6\times 10^{37}$ erg/s \citep{halpern2012}.

We re-analyzed the EPIC pn data of the 98 ks \xmm\ observation (obs. id. 0552790101), performed in small-window mode, with as main goal to derive the pulsed spectrum (not published so far) and from this the pulsed fraction. The pulsed signal was clearly visible at the predicted frequency \citep{gotthelf2009} and the pulse profile for the 2--10 keV energy band is shown in the upper panel of Fig. \ref{psrj1813pulseprofiles}. Below $\sim 2$ keV no pulsed signal can be detected. The $21.1\sigma$ 2--10 pulse profile was fitted with a truncated Fourier-series using the fundamental and one harmonic, and the model has subsequently been used in the extraction process of pulsed excess counts. These pulsed excess counts have been converted to pulsed flux values adopting an absorbed power-law model in a forward folding fit procedure using response information taking into account the $15\arcsec$ extraction radius. 

For comparison purposes with the spectral results determined by \citet{helfand2007} for the total point source emission we fixed the Hydrogen column density N$_{\hbox{\scriptsize H}}$ to $9.8\times 10^{22}$ cm$^{-2}$. We derived a photon index of $-1.30\pm 0.03$ and an unabsorbed/absorbed 2--10 keV pulsed flux of $(9.2 \pm 0.45)\times 10^{-13} / (5.8 \pm 0.82)\times 10^{-13}$ erg/cm$^2$s. This translates, using the absorbed total 2--10 keV flux value of $1.3\times 10^{-12}$ erg/cm$^2$s as derived by \citet{helfand2007}, to a (2--10 keV) pulsed fraction of $0.45 \pm 0.06$, consistent with that estimated by \citet{gotthelf2009,halpern2012}. The EPIC pn pulsed flux measurements and its best fit are shown in the spectral compilation depicted in Fig. \ref{psrj1813_spc} as dark-orange/red datapoints and solid line, respectively.

We also performed a spatial analysis of the EPIC pn small-window mode data by fitting (adopting Poissonian statistics) a model composed of the EPIC pn PSF, centered on the pulsar's X-ray counterpart, and a (locally) flat background model to the measured 2d-event distribution (extraction radius $30\arcsec$) for every required energy band. Because the EPIC pn PSF is worse than that of \cxo\ ACIS-I the extracted total counts are  composed of the total emission component of the pulsar and a contribution from the inner pulsar wind nebula, which can not be resolved by XMM.   

Initially, we fitted the measured count spectrum with an absorbed power-law model with N$_{\hbox{\scriptsize H}}$ fixed to $9.8\times 10^{22}$ cm$^{-2}$. However, this rendered a rather poor fit ($\chi_{r,17-2}^2 = 28.90/15=1.93$; 1.5\% probability that a good model gives such a $\chi^2$). The flux measurements and its best fit, adopting N$_{\hbox{\scriptsize H}} \equiv 9.8 \times 10^{22}$ cm$^{-2}$, are shown in Fig. \ref{psrj1813_spc} as red datapoints and solid line, respectively.   We also left N$_{\hbox{\scriptsize H}}$ free in the absorbed power-law fit and obtained a good fit with $\chi_{r,17-3}^2 = 18.34/14=1.31$ yielding a N$_{\hbox{\scriptsize H}}$ of $(11.7 \pm 0.35)\times 10^{22}$ cm$^{-2}$, slightly higher, but still consistent with 
the value derived by \citet{helfand2007} using \cxo\ ACIS-I data. The unabsorbed / absorbed 2--10 keV flux was $(3.30 \pm 0.05)\times 10^{-12} / (1.94 \pm 0.11)\times 10^{-12}$ erg/cm$^2$s and the photon-index $-1.31 \pm 0.01$. The absorbed 2--10 keV flux value is comparable with the sum of the point source - and inner nebula flux of $\sim 1.7 \times 10^{-12}$ erg/cm$^2$s as given in \citet{helfand2007}.

We also revisited PCA/HEXTE data from a dedicated \rxte\ observation of HESS J1813-178 (observation identifier 93022) performed during Nov. 16--20, 2007 (about 500 days before the detection of the pulsed nature) with only one operational PCU and at a $28\farcm 3$ offset angle from \psri, thereby considerably reducing the PCA overall sensitivity. The availability of an (incoherent) ephemeris \citep{halpern2012} with a sufficiently accurate period derivative made a restricted period search around the predicted (backwards extrapolated) pulse frequency value feasible.

Selecting PCA events with PHA values in the range 5--50 ($\sim 2-20$ keV) using all detection layers from PCU-2, the only operating unit during the 4-days covering observation (MJD 54420.101-54424.332) of about 125 ks, this search yielded a $4.6\sigma$ single trial peak shifted 6.8 Independent Fourier Steps (IFS$=1/\tau$, in which $\tau$ is the full time stretch of the observation) from the prediction. Taking into account the number of search steps (50 IFS) this results in an overall detection significance of $3.7\sigma$. The periodogram covering 50 IFS and showing $Z_1^2$ versus trial frequency near the predicted frequency, is depicted in Fig. \ref{psrj1813pcaperiodogram}. We oversampled, for display purposes, one IFS by a factor of 20 to follow the peak structure in detail. This detection of pulsed emission by the PCA at a frequency shifted marginally\footnote{The {\it incoherent} ephemeris given by \citet{halpern2012} does not contain a $\ddot\nu$ value, and young pulsars often show glitches making 
precise predictions impossible given the 500 days back extrapolation.} from the prediction is very plausible, because 
(W3)PIMMS\footnote{Portable, Interactive Multi-Mission Simulator; a HEASARC tool to convert count rates/fluxes observed for one high-energy instrument into that for another one, 
adopting a (photon) spectral model and absorbing Hydrogen column. Online documentation can be found at {\it http://heasarc.gsfc.nasa.gov/docs/software/tools/pimms.html}. } predicts for the PCA with {\it one\/} operational PCU at an offset of about $30\arcmin$, reducing the source count rate by $\sim 50\%$, an observation time of 111 ks to reach a $5\sigma$ signal in the 0-49 PHA range given the 2--10 keV pulsed flux measured by XMM EPIC pn. 

We derived pulsed excess counts for three broad PCA energy bands (see e.g. panel b of Fig.\ref{psrj1813pulseprofiles} for the 8.1-27.5 keV lightcurve) by (XMM EPIC pn) template fitting and converted these to flux values (see Fig. \ref{psrj1813_spc}; aqua data points) using proper response information and adopting an absorbed power-law model with  N$_{\hbox{\scriptsize H}}$ fixed to $9.8\times 10^{22}$ cm$^{-2}$. In HEXTE ($\ga 15$ keV) data the pulsed emission could not be detected.

\begin{figure}
  \begin{center}
     \includegraphics[width=8.2cm,bb=50 155 550 660,clip=]{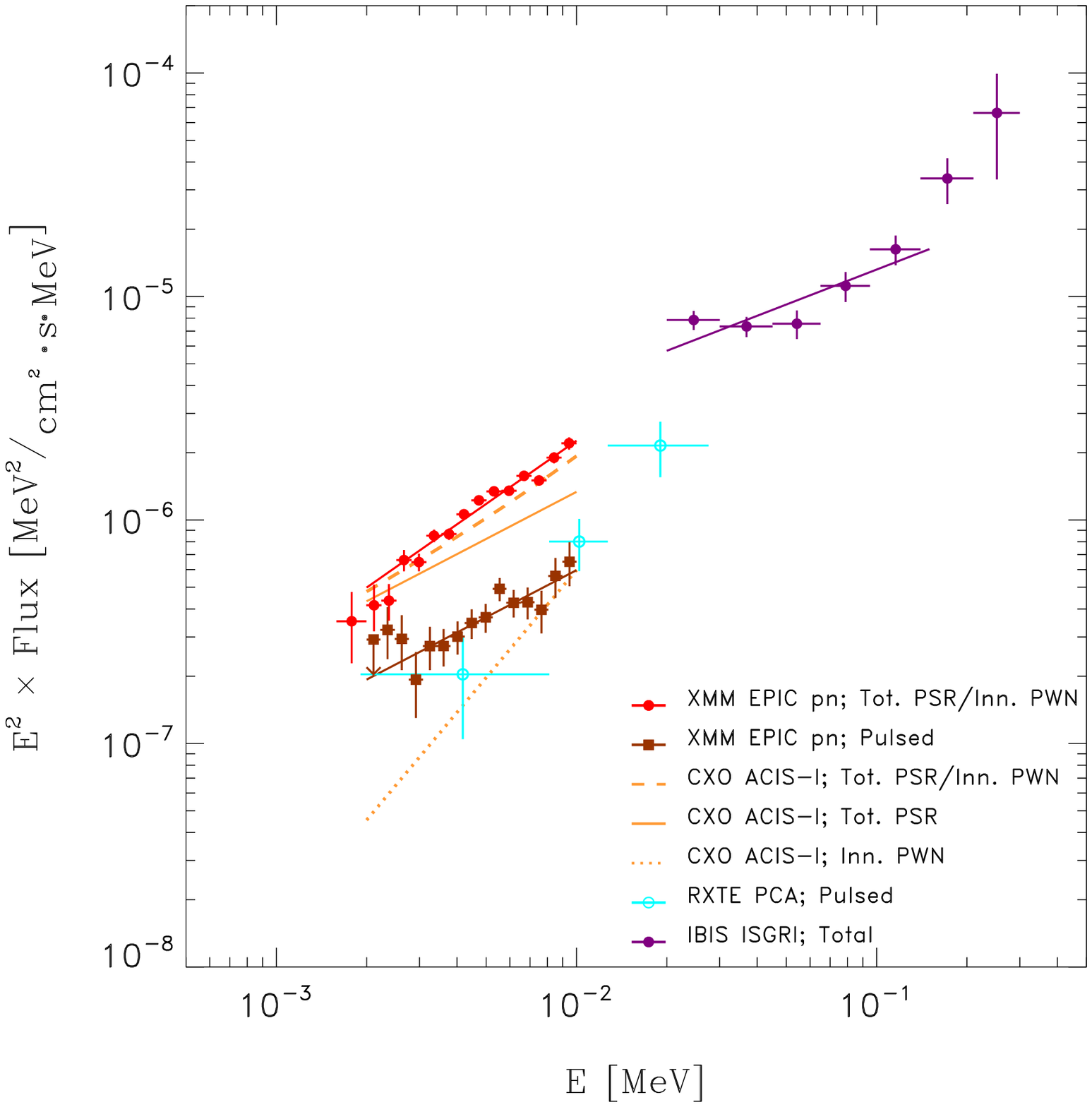}
     \caption{\label{psrj1813_spc}{\bf \psri;} unabsorbed high-energy spectrum of the pulsed emission along with that of its inner PWN as obtained by CXO ACIS-I \citep[2--10 keV; inner PWN, total pulsar and its combined emission;][]{helfand2007}, \xmm\ EPIC pn (2--10 keV; pulsed, total pulsar plus inner PWN; this work), \rxte\ PCA (2--30 keV; pulsed; this work) and \integral\ ISGRI (20--300 keV; total=total pulsar plus inner PWN and whole PWN; this work). An absorbing Hydrogen column density of $9.8\times 10^{22}$ cm$^{-2}$ has been assumed for all X-ray measurements to facilitate the comparison of CXO ACIS-I and \xmm\ EPIC pn spectral results.}
  \end{center}
\end{figure} 

In the soft \gr-ray band we obtained spectral information on the total emission of \psri\ from the \integral\ ISGRI skymaps produced in seven differential energy bands. These skymaps were centered on \psrb\ (see Section \ref{psrj1811_section}), which is only $1\fdg 7$ away from \psri. 
The flux measurements are shown as purple data points in Fig. \ref{psrj1813_spc}. Fitting a power-law model we derived a photon index of $-1.48 \pm 0.08$ and a 20--100 keV flux of $(2.30 \pm 0.13)\times 10^{-11}$ erg/cm$^2$s, consistent with the value given by \citet{ubertini2005}. Note, that the ISGRI fluxes, due to the lack of resolving power at arcsecond scales, represent the combined emission from the pulsar (pulsed and unpulsed), the inner PWN and outer PWN.

In the hard \gr-ray band no pulsed emission from \psri\ has been detected by the \fermi\ LAT instrument \citep{abdo2013}.

Finally, in the radio-band no pulsed emission has been detected so far at 1374 MHz \citep{helfand2007} using the ATNF Parkes telescope and at 2 GHz \citep{halpern2012} using the NRAO Green Bank Telescope
(GBT) telescope, rendering one of the deepest flux upper limits obtained so far for a young radio pulsar ($< 0.006$ mJy at 2 GHz, equivalent to $< 0.01$ mJy at 1.4 GHz).


\subsection{\psraxj\ / \axj}

The detection of soft \gr-ray emission up to $\sim 300$ keV from the \asca\ source \axj\ has been reported by \citet{malizia2005} using \integral\ ISGRI data. 
Its location made an association with TeV source HESS J1837-069 \citep{aharonian2005} plausible, and thus suggests a pulsar/PWN origin. 
This was indeed confirmed by \citet{gotthelf2008}, who detected a young (23 kyr) 70.5 ms pulsar in the \integral\ ISGRI error circle using \rxte\ PCA data. Its spin-down rate, energetics and spectra were reported by \citet{kuiper2008}. So far, no radio emission associated with the pulsar in \axj\ has been reported  \citep[see e.g. Sect. 5 of][who report a flux limit of 1-2 mJy at 1.4 GHz]{malizia2005}.

Since its identification as a pulsar early 2008 with \rxte\, \axj\ has been monitored up to December 6, 2010. We have made accurate phase-coherent timing models (see Table \ref{eph_table}) for the period February 17, 2008 -- December 6, 2010, during which a large timing glitch has been detected, occuring somewhere between MJD 55002 and MJD 55018 (June 20 -- July 6, 2009) with a fractional frequency ($\Delta\nu/\nu$) jump size of $(1.55\pm 0.07)\times 10^{-6}$ \citep{kuiper2010a}. 

\begin{figure}
  \begin{center}
     \includegraphics[width=8cm,height=14cm,angle=0,bb=125 95 465 695 ]{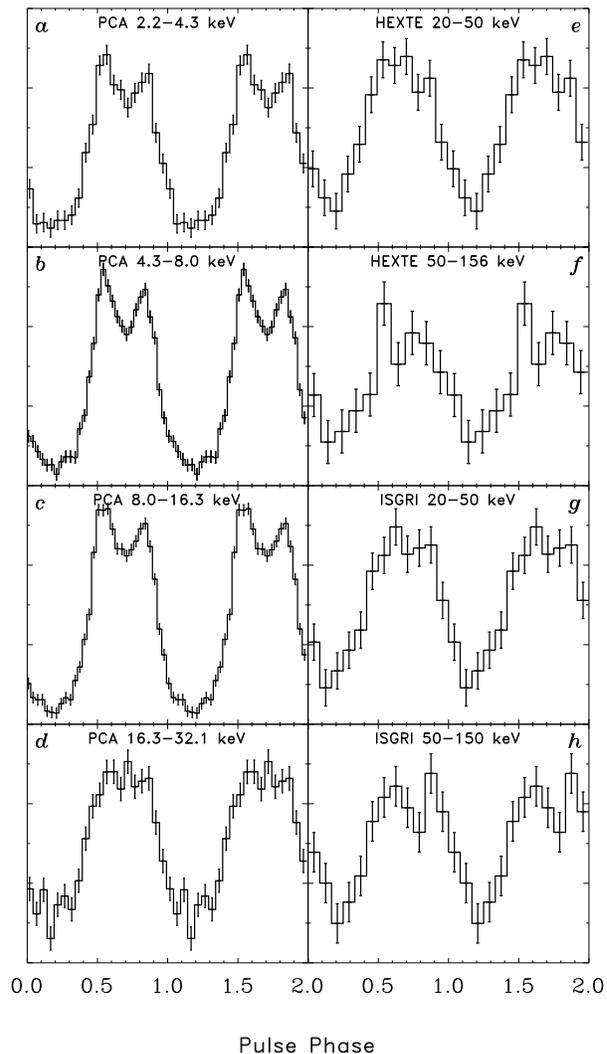}
     \caption{\label{axj1838_prof}{\bf \axj;} pulse profiles in various energy bands for \rxte\ PCA (panels a--d), \rxte\ HEXTE 
                                  (panels e--f) and \integral\ ISGRI (panels g--h). Pulsed emission has been detected up to $\sim 150$ keV.}
  \end{center}
\end{figure}

Applying these ephemerides in a timing study of \rxte\ PCA and HEXTE data collected during observation programs P93429 and P94303 (February 17, 2008 -- December 5, 2009; MJD 54513 -- 55170) resulted in the pulse phase diagrams shown in Fig. \ref{axj1838_prof}a--f. The screened (and PCU-detector break-down cleaned) exposure times for units 0--4 are: 60.34, 69.24, 256.60, 45.45 and 72.21 ks, respectively. For HEXTE the deadtime-corrected on-exposure times for the same period are 163.13 and 85.51 ks, for cluster-0 (staring on-source) and 
cluster-1 (rocking), respectively. Below $\sim 16$ keV a structured broad pulse is apparent.

Application of the same timing models in a timing analysis of ISGRI data, spanning \integral\ revolutions 592 - 910 (August 19, 2007 -- March 29, 2010; MJD 54331 -- 55284\footnote{For the period August 19, 2007 -- February 17, 2008 (MJD 54331-54513) the first \rxte\ PCA ephemeris of \axj\ (see Table \ref{eph_table}) is used for ISGRI data taken before the monitoring of \axj\ commenced with \rxte\ (thus in backwards extrapolation).
The pulsed signal (with proper alignment) could easily be recognized in the ISGRI data during this period not covered by the validity interval of the first ephemeris. This proves that the ephemeris is also valid across a larger (backwards extended) time period.}; 3.3 Ms of screened vignetting corrected effective on-axis exposure), revealed a $12.1\sigma$ pulsed signal in the 15-150 keV band. 
The significance of the pulsed signal above 50 keV is still $7.2\sigma$ (see Fig. \ref{axj1838_prof}h). 

\begin{figure}
  \begin{center}
     \includegraphics[width=8cm]{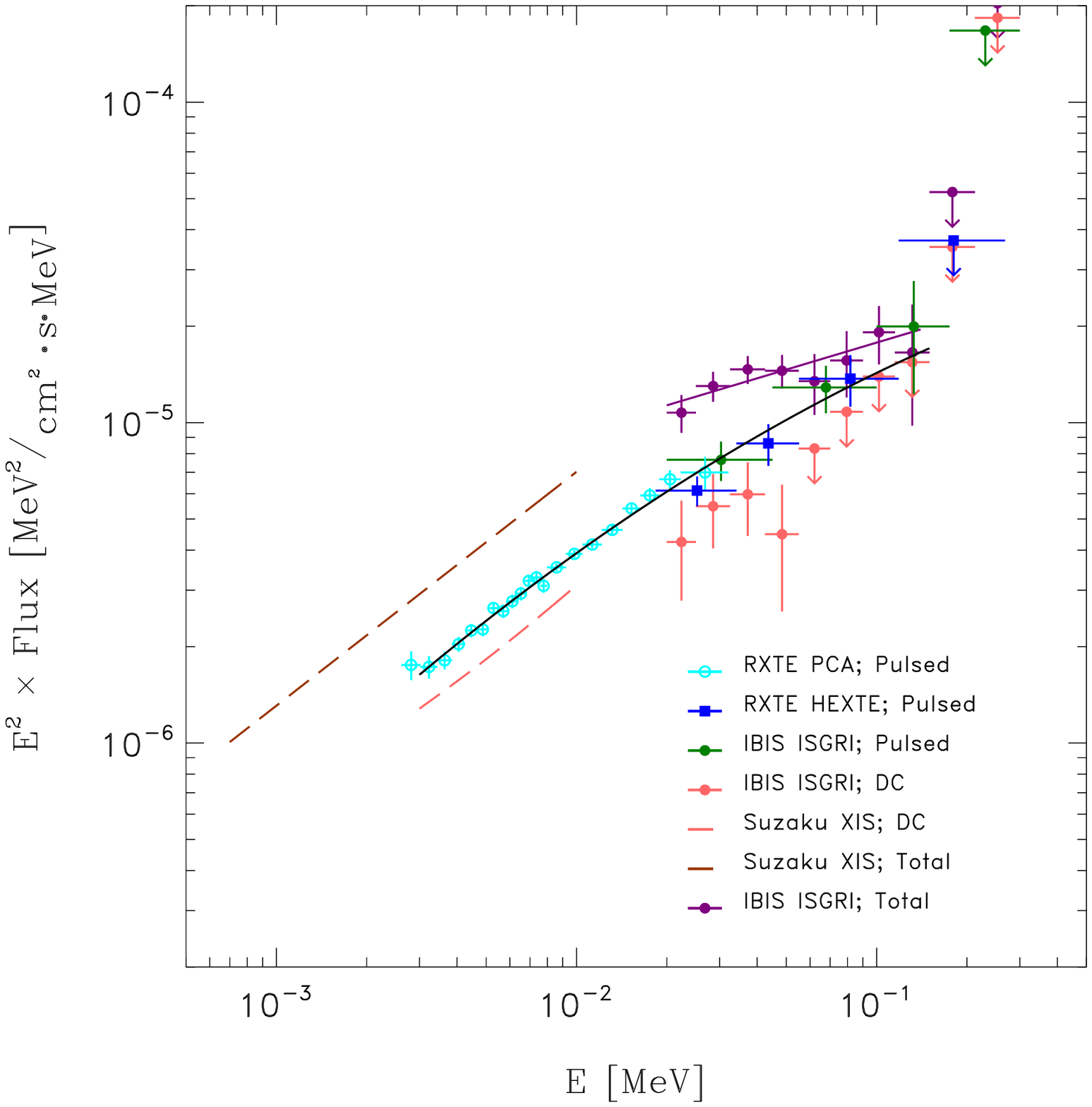}
     \caption{\label{psraxj_spc}{\bf \axj;} high-energy (0.7-300 keV) total and pulsed spectra as derived from
                                measurements by \Suzaku\ XIS (total 0.7-10 keV; dashed brown), \rxte\ PCA 
                                (pulsed 2-30 keV; aqua), \rxte\ HEXTE (pulsed 15-150 keV; blue) and \integral\ ISGRI 
                                (20-300 keV; total (purple) / pulsed (green)). The best fit model to the pulsed 
                                flux measurements is superposed as solid black line. The red dashed
                                line and data points are flux estimates for the underlying PWN, which suggest a
                                break/bend near 50 keV.}
  \end{center}
\end{figure}

From the \rxte\ PCA/HEXTE and \integral\ ISGRI pulse-phase histograms, we have extracted the pulsed excess counts through template fitting, and converted these to photon-flux values adopting (for the PCA range) an absorbing Hydrogen column N$_{\hbox{\scriptsize H}}$ of $5.4 \times 10^{22}$ cm$^{-2}$ \citep[see e.g.][]{anada2009,kargaltsev2012}. These pulsed flux measurements are shown in Fig. \ref{psraxj_spc}.
The combined \rxte\ PCA/HEXTE and \integral\ ISGRI pulsed flux measurements could be accurately fitted across the 3-150 keV band by a ``curved'' power-law model (a parabola in a $(\log(\nu F_{\nu})$ vs. $\log(\nu))$ spectral representation) with a photon-index
of $1.36(2)$ at 14.2 keV (this model is indicated by a black solid line in Fig.\ref{psraxj_spc}). Pulsed flux estimates (unabsorbed) are $(6.00\pm 0.33)\times 10^{-12}$ erg/cm$^2$s for the 2--10 keV, and $(2.53\pm 0.18)\times 10^{-11}$ erg/cm$^2$s for the 20--100 keV band, respectively. 

Our derived value for the 2--10 keV unabsorbed flux for the pulsed emission is significantly less than the value of $9.00\times 10^{-12}$ erg/cm$^2$ determined by \citet{gotthelf2008}, who analysed much less \rxte\ PCA data. However, the spectral shape the latter authors derived for the PCA band is compatible with our ``curved'' spectral model in the overlapping region.

We also derived total flux measurements of \axj\ for energies above 20 keV from an \integral\ ISGRI imaging analysis of the Scutum region (\axj\ was in the field 
of view of \psra\ (see Sect. \ref{psrj1846_section}) for which we reported the details in \citet{kuiper2009}), combining observations performed between \integral\ revolutions 49 and 441 (March 10, 2003 -- May 27, 2006; covering a considerably longer time span than \citet{malizia2005}, who used data from Revs 46 to 186). Note that for this period no valid ephemerides exist and thus timing results are lacking. The ISGRI total flux measurements (see Fig.\ref{psraxj_spc} for data points and best fit) can be described by a power-law with a photon-index of $-1.72\pm 0.07$, consistent with the value of $-1.66\pm 0.23$ obtained by \citet{malizia2005}, and a 20--100 keV (unabsorbed) energy flux of $(3.69\pm 0.21)\times 10^{-11}$ erg/cm$^2$s. The 20--300 keV flux is $(7.35\pm 0.39)\times 10^{-11}$ erg/cm$^2$s, slightly less than the value of $9\times 10^{-11}$ erg/cm$^2$s from \citet{malizia2005}. The pulsed fraction in the 20--100 keV band is $69\pm 6\%$.

In the soft X-ray band ($<10$ keV) spectral information for \axj\ and its PWN has been derived from \cxo\ ACIS \citep{gotthelf2008}, \Suzaku\ XIS \citep{anada2009} and \xmm\ EPIC MOS-2 \citep{kargaltsev2012} data. The \Suzaku\ XIS spectral analysis of the pulsar plus its surrounding PWN used an extraction radius of $3\arcmin$ encompassing $\sim 90\%$ of the source photons, while the \xmm\ EPIC MOS-2 spectral analysis utilized a $40\arcsec$ extraction radius centered on the source. Both (unabsorbed) flux measurements in the 2--10 keV band, converted from the 0.7--10 keV flux value of $(13.2_{-0.6}^{+0.8})\times 10^{-12}$ erg/cm$^2$s for the XIS and from the 1--11 keV flux of $(12.4\pm 2)\times 10^{-12}$ erg/cm$^2$s for the EPIC MOS-2, are consistent within uncertainties, $(10.7_{-0.5}^{+0.6})\times 10^{-12}$ erg/cm$^2$s and $(9.7\pm 1.6)\times 10^{-12}$ erg/cm$^2$s, respectively. 
Moreover, the derived photon indices for the XIS and MOS-2 measurements of $-1.27\pm 0.11$ and $-1.25_{-0.14}^{+0.30}$, respectively, and absorbing Hydrogen column densities of $(5.4\pm 0.5)\times 10^{22}$ cm$^{-2}$ and $(5.2_{-0.8}^{+1.0})\times 10^{22}$ cm$^{-2}$, respectively, are consistent.    

In a 20 ks \cxo\ ACIS-I observation of HESS J1837-069, as reported by \citet{gotthelf2008}, \axj\ and its PWN were detected $5\farcm25$ off-axis (from I3 aim-point) degrading 
considerably the image quality. These authors performed spatially resolved spectral analyses for the compact central source - the pulsar - selecting events from a 
$5\arcsec \times 7\arcsec$ elliptical aperture centered on the source peak, and the PWN selecting events from a $1\arcmin$ radius aperture centered at a location slightly 
offset from the compact source, while excluding a $7\arcsec \times 9\arcsec$ elliptical aperture around the compact source. They obtained an {\it extremely} hard spectrum 
for the compact source with a photon index of $-0.5\pm0.2$ and unabsorbed 2--10 keV flux of $8.8 \times 10^{-12}$ erg/cm$^2$s, and for the PWN a photon index of 
$-1.6\pm0.4$ and unabsorbed 2--10 keV flux of $1.0 \times 10^{-12}$ erg/cm$^2$s, both absorbed through a Hydrogen column of $(4.5\pm 0.8)\times 10^{22}$ cm$^{-2}$. 
While the combined compact source and PWN unabsorbed flux of $9.8 \times 10^{-12}$ erg/cm$^2$s is consistent with that derived by both \Suzaku\ XIS and \xmm\ EPIC MOS-2, the spectral shape does not! The cause of this discrepancy is not yet clear.

The spectral model of the total (pulsed plus DC) emission from \axj\ {\it and} its PWN, as derived by \citet{anada2009}, using \Suzaku\ XIS data for energies below 10 keV (=consistent with \xmm\ result) is shown in Fig.\ref{psraxj_spc} as a dashed brown line. The pulsed fraction in the 2--10 keV band is $0.56\pm 0.04$, including the PWN contribution in the total emission. Thus the genuine pulsed fraction of \axj\ is larger than 56\%. 

We can estimate the spectrum of the underlying DC-component (0.7-150 keV), which (mainly) originates from the PWN, by subtracting the flux of the pulsed component from that of the total emission. These DC-component flux estimates are also shown in Fig. \ref{psraxj_spc} as red line ($<10$ keV) and data points ($>20$ keV), and suggest that the PWN spectrum breaks/bends near 50 keV, providing clues on the magnetic field in the PWN (e.g. its averaged strength).

At GeV energies ($> 100$ MeV) no pulsed \gr-ray emission has been detected so far from \axj\ in Fermi LAT data \citep{abdo2013}. \citet{lande2012}, however, reported recently the detection of spatially extended emission in the 10-100 GeV band from a location positionally consistent with HESS J1837-069.

\subsection{\psra}
\label{psrj1846_section}
\citet{gotthelf2000} reported the discovery of \psra\ in \asca\ and \rxte\ data. It is a ``slow" radio-quiet
pulsar, $P \sim 324$ ms, but it has the smallest characteristic age, $\tau \sim 723$ y, of all known pulsars.
Its surface magnetic field strength of $4.9 \times 10^{13}$ G is above the quantum critical field strength of 
$4.413 \times 10^{13}$ G and this classifies the pulsar as a high-B-field pulsar. 
It is located at the centre of SN-remnant Kes 75 \citep{helfand2003} and shows up as a bright hard X-ray source 
surrounded by a diffuse PWN, also emitting hard X-rays. From \rxte\ monitoring data it was found that \psra\/ 
behaves as a very stable rotator \citep{livingstone2006}.
With \integral\ point-source emission has been detected up to $\sim$ 200 keV \citep{mcbride2008,kuiper2009}. 
Pulsed emission has been detected up to $\sim 150$ keV \citep{kuiper2009}. For details about the pulsed and total 
high-energy spectrum across the $\sim$ 2-300 keV band we refer to \citet{kuiper2009}.  

\citet{kumar2008} reported the detection of a dramatic brightening of the pulsar in
\cxo\ observations of Kes 75 during June, 7-12, 2006. The pulsar's spectrum softened considerably from a power-law 
spectrum with index $\Gamma \sim 1.32$ to $1.97$. 
\citet{gavriil2008} showed that during this radiative event, lasting for about 55 days, phase coherence in pulsar
timing was lost due to an unprecedented increase of the timing noise. They also discovered five short magnetar-like 
bursts during the outburst. \citet{kuiper2009} discovered that the onset of the radiative event was accompanied by 
a strong glitch in the rotation behaviour of the pulsar probably triggering the sudden release of magnetic energy.
Thus, this over many years very stably behaving source, both temporally and spectrally, exhibited suddenly
magnetar-like behaviour during the outburst, after which it continued again as a young energetic rotation-powered pulsar. 
Therefore, this high-B-field pulsar \psra\ might play a crucial role in revealing the connection between rotation-powered 
and magnetically powered pulsars (magnetars).

So far, the coherent signal from \psra\ remains undetected at radio-frequencies with a $4\sigma$ upper-limit at 1.95 GHz 
ranging between 4.9 and 43 $\mu$Jy, depending on the assumed duty cycle, and at 1400 MHz 27 $\mu$Jy around the time of the X-ray bursts \citep{archibald2008}. Also, \citet{parent2011} 
have not detected the pulsed signal of \psra\ for energies above 100 MeV, analyzing about 20 months of \fermi\ LAT data. 
However, the source HESS J1846-029 has been detected at TeV energies positionally consistent with Kes 75 with a (post-trial) 
significance of $8.3\sigma$ \citep{djannati2008}. Its TeV flux is at the level of $\sim 2\%$ of that of the Crab nebula, 
and its intrinsic extension is compatible with a point-like source.


\begin{figure}
  \begin{center}
     \includegraphics[width=8.25cm,bb=25 160 540 655,clip=]{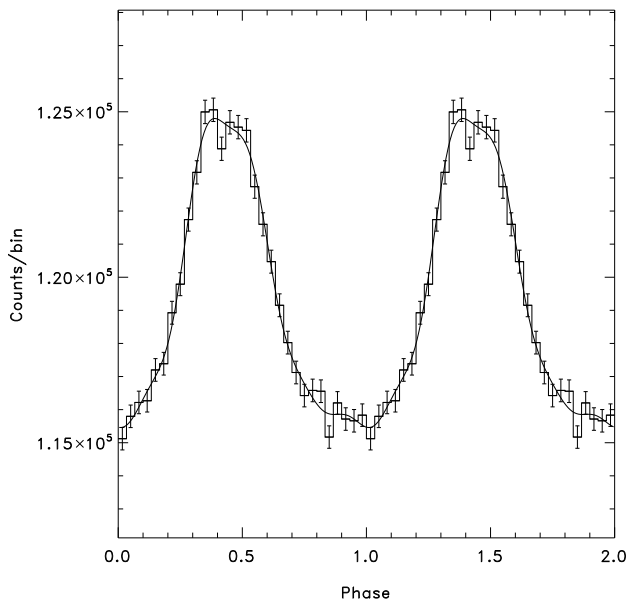}
     \caption{\label{igr_pca_prof}{\bf \psrigrb;} \rxte\ PCA 30 bins pulse profile for the $\sim 2-28$ keV band (PHA 5--65). 
     Two cycli are shown for clarity. A truncated Fourier-series fit adopting 5 harmonics is superposed.}
  \end{center}
\end{figure}

\subsection{\psrigrb\ / \igrb}
\label{sectigrj1849}
\igrb\ was discovered by \citet{molkov2004} in a survey of the Scutum region, followed somewhat later by the detection of a faint TeV
source, HESS J1849-000, coincident with the hard X-ray source \citep{terrier2008}, suggesting the source to be a PWN powered by an energetic pulsar. 
\swift-XRT \citep[][12.3 ks]{rodriguez2008} and \xmm\ \citep[][10 ks]{terrier2008} observations, both performed early 2006, 
revealed the X-ray counterpart of \igrb\ at soft X-rays and spectral analysis indicated a (very) hard spectrum from the somewhat 
blurred counterpart.
\citet{ratti2010} nailed down the location of \igrb\ to sub-arcsecond level by analyzing a 1.2 ks \cxo\ HRC observation performed 
on February 16, 2006. No obvious counterparts have been found at lower-energies in the {\it i'} and {\it K$_s$} bands \citep{ratti2010}. 
Also, no radio counterpart could be detected in the 610 MHz band of the GMRT in a systematic follow-up on \integral\ sources, resulting 
in an upper-limit of 3.5 mJy \citep{pandey2006}. 

\begin{figure*}
  \begin{center}
     \includegraphics[width=8.25cm]{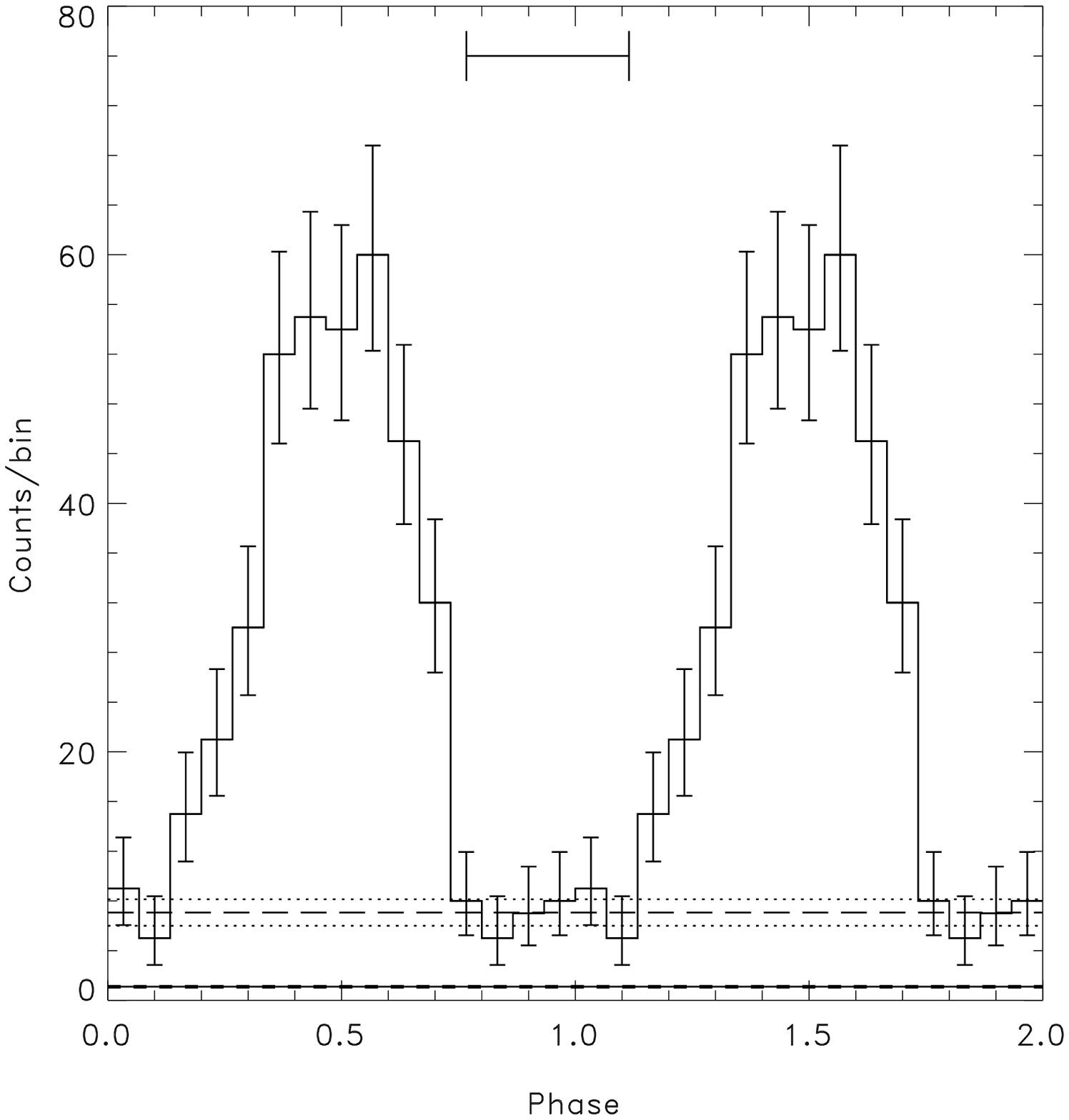}
     \includegraphics[width=8.25cm]{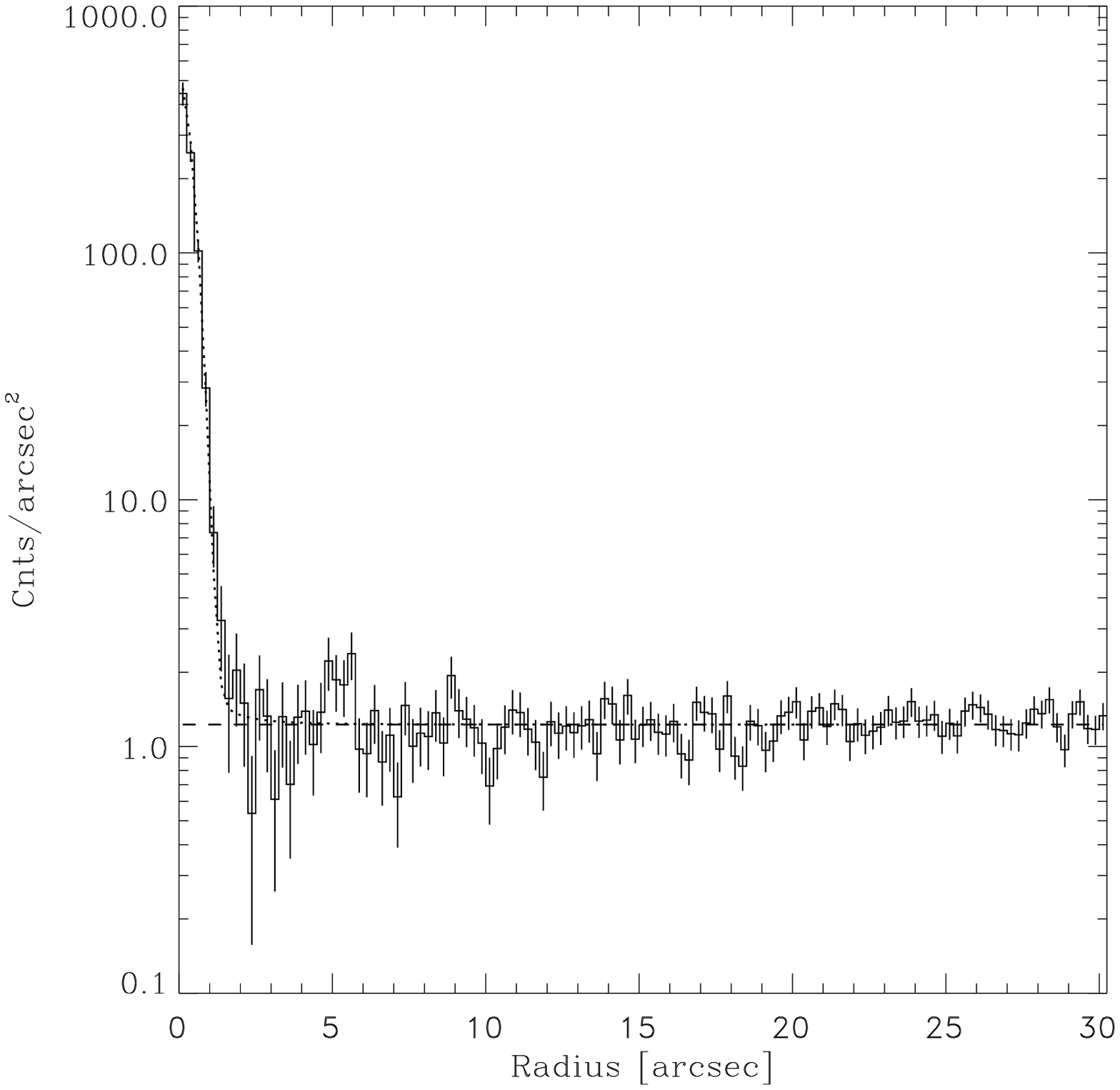}
     \caption{\label{igrj18490cxohrc}{\bf \psrigrb;} the left panel shows the (15 bins) HRC-S pulse profile (0.06--10 keV), selecting events within a 
     2\arcsec\ aperture around its centroid. The background level at 1.1 counts/bin and its $1\sigma$ uncertainty levels are indicated by the 
     horizontal solid and dashed lines, respectively. The unpulsed (DC) level and its $1\sigma$ uncertainty levels, as derived from the bootstrap method 
     outlined by \citet{swanepoel1996}, are represented by the long dashed and dotted lines, respectively. The solid horizontal line segment 
     given at the top indicates the unpulsed ``region'' as estimated by this bootstrap method. The intrinsic pulsed fraction is $0.77\pm 0.04$. 
     The right panel shows the 
     radial distribution of HRC-S events centered on \igrb\ up to 30\arcsec\ along with a model (dotted line) composed of a flat 
     background (dashed horizontal line) and a point source with 406 counts. No indication for structured extended emission is 
     found in the near vicinity of \igrb.}
  \end{center}
\end{figure*}

The true nature of \igrb\ was found by \citet{gotthelf2011}, who detected, pointing at the \cxo\ location, a young and energetic 
38.5 ms pulsar using 112.3 ks of \rxte\ PCA exposure time, collected during dedicated observations spanning Nov. 25 - Dec. 15, 2010. 
Analysing also \xmm\ EPIC pn and MOS data from an 11 ks observation performed on April 3, 2006 these authors estimated a pulsed 
fraction of $\sim$ 25\% for the 2--10 keV band. \citet{gotthelf2011} found also evidence for extended emission around \igrb\ between 
$20\arcsec$ and $150\arcsec$ radius. The spectrum of this diffuse component was much softer, with photon index of $-2.1 \pm 0.3$, 
than that of the point-source emission, with photon index of $-1.1 \pm 0.2$, while it represents about 25\% of the point source emission. 
Given the large Hydrogen column density N$_{\hbox{\scriptsize H}}$ of $(4.3\pm 0.6) \times 10^{22}$ cm$^{-2}$ \citep{gotthelf2011} the 
diffuse emission could also be interpreted as a dust-scattered halo around a bright point source. 

We (re)analysed the \rxte\ PCA/HEXTE data, collected during the dedicated observations, to obtain a phase coherent timing model and to 
derive the pulsed emission spectrum over the $\sim 3 - 150$ keV band. Through template cross-correlation we derived pulse arrival times (ToA's) 
and from these we determined a phase coherent timing solution, which is shown in Table \ref{eph_table}. We produced pulse profiles 
of \igrb\ for all 255 PHA channels of the PCA, allowing contrary to \cite{gotthelf2011} all detector layers, by phase folding 
(using the coherent timing model) the barycentered event arrival times that passed our selection criteria, and subsequently sort 
these on PHA. The pulse profile for the PHA band 5-65 in 30 bins ($\sim 2-27$ keV; $Z_1^2=2791.9 \equiv 35.2 \sigma$) is shown 
in Fig. \ref{igr_pca_prof} and consists of one broad (somewhat structured) single pulse. A smoothed version of this profile 
(a truncated Fourier series adopting 5 harmonics; shown in Fig. \ref{igr_pca_prof} as solid line) has been used in the 
extraction procedure of the pulsed excess counts, because we do not see pulse morphology changes as a function of energy.  
The pulsed excess counts have subsequently been converted to pulsed fluxes in a forward folding procedure assuming an underlying 
power-law model absorbed through a column density N$_{\hbox{\scriptsize H}}$ of $4.5 \times 10^{22}$ cm$^{-2}$ (see further in 
this section) yielding as best fit parameters an unabsorbed 2--10 keV pulsed flux of $(4.21 \pm 0.08)\times 10^{-12}$ erg/cm$^{-2}$s 
and a photon index of $-1.37 \pm 0.01$. While the latter value is consistent with that derived by \cite{gotthelf2011}, our derived 
flux value is about 4 times larger than the value quoted by \cite{gotthelf2011}.

The HEXTE data from the (only active) staring cluster-A detectors have been screened adopting default screening conditions. 
The barycentered events have subsequently been folded using the coherent timing model to yield pulse profiles for all HEXTE 
PHA channels. Pulsed emission has been detected significantly up to $\sim 60$ keV (HEXTE band 31.0--60.1 keV; $4.8\sigma$). 
Pulsed excess counts in various energy bands have been determined through template fitting. These counts have been converted 
to flux values using proper HEXTE cluster-A response information and the deadtime corrected exposure for cluster-A, adopting 
the photon power-law model derived from PCA data. 
The PCA and HEXTE pulsed flux values are shown in the spectral compilation depicted in Fig. \ref{igrj18490acisspc} as aqua (PCA) 
and blue (HEXTE) data points.

Because our own analysis of the pulsed X-ray spectrum of \igrb\ using \rxte\ PCA data yielded a $\sim 4\times$ larger pulsed 
flux than that given by \citet{gotthelf2011}, we obtained also a higher pulsed fraction of $88\pm 8$ \%. To shed light on this discrepancy 
and on the extended emission around \igrb\, we proposed 2 \cxo\ observations of 25 ks each, one with the HRC-SI (timing) and 
one with the ACIS-S (spatially resolved spectral analysis). Our aim was to determine the genuine intrinsic pulsed fraction in 
the X-ray band and to perform accurate spatially resolved spectral analysis at arcsec scales.

The HRC-S observation in timing mode (6\arcmin $\times$ 30\arcmin field of view; time resolution $16 \mu$s) was performed on 
Nov. 20, 2011 for an effective exposure time of 25.1 ks (CXO observation id. 13292). An imaging analysis using Maximum 
Likelihood (ML) techniques  for Poissonian distributed pixel content \citep[see e.g. the Supplementary Material of][for a 
similar approach]{hermsen2013} of the HRC-S data showed a clear point-source at 
$(\alpha_{2000},\delta_{2000})=$(18\ 49\ 1\fs632,-00\ 01 17\farcs45) with a formal systematic uncertainty of 0\farcs6 
(in radius; 90\%) and a negligible statistical uncertainty of 0\farcs02 ($1\sigma$) in each coordinate. Selecting (barycentered) 
events within 2\arcsec\ from the centroid location of \igrb\ yielded a number of 406 events of which $16.4 \pm 1.2$ events are 
expected to be background events as estimated in an annulus of 4\arcsec\ inner radius and 8\arcsec\ outer radius centered on \igrb. 

A restricted period search around the predicted value using the $Z_2^2$-test statistic yielded a very significant signal at 
$\nu_{\max}=25.9609614(25)$ Hz, exactly at the extrapolated value from the (\rxte\ PCA based) coherent ephemeris (this work) as 
given in Table \ref{eph_table}. Subsequent phase folding yielded the CXO HRC-S pulse profile (energy integrated; 15 bins) of 
\igrb\ as shown in the left panel of Fig. \ref{igrj18490cxohrc}. The pulsed fraction, not corrected for (the small) background, 
is $0.74\pm 0.04$, which translates in a (background corrected) intrinsic pulsed fraction of $0.77 \pm 0.04$, 
inconsistent with the value of $\sim 0.25$ derived by \citet{gotthelf2011}, but consistent with the value of $0.88\pm 0.08$ 
we derived earlier from our own \rxte\ PCA spectral analysis of the pulsed emission and the \xmm\ total source flux of 
\igrb\ as reported by \citet{gotthelf2011}.

\begin{figure}
  \begin{center}
     \includegraphics[width=8.25cm]{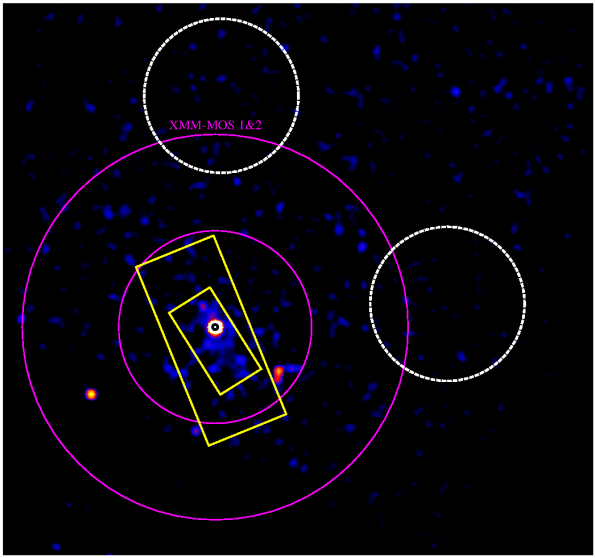}
     \caption{\label{igrj18490cxoacis} {\bf \psrigrb;} ACIS-S3 2--10 keV image of the near environment of \igrb\ (black circle). 
                     The image is Gaussian smoothed and scaled logarithmically to emphasize diffuse emission, which is clearly 
                     visible in a ``rectangular'' region of size
                     75\arcsec $\times$ $37\farcs5$ (small yellow box) around \igrb. Just outside the larger yellow box of
                     size 150\arcsec $\times$ 65\arcsec, used for the determination of the X-ray spectrum of the extended emission,
                     a cluster of three point sources are popping-up at about $60\arcsec$ from \igrb\ (reddish spot).  
                     The two circular apertures (white dashed circles) each with a 60\arcsec\ radius indicate sky regions used to 
                     estimate the background. The purple annulus with radii at 75\arcsec\ and 150\arcsec\ represents the region selected 
                     in a study of the diffuse PWN emission around \igrb\ using \xmm\ MOS 1\&2 data. }
  \end{center}
\end{figure}

To study the extended emission in the vicinity of \igrb\ we applied a (deep) Gaussian smoothing to the central part of the 
raw image centered on \igrb\ and displayed the image on a log scale to emphasize possibly extended emission. 
The resulting map (PHA integrated) did not show evidence for the presence of diffuse emission near \igrb. It should be noted, however, 
that given the response of the HRC-S with maximum sensitivity between 1 -- 2 keV (strongly reduced beyond 2 keV) this image mainly 
shows the spatial information at soft X-rays (0.1-2 keV). Moreover, the genuine underlying emission around \igrb\ is strongly 
suppressed by the highly absorbing intervening Hydrogen column. 

Analyzing the HRC-S data, the more quantitative ML-imaging approach of the near environment of \igrb\ (in this case a region with 
a radial extent of 10\arcsec\ centered on the centroid scanned at $0\farcs2$ scale) also yielded no evidence for structured diffuse 
emission. This method assigns $406 \pm 21$ counts to the point source at a background rate of $1.224 \pm 0.066$ counts/arcsec$^2$. 
The projection of the 2d-imaging information into a radial event distribution ($0\farcs25$ binsize) is shown in the right panel 
of Fig. \ref{igrj18490cxohrc} with superposed the point source model for \igrb\ with 406 source counts on top of the local 
background, estimated in an annulus with 10\arcsec\ inner- and 25\arcsec\ outer radius. The latter background value is fully 
consistent with that derived from the 2d-ML method fitting the background and source simultaneously. No significant deviations 
between measured distribution and model can be found, indicating that there is no evidence for structured extended emission in 
the near environment of \igrb.

Our second \cxo\ observation (CXO observation id. 13291) with \igrb\ at the aim point of the back-illuminated ACIS-S3 chip 
was performed on Nov. 16, 2012 for 25.6 ks yielding an effective exposure of 22.7 ks. A Gaussian smoothed and logarithmically 
scaled (to emphasize diffuse emission) ACIS-S3 map for energies in the range 2--10 keV is shown in Fig. \ref{igrj18490cxoacis}. 
Diffuse emission is now clearly visible confined more or less in a ``rectangular'' region of size 75\arcsec $\times$ $37\farcs5$ 
(small yellow solid rectangle/box) around \igrb. A similar map for events with energies in the range 0.5--2 keV does not show 
evidence for diffuse emission, consistent with the HRC-S findings, while the point source emission from \igrb\ is still detected.

\begin{figure}
  \begin{center}
     \includegraphics[width=8.25cm,bb=50 150 550 660,clip=]{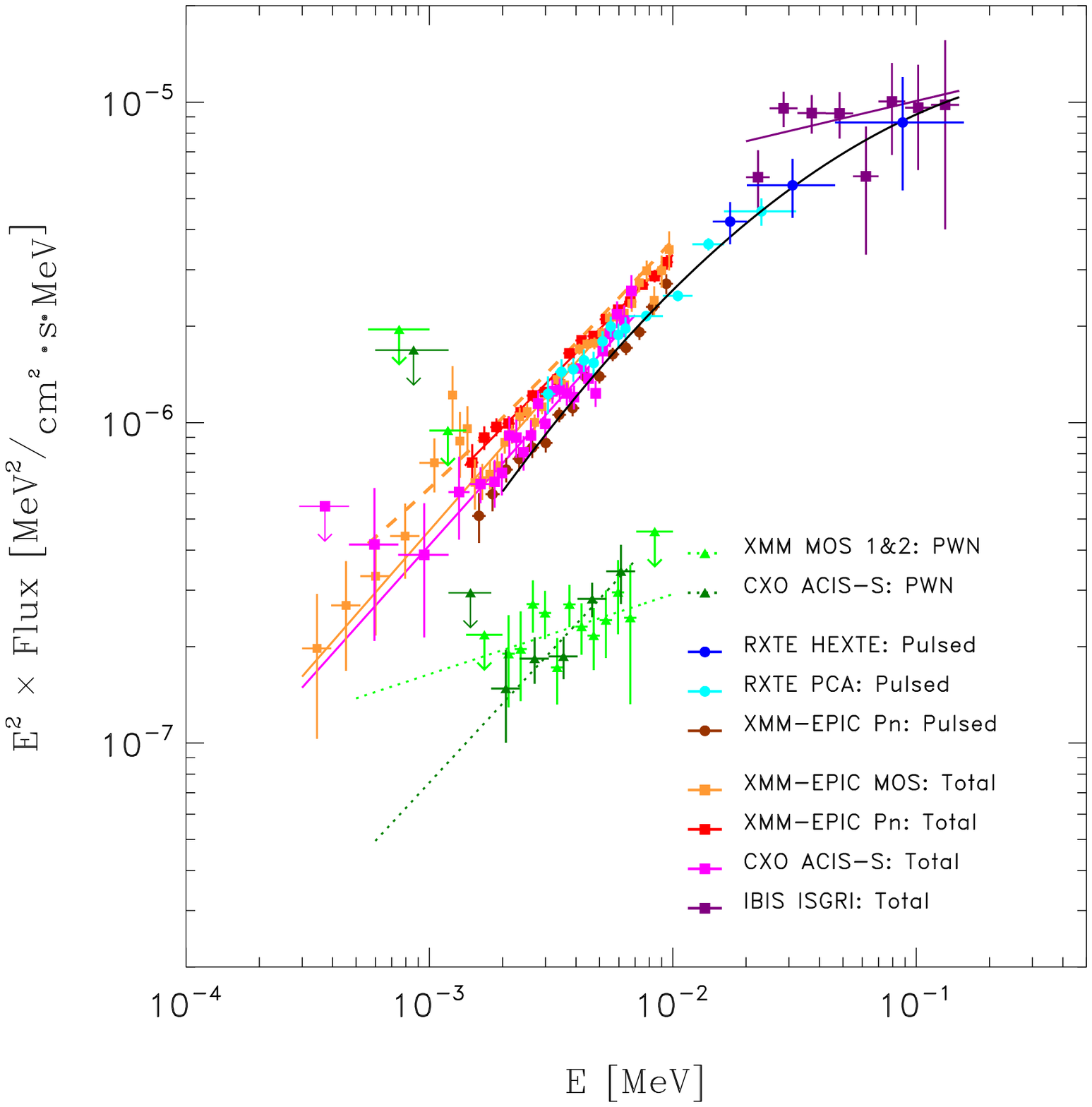}
     \caption{\label{igrj18490acisspc}{\bf \psrigrb;} unabsorbed (N$_{\hbox{\scriptsize H}} = 4.5 \times 10^{22}$ cm$^{-2}$) total and pulsed 
     spectra from 0.3-150 keV combining data from \rxte\ PCA and HEXTE (pulsed), \integral\ ISGRI (total), \cxo\ 
     ACIS-S (total), \xmm\ EPIC pn (pulsed \& total) and MOS (total). Total flux measurements have filled square symbols, while pulsed
     fluxes have filled circles. A fit (``curved'' power-law model) through all pulsed flux measurements is shown as black solid line. 
     The emission of \igrb\ is highly pulsed with a pulsed fraction of about 80\%. The spectrum of the diffuse emission is also shown below 
     that of the pulsar (a) lime triangles ACIS-S; $150\arcsec \times 65\arcsec$ extraction box minus $5\arcsec$ circular source region; 2) 
     green triangles \xmm\ EPIC MOS 1+2; annulus of $75\arcsec$ inner- and $150\arcsec$ outer radius).}
  \end{center}
\end{figure}

More quantitatively, we produced a Maximum Likelihood ratio map selecting events with energies within 0.5--10 keV for a 
$2\arcmin \times 2\arcmin$ region centered on \igrb\ to search for point sources in the near environment of \igrb,
polluting the determination of the genuine diffuse emission. This method assigns $2359\pm 49$ counts to the point source 
\igrb\ in this band, yielding a source count rate $C_R^{\hbox{\scriptsize 0.5-10 keV}}$ of $0.104\pm 0.002$ c/s. In addition, five 
sources are popping-up with detection significances $>5 \sigma$ of which three (with relatively high significances in the 
range $6.7 - 15.2\sigma$) cluster in a small region at about $\sim 60\arcsec$ from \igrb\ (see Fig.\ref{igrj18490cxoacis}, 
just outside the large dashed yellow box). In the estimation of the spectrum of the diffuse emission we chose an extraction 
region for the diffuse emission excluding these sources. In fact, these sources clearly polluted the diffuse spectrum derived 
from \xmm\ EPIC data by \citet{gotthelf2011}, who used an annulus of $30\arcsec$  
inner- and $150\arcsec$ outer radius centered on \igrb\ as extraction region \citep[see also Fig. 2 of][]{terrier2008}.

We used two rectangular extraction regions encompassing \igrb, excluding a circular region of $5\arcsec$ radial extend 
centered on \igrb, to evaluate the spectrum of the diffuse emission: 1) a $75\arcsec \times 37\farcs5$ box 
(small yellow box in Fig.\ref{igrj18490cxoacis}) and 2) a $150\arcsec \times 65\arcsec$ box (large yellow box in Fig.\ref{igrj18490cxoacis}). 
We also carefully chose two $60\arcsec$ circular background regions, free of sources, in the vicinity of \igrb\ (see also 
Fig.\ref{igrj18490cxoacis}; the two white dashed circles). The background subtracted count spectrum (0.6--7 keV band), 
properly taking into account the area difference of the diffuse- and background extraction regions, has been fitted in a 
forward folding procedure adopting an absorbed power-law model with a fixed Hydrogen column density of 
N$_{\hbox{\scriptsize H}}$ of $4.5 \times 10^{22}$ cm$^{-2}$ (see further this section) and proper response information. 
We found for the large extraction box a photon index of $-1.18 \pm 0.05$ and an unabsorbed 2--10 keV flux of 
$(7.1 \pm 0.5) \times 10^{-13}$ erg/cm$^2$s (see Fig. \ref{igrj18490acisspc} for the ACIS-S PWN fluxes and fit; lime coloured), about 5--6 
times smaller than that of the \igrb\ point source. For the smaller extraction box we found a similar photon index of 
$-1.13 \pm 0.06$ and a $\sim 1.6$ times lower flux. Thus, our CXO ACIS-S derived spectrum of the diffuse emission is much 
harder than that derived by \citet{gotthelf2011} using \xmm\ EPIC data, while our flux estimates are compatible given 
the different extraction regions and methods used.  

Next, we derived the total ACIS-S (point source) count spectrum of \igrb\ by fitting simultaneously a point-source model and a 
flat background model to the spatial event distribution (using the ML-method adopting Poissonian statistics) for each chosen 
energy band between 0.3 and 7 keV. These spatial event distributions have been sampled in a 5\arcsec\ circular region centered 
on \igrb\ in $0\farcs5 \times 0\farcs5$ bins. The resulting source counts per energy slice have subsequently been converted 
to photon flux values using proper energy response information and an effective exposure value of 22.67 ks adopting an 
absorbed power-law model. A model fit with N$_{\hbox{\scriptsize H}}$ free yielded a column density of 
$(4.30 \pm 0.16)\times 10^{22}$ cm$^{-2}$ and a photon index of $-1.08 \pm 0.02$. 
Fixing N$_{\hbox{\scriptsize H}}$ to $4.5 \times 10^{22}$ cm$^{-2}$ resulted in a slightly softer photon index of 
$-1.15 \pm 0.02$ and an unabsorbed 2--10 keV flux of $(4.11 \pm 0.09)\times 10^{-12}$ erg/cm$^2$s. These flux 
measurements are shown in Fig. \ref{igrj18490acisspc} as magenta data points.

We studied the effects of pile-up on the spectral parameter estimation for this CXO ACIS-S observation, because the measured 
count rate (0.5-10 keV) per 3.2 s frame time is $\sim 0.33$ cnts/frame which is sufficiently large for a moderate pile-up 
impact (see e.g. the middle panel of Fig. 3 of ``The \cxo\ ABC Guide to pileup" which is referenced from 
{\it http://cxc.harvard.edu/ciao/ahelp/acis\_pileup.html}).

The pile-up fraction $f_f$ (i.e. the fraction of frames that have detected events containing two or more events) is 
about 15\%, while the total fraction $f_t$ of events lost, either through grade or energy migration, with respect to the 
expect rate is about 28\%. To study the impact of pile-up on our spectral results in more detail we simulated spectra 
using {\it webspec} adopting different pile-up model parameters (no pile-up, pile-up with grade migration parameter 
$\alpha=(0,0.5,1)$ for an input photon flux model with photon-index -1.08 and normalization at 1 keV of 
$4\times 10^{-4}$ ph/cm$^2$s keV absorbed by a Hydrogen column of $4.5 \times 10^{22}$ cm$^{-2}$. 
We found for all the three different pile-up models a photon flux reduction of about 22\% with respect to the unpiled case. 
The pile-up impact on the reconstructed photon index was marginal and its size is about 0.1.

\xmm\ has observed \igrb\ for a second time on March 23-24, 2011 for about 54 ks. EPIC MOS operated in Prime Full 
Window mode using a medium filter for an effective exposure of 52.635 ks and 52.664 ks for MOS 1 and MOS 2, respectively. 
EPIC pn operated in Small Window mode (medium filter) for an effective exposure of 37.76 ks allowing studies of the 
pulsed signal given the $\sim 5.7$ ms time resolution in this mode. These \xmm\ observations do not suffer from pile-up 
effects, and allow us to reconstruct the unpiled total (MOS 1 \& 2,pn) and pulsed (pn, only) spectrum of the source.

To extract the total emission spectrum of \igrb\ (MOS 1 \& 2; pn) we applied a similar procedure as used for the ACIS-S, 
now sorting the events with energy between 0.3-10 keV and within $60\arcsec$ from the \igrb\ centroid in 
$2\arcsec \times 2\arcsec$ spatial bins for each energy band.
For MOS 1 \& 2 we derived, adopting an absorbed power-law model, an Hydrogen column density 
N$_{\hbox{\scriptsize H}}$ of $(4.51\pm 0.10)\times 10^{22}$ cm$^{-2}$ and a photon index $\Gamma$ of 
$-1.13 \pm 0.01$ ($\chi^2_{r,35-3}=1.20$). For the pn we found for N$_{\hbox{\scriptsize H}}$ a value of 
$(4.54\pm 0.10)\times 10^{22}$ cm$^{-2}$ and for the photon index $-1.23 \pm 0.01$, somewhat softer than determined 
from MOS 1 \& 2 data ($\chi^2_{r,17-3}=0.86$ ; 1.42-10 keV). The derived Hydrogen column density is consistent with 
the value found by \citet{gotthelf2011}, and in all further spectral analysis of X-ray data we fixed 
N$_{\hbox{\scriptsize H}}$ to $4.5 \times 10^{22}$ cm$^{-2}$. The unabsorbed 2--10 keV total flux of \igrb\ is 
$(4.72 \pm 0.05)\times 10^{-12}$ and $(4.92 \pm 0.05)\times 10^{-12}$ erg/cm$^2$ s, for MOS 1 \& 2 and pn, respectively, 
about 17\% higher than the ACIS-S estimate, which suffered from pile-up effects. The flux measurements are shown in 
Fig. \ref{igrj18490acisspc} as red (EPIC-pn) and orange (MOS 1 \& 2) datapoints.

We also studied the diffuse emission around \igrb\ using MOS 1 \& 2 data. We chose a source region consisting of an annulus 
with an inner radius of $75\arcsec$ (just outside the regions studied using ACIS-S data) and an outer radius of $150\arcsec$ 
(purple annulus in Fig. \ref{igrj18490cxoacis}), well outside the ``source cluster" at an angular distance of about $60\arcsec$ 
from \igrb. The background region consisted 
of an annular region with $150\arcsec$ inner and $225\arcsec$ outer radius. The difference count spectrum, taking into 
account the difference in area of the source- and background extraction regions, has been fitted with an absorbed 
power-law model (N$_{\hbox{\scriptsize H}}$ fixed to a value of $4.5\times 10^{22}$ cm$^{-2}$) utilizing response functions 
produced for the annular source region. The fit yielded an unabsorbed 2--10 keV energy flux of 
$(6.23\pm 0.42)\times 10^{-13}$ erg/cm$^2$s and a photon index of $-1.75\pm 0.05$, considerably softer than found by the 
ACIS-S for the inner PWN region. This can best be explained as a manifestation 
of Synchrotron cooling, which makes the outskirts of a PWN softer than the inner parts. 
The MOS 1 \& 2 PWN flux measurements are shown in Fig. \ref{igrj18490acisspc} as green datapoints.

We also performed a timing analysis of the EPIC-pn data, which, sampled in Small Window Mode, has a time resolution of $\sim 5.7$ ms, 
amply sufficient to study the 38.5 ms pulsed signal. For this purpose we selected barycentered events within $15\arcsec$ from 
the \igrb\ centroid. A restricted period search around the predicted pulsed period extrapolating the timing parameters of 
the \igrb\ entry listed in Table \ref{eph_table} yielded a very significant peak in the periodogram exactly atop the prediction. 
Subsequent phase folding adopting again the timing model listed in Table \ref{eph_table} for \igrb\ yielded significant pulsed 
emission down to about 1 keV.

To derive the pulsed spectrum we first applied for each selected energy band, adopting Poissonian statistics, a Maximum Likelihood template fit 
procedure consisting of optimizing the number of pulsed excess counts and unpulsed (flat) level simultaneously assuming that the 
pulse shape is described by the pulse profile (=template) as shown in Fig. \ref{igr_pca_prof}. 

These excess counts are converted to flux values in a forward folding procedure adopting an absorbed power-law model 
(N$_{\hbox{\scriptsize H}} \equiv 4.5 \times 10^{22}$ cm$^{-2}$) using instrument response functions determined for the 
$15\arcsec$ extraction radius. We derived an unabsorbed 2--10 keV pulsed flux of $(3.65 \pm 0.06)\times 10^{-12}$ erg/cm$^{-2}$s 
and a photon index of $-1.14 \pm 0.01$. The pulsed flux measurements by EPIC-pn are shown as dark-red datapoints in Fig. \ref{igrj18490acisspc}.
Given the total and pulsed 2--10 keV (XMM) fluxes we determined a pulsed fraction 
of $0.76 \pm 0.02$ for the 2--10 keV band, in full agreement with the HRC-S estimate for its integral energy band and the 
PCA based value.

The sky region near \igrb\ was also frequently in the field of view of the soft \gr-ray imager ISGRI (15-300 keV) aboard the 
\integral\ satellite. A deep image centered on rotation-powered pulsar \psra\ (see Section \ref{psrj1846_section}) and enclosing 
\igrb\ has been shown by \citet{kuiper2009}. The total spectrum of \igrb\ for the 20-300 keV band has been derived from maps 
produced in 10 logarithmically binned energy bands covering the 20-300 keV window. The selected \integral\ observations covered 
\integral\ revolutions 49 -- 603 (March 10, 2003 -- Sept. 23, 2007) and represent a total GTI exposure of 7.35 Ms. 
For more details we refer to Section 3.1 of \citet{kuiper2009}. Fitting a power-law model we derived a photon index of 
$-1.82 \pm 0.14$ and a 20--100 keV energy flux of $(2.25\pm 0.16)\times 10^{-11}$ erg/cm$^2$s, consistent with the values 
determined by \cite{terrier2008}, who used less \integral\ exposure time. These ISGRI flux measurements (sum of pulsed/unpulsed 
emission from the pulsar and the diffuse emission) and best fit are shown in Fig. \ref{igrj18490acisspc} as purple data points and solid line, respectively. 
Finally, in the GeV \gr-ray band no pulsed emission from \igrb\ has been detected by \fermi\ LAT \citep{abdo2013}.

In summary, we found hard pulsed emission for \igrb\ from $\sim 1$ keV to $\sim 60$ keV, which represents about 80\% of the 
total emission. The diffuse emission, 5--6 times weaker than the point source emission of \igrb, is also hard and softens 
towards the outer parts of the PWN. This is characteristic for Synchrotron cooling.  


\begin{figure}
  \begin{center}
     \includegraphics[width=8cm,height=8cm,angle=0,bb=150 250 445 560]{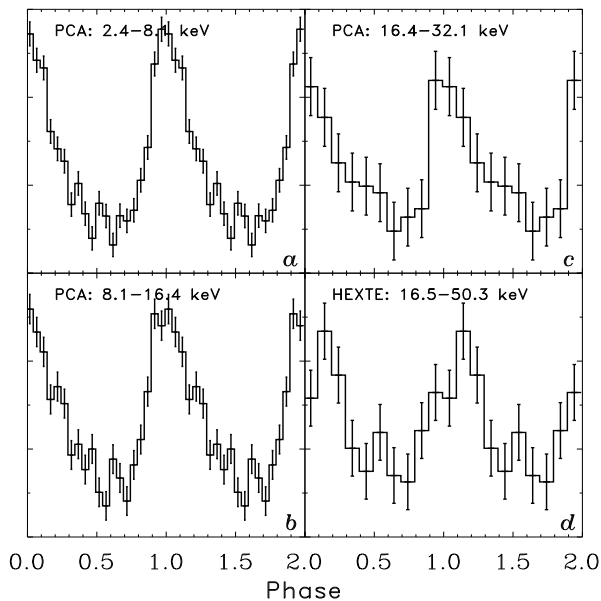}
     \caption{\label{psrj1930_prof}{\bf \psrd;} pulse profiles in various energy bands for \rxte\ PCA (panels a--c) and \rxte\ HEXTE 
                                  (panel d). Pulsed emission has been detected up to $\sim 50$ keV ($4.5\sigma$; panel d).}
  \end{center}
\end{figure}

\subsection{\psrd}
\citet{seward1989} reported the detection of the Crab-like SNR G54.1+0.3 at X-rays using \einstein\ IPC (0.5-4 keV) data  from an 
observation performed on May 7, 1980 collecting about 50 photons, too faint to determine accurate spectral information. More detailed 
X-ray spectral information was obtained by \citet{lu2001} using data from the \rosat\ PSPC (0.1-2.5 keV; $\sim 20$ ks) and the \asca\ GIS 
(16.5 ks) and SIS ($\sim 20$ ks) detectors, both operating in the 0.5-10 keV range. Spatially resolved spectral analysis in the X-ray 
band became possible using the superb imaging quality of the \cxo\ ACIS instrument, revealing in a 30.9 ks exposure a central bright 
hard point-source - the putative pulsar - and several diffuse structures surrounding it \citep{lu2002}. 

Targeting at the central source \citet{camilo2002} discovered the energetic pulsar \psrd\ at radio frequencies. 
Its period of $\sim 136$ ms was also detected at soft X-rays in the archival \asca\ GIS data. \citet{lu2007} showed, analysing additional 
\cxo\ ACIS-I data taken on June 30, 2003 for a total exposure time of 58.4 ks in CC-mode, that the pulsed emission is hard with a 
photon-index $\Gamma$ of $1.2(2)$ and that the source has a high pulsed fraction of $71\pm 5\%$. The absorbed pulsed flux in the 
2--10 keV band was $1.2\times 10^{-12}$ erg/cm$^2$s, which translates to an unabsorbed flux of $\sim 1.33\times 10^{-12}$ erg/cm$^2$s 
in the same band taking into account the $1.6\times 10^{22}$ cm$^{-2}$ absorbing Hydrogen column. These authors also analyzed data from 
a $\sim 80$ ks \rxte\ observation taken during September 12--14 and December 23--25, 2002, but only for timing purposes. 

\begin{figure}
  \begin{center}
     \includegraphics[width=8cm,bb=55 160 545 655]{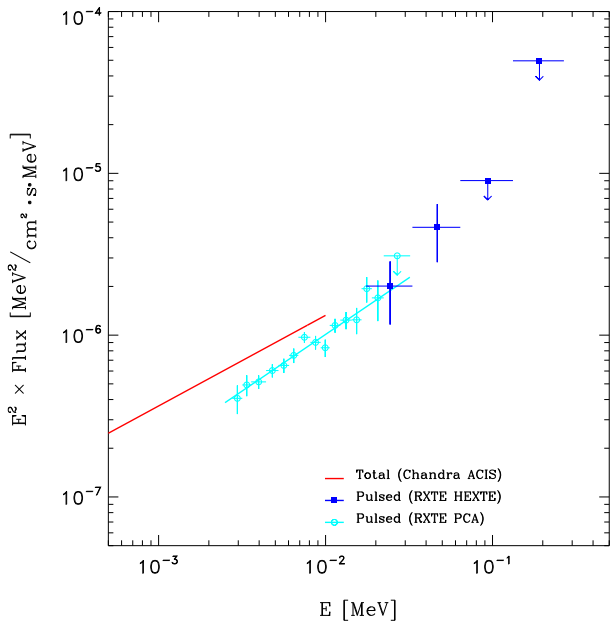}
     \caption{\label{psr1930_spc}{\bf \psrd;} high-energy (0.5-250 keV) total and pulsed spectra as derived from
                                measurements by \cxo\ ACIS (total pulsar 0.5-10 keV; solid red; \citet{temim2010}),
                                \rxte\ PCA (pulsed 2.5-30 keV and best fit; aqua; this work) and \rxte\ HEXTE (pulsed 15-250 keV; 
                                blue; this work).}
  \end{center}
\end{figure}

Guided by the available radio ephemeris \citep{camilo2002} we generated a phase coherent ephemeris based on these X-ray data for the 
2002 September 12 -- December 25 period, listed in Table \ref{eph_table}. We re-analysed the \rxte\ PCA and HEXTE data using template 
fitting to obtain the pulsed emission spectrum over a much broader bandpass than \cxo. Pulse profiles for various energy bands are 
shown in Fig.\ref{psrj1930_prof} and pulsed emission has been detected up to $\sim 50$ keV (see Fig. \ref{psrj1930_prof}d). 
The reconstructed photon spectrum of the pulsed emission across the 2.8-32.1 keV range, adopting a Hydrogen column density 
N$_{\hbox{\scriptsize H}}$ of $1.6 \times 10^{22}$ cm$^{-2}$ \citep[see][]{lu2002}, has a power-law index of $-1.30 \pm 0.02$ 
and an unabsorbed 2-10 keV flux of $(1.57 \pm 0.07)\times 10^{-12}$ erg/cm$^2$s, slightly higher than the value obtained by \citet{lu2007} 
who used \cxo\ ACIS data. Using \cxo\ ACIS-S data of observations performed over the period 8--15 July 2008, totaling 290.77 ks of 
exposure time, \citet{temim2010} derived for the total pulsar spectrum a photon index of $-1.44 \pm 0.04$ and an unabsorbed 0.3-10 keV 
flux of $3.26\times 10^{-12}$ erg/cm$^2$s, which translates to a 2-10 keV flux of $2.25\times 10^{-12}$ erg/cm$^2$s. From this total 
2--10 keV flux value and our pulsed 2-10 keV flux we derive a pulsed fraction of $\sim 70\%$, consistent with the value derived by using 
only \cxo\ data. The pulsed (\rxte\ PCA and HEXTE) and total \citep[\cxo;][]{temim2010} 
emission spectra of \psrd\ are shown in Fig. \ref{psr1930_spc}. 

Soft \gr-ray emission in the 20-60 keV band from a location consistent with \psrd\ has also been detected in an imaging study 
of ISGRI data  \citep{segreto2010}. 

\psrd\ is not listed as high-energy \gr-ray pulsar in the \fermi\ second pulsar catalogue 
\citep{abdo2013}, however, at TeV energies a pointlike source, VER J1930+188, has been detected by the VERITAS ground-based gamma-ray observatory
from the direction of SNR G54.1+0.3 with an integral flux above 1 TeV of 2.5\% of the Crab nebula, very likely the PWN associated 
with \psrd\ \citep{acciari2010}.


\begin{figure}
  \begin{center}
     \includegraphics[width=6cm,height=12cm,angle=0,bb=170 95 420 695]{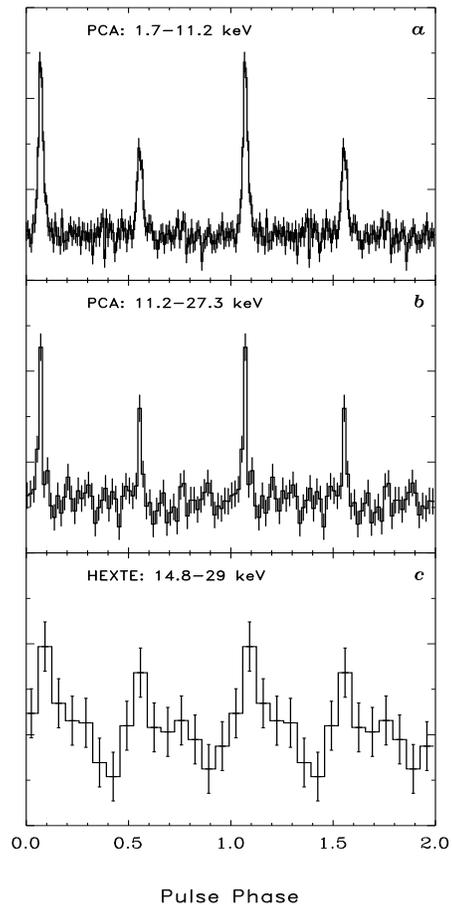}
     \caption{\label{psrj2022_prof}{\bf \psrh;} pulse profiles for \rxte\ PCA (panels a \& b) and \rxte\ HEXTE 
                                  (panel c). Pulsed emission has been detected up to $\sim 30$ keV. Note the sharpness of the pulses
                                  of about 1.2 ms (FWHM).}
  \end{center}
\end{figure}

\subsection{\psrh}

\psrh\ was discovered at 1.95 GHz with the GBT SPIGOT spectrometer targeting at the point source detected in the X-ray band
at the center of SNR G76.9+1.0 \citep[see e.g.][]{wendker1991,landecker1993} in a 54 ks \cxo\ ACIS observation \citep{arzoumanian2011}. 
The radio pulsations of \psrh\ are severely affected by dispersion (DM $=429.1 \pm 0.5$) and scattering in the interstellar 
medium, and these effects explain the non-detections of pulsed emission in targeted searches at lower radio frequencies, 
618 and 1170 MHz, with the GMRT \citep{marthi2011}.
Subsequent radio-observations with the GBT of this weak pulsar (60 $\mu$Jy at 2 GHz with a very flat spectrum, index $\alpha=-0.33$) settled the timing properties of 
\psrh\ \citep{arzoumanian2011}. A spin-up glitch occuring between MJD 54400 and 54950 was also evident in  these radio observations.

Dedicated \rxte\ monitoring observations performed between Jan. 27, 2010 and Febr. 4, 2010 revealed the pulsed signal (a single sharp 
pulse at pulse period 24.3 ms) also in the 2-20 keV PCA band \citep{arzoumanian2011}. These authors found a hard spectral index of 
$-1.1 \pm 0.2$ and an absorbed 2--10 keV pulsed flux of $5.4 \times 10^{-13}$ erg/cm$^2$s, which is comparable to the total measured absorbed 2--10 keV flux by \cxo, indicating a pulsed fraction of $\sim 100\%$.

An \xmm\ observation of \psrh\ of 116 ks performed on April 14, 2011 revealed that the pulse period is actually twice (48.6 ms) the initially found period due to the presence of an interpulse at about $\sim 0.5$ phase separation of the main pulse \citep{arumugasamy2014}. The adapted spin-down luminosity and characteristic age are now: $\dot{E}_{sd} \sim 2.97 \times 10^{37}$ and  8.9 kyr, respectively.

We re-analyzed the \rxte\ PCA data (88.73 ks of screened PCU-2 exposure) in order to deal with this ``change'' from single peaked to double peaked nature and to better characterize the hard X-ray spectrum by including all PCA detection layers, being more sensitive to photons with energies in excess of $\sim 10$ keV. We also analyzed the HEXTE data. A high-statistics template pulse profile (double peaked) was used in the correlation analysis to derive the pulse time-of-arrivals from which we constructed a phase coherent timing model (see Table \ref{eph_table}). This ephemeris has subsequently been used in a folding procedure creating the event distributions as a function of pulse phase and energy for both the PCA and HEXTE. Pulsed emission has been detected up to $\sim 30$ keV (see Fig. \ref{psrj2022_prof} panels a \& b for the PCA 1.7-11.2 ($28.2\sigma$) and 11.2--27.3 keV ($10.8\sigma$) and panel c for the HEXTE 14.8-29 keV ($3.4\sigma$) lightcurves, respectively). 

The broad-band PCA pulse profile can satisfactorily be fitted by a model composed of two Gaussians plus background yielding a pulse phase separation of $0.484\pm 0.001$ between main- and interpulse and pulse widths (FWHM) of $0.0254\pm 0.0014$ (main pulse) and $0.0273\pm 0.0024$ (interpulse). The X-ray pulse profile of \psrh\ bears strong similarities with the pulse profile of \psrf\ \citep[see e.g.][]{kuiper2010}.

The pulsed excess counts for the main pulse (P1), interpulse (P2) and its sum (total pulsed; TP=P1+P2) have been determined across the 2.5-32 keV PCA band (for which a proper
energy calibration exists) through fitting the measured pulse profiles for various differential energy bands in terms of the previously derived best Gaussian models (fixed position and shape) and background. Subsequently, these excess counts have been converted to pulsed fluxes in a forward folding spectral fitting procedure assuming a power-law model and adopting an absorbing Hydrogen column density N$_{\hbox{\scriptsize H}}$ of $1.7 \times 10^{22}$ cm$^{-2}$ \citep[][]{arzoumanian2011}. 

We found for the 2.5-32 keV band a photon index of $-1.31 \pm 0.02$, $-1.04 \pm 0.03$ and $-1.20 \pm 0.014$ for the main pulse, inter pulse and total pulse, respectively. The corresponding unabsorbed 2--10 keV band fluxes are $(3.60\pm 0.12)\times 10^{-13}$, $(1.91\pm 0.11)\times 10^{-13}$ and $(5.50\pm 0.17)\times 10^{-13}$ erg/cm$^2$s, for P1, P2 and TP, respectively. The main pulse thus comprises $\sim 65\%$ of the total pulsed flux. The absorbed/unabsorbed 2--10 keV total pulsed flux of $(4.94\pm 0.37)\times 10^{-13}$/$(5.50\pm 0.17)\times 10^{-13}$ erg/cm$^2$s is somewhat less than found by \citet{arzoumanian2011}, but is still comparable (93 \%) to the total absorbed 2--10 keV flux of the point source derived from \cxo\ (imaging) data. 

\psrh\ is not listed as high-energy \gr-ray pulsar in the \fermi\ second pulsar catalogue \citep{abdo2013}, however, the \fermi\ LAT second source catalogue \citep{nolan2012} reports on a $10\sigma$ \gr-ray source, 2FGL J2022.8+3843c, positionally consistent with G76.9+1.0. 
At TeV energies (100 GeV - 50 TeV) the VERITAS survey of the Cygnus region, reaching a VHE sensitivity of $\sim 4\%$ of the Crab nebula flux for an energy threshold of 200 GeV, did not result in a detection of a TeV source at the location of \psrh / G76.6+1.0 
\citep{ward2010}.


\begin{figure}
  \begin{center}
     \includegraphics[width=6cm,height=10cm,angle=0,bb=170 175 420 630]{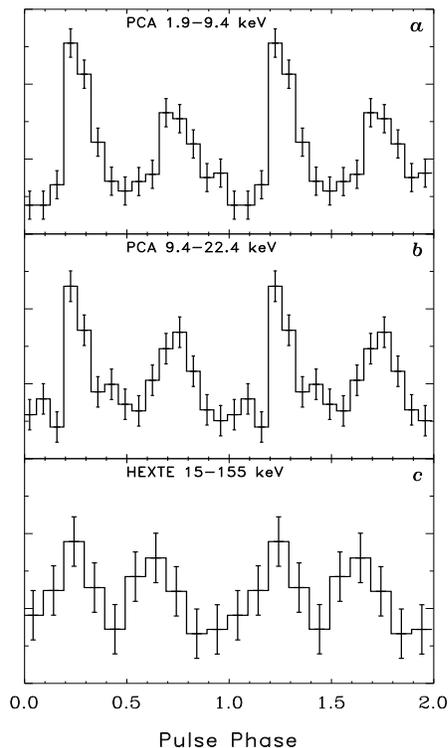}
     \caption{\label{psrj2229_prof}{\bf \psre;} pulse profiles in various energy bands for \rxte\ PCA (panels a--b), \rxte\ HEXTE 
                                  (panels c). Pulsed emission has been detected up to $\sim 22$ keV.}
  \end{center}
\end{figure}

\subsection{\psre}

\psre\ was discovered at radio frequencies in the error region of the unidentified EGRET source 3EG J2227+612 after the end of the \cgro\ mission \citep{halpern2001b}. The pulsed signal ($P \sim 51.6$ ms) was also detected in the X-ray band using archival \asca\ GIS (0.7-10 keV) data, taken 1.6 year before the detection of the radio pulsations, showing 2 pulses separated about 0.5 in phase. 

A non-thermal incomplete radio shell, G106.6+2.9, surrounds \psre. This compact radio nebula of size $\sim 3\arcmin$, called the ``Boomerang", has a flat spectrum and is polarized at a level of $\sim 25\%$, and represents the PWN associated with \psre. It is located at the northeastern tip of the radio continuum source G106.3+2.7 of size $0\fdg5 \times 1\degr$ which is classified as the remnant of a supernova \citep[see e.g.][]{kothes2006}, that gave birth to \psre. Diffuse emission from the near field of \psre\ has also been clearly detected at X-rays by \citet{ng2004} in a 95 ks \cxo\ ACIS-S observation (taken March 15, 2002), who fitted the observed morphology with their pulsar wind torus model.

In recent years the pulsed fingerprint of \psre\ has also securely been found in \agile\ and \fermi\ LAT data at energies above 100 MeV \citep{pellizzoni2009,abdo2009b}, showing a single asymmetric peak at a phase separation of
0.49 from the (single) radio pulse.
At TeV energies \citet{acciari2009} reported the detection of extended emission from G106.3+2.7, as measured by VERITAS.

We analyzed the - so far unpublished - \rxte\ PCA and HEXTE data from \rxte\ observation runs 60130 and 80092 (performed
on Dec. 7--8, 2001 and Nov. 25--29, 2003 for 90.4 ks and 129.8 ks, respectively) in order to characterize the pulsed emission
spectrum of \psre\ over the 3--250 keV band. In both observation runs the pulsed signal could easily be detected in the PCA $\sim$ 2--10 keV band (PHA range 4--27) taking into account the phase smearing due to the frequency derivative by including the $\dot\nu$ value of $-2.92006(5) \times 10^{-11}$ Hz/s, found at radio frequencies for a later epoch. The event cubes (2d distribution of events versus pulse phase (60 bins) and energy (PHA; 256 bins)) for both runs have been made and combined after applying an appropriate phase shift to align both distributions in phase. The combined (cleaned) exposure for PCU-2 amounts 220.144 ks. Similar folding and alignment (with known phase shift) procedures have been followed for HEXTE data. The pulse profiles of \psre\ obtained for the PCA bands 1.9-9.4 keV (PHA 5--22) and 9.4-22.4 keV (PHA 23--53) are shown in Fig. \ref{psrj2229_prof}a ($12.3\sigma$) and b ($8.3\sigma$), respectively. The double peaked morphology, already visible in the \asca\ GIS data for the 0.8-10 
keV 
band \citep{halpern2001b}, is evident, now also for energies above 10 keV. Pulsed emission has been detected up to $\sim 22$ keV.
In HEXTE data we found only indications for the pulsed signal for energies above $\sim 15$ keV (see Fig. \ref{psrj2229_prof}c; $2.4\sigma$).

We also analyzed archival \asca\ GIS \citep{ohashi1996} data obtained during an observation run (57038000) on August 4--7, 1999 with an effective exposure of about 113.8 ks. We confirmed the findings of \citet{halpern2001b}, but derived a more significant pulse profile for the GIS 0.8-10 keV range with a more pronounced double peak morphology by using $2\arcmin$ as optimum extraction radius instead of $4\arcmin$. We fitted this lightcurve with 2 symmetric Lorentzians on top of a flat background and derived a phase separations of
$0.49\pm 0.02$. The individual Lorentzian profile shapes are subsequently used as templates in the derivation of the pulsed excess counts for both \asca\ GIS and \rxte\ PCA pulse profiles for user defined energy bands. Applying proper response information, (effective) exposure times, correction factors for the missing fraction of the PSF (\asca\ GIS; the extraction region was a circular aperture of 2\arcmin, which covers about 57\% of the PSF) these pulsed excess counts have been converted to pulsed fluxes in a forward folding spectral analysis procedure, adopting an absorbed power-law model assuming an absorbing Hydrogen column N$_{\hbox{\scriptsize H}}$ of $6.3 \times 10^{21}$ cm$^{-2}$ \citep{halpern2001a}.

For the PCA 2.8-32.2 keV band the pulsed emission (sum of the two pulses) is properly described by a power-law with (hard) photon index of $-1.11\pm 0.03$ and an unabsorbed 2--10 keV flux of $(5.2\pm 0.3) \times 10^{-13}$ erg/cm$^2$s, consistent with the value 
derived by \citet{halpern2002} using \asca\ GIS data. We found no evidence for different spectral behaviour of the pulses in the X-ray band. However, between $\sim 20$ keV and 100 MeV the pulse morphology drastically changes from a double peaked to an asymmetric single
peaked profile \citep{abdo2009b}, indicating different spectral behaviour as a function of pulse phase. Unfortunately, there are no sensitive measurements in this energy band to further constrain the pulsed emission properties. The total pulsed emission spectrum of \psre\ from soft X-rays to high-energy \gr-rays is shown in Fig. \ref{spectralcompilation} including \asca\ GIS (0.8-10 keV), \rxte\ PCA (2.7-32 keV) and \fermi\ LAT measurements as orange data points/symbols and hatched region.

\section{Candidate soft \gr-ray pulsars}
\label{sect_cand_pul}
The sample size of detected soft \gr-ray pulsars is still small, despite efforts to increase this sample by searching for timing signatures
for energies above 20 keV from very promising candidate RPPs. Such strong candidates for detection at hard X-rays
are \fermi\ LAT detected very young ($\la 15$ kyr) energetic ($L_{sd} \ga 10^{36}$ erg/cm$^{2}$ s) pulsars, 
often coincident with a TeV source, but also similarly young and energetic pulsars which are {\it not} detected by the LAT. 
In this section we will discuss the status (from literature and our work) for the candidate soft \gr-ray pulsars as defined above.

\subsection{PSR J1833-1034 in SNR G21.5-0.9}
An interesting example is the weak ($S_{1400} = 71\mu{Jy}$ ) radio-loud and \fermi\ LAT pulsar PSR J1833-1034 ($\tau\simeq 4.8$ kyr; $L_{\hbox{\scriptsize sd}}\simeq 3.4\times10^{37}$ erg/s) at the centre of SNR G21.5-0.9, associated with TeV source HESS J1833-105 \citep{djannati2008}, which has a bright \integral\ soft \gr-ray counterpart \citep{derosa2009}. Therefore, it evidently is a soft \gr-ray pulsar candidate. 
However, deep timing searches in \nustar\ data (3--79 keV) did not reveal a pulsed signal from this pulsar \citep{nynka2014}. This suggests that the soft \gr-ray beams are for this pulsar not pointed in the observer's direction, contrary to the high-energy \gr-ray beams. 

\subsection{PSR J1119-6127, PSR J1124-5916 and PSR J1357-6429}
Other puzzling examples are the three, very young, energetic pulsars PSR J1119-6127 ($\tau\simeq 1.6$ kyr; $L_{\hbox{\scriptsize sd}}\simeq 2.3\times10^{36}$ erg/s), 
PSR J1124-5916 ($\tau\simeq 2.8$ kyr; $L_{\hbox{\scriptsize sd}}\simeq 1.2\times10^{37}$ erg/s) and PSR J1357-6429 ($\tau\simeq 7.3$ kyr; $L_{\hbox{\scriptsize sd}}\simeq 3.1\times10^{36}$ erg/s), all radio and \fermi\ LAT pulsars associated with SNRs. From these pulsars, pulsed {\it thermal} X-ray emission has been detected (well below 2 keV and with a broad single pulse), but not the expected strong non-thermal ($\ga 2$ keV) pulsed emission \citep[see e.g.][respectively]{gonzalez2005,hughes2003,chang2012}. 
For two of the three pulsars also TeV counterparts have been identified, HESS J1119-614 for PSR J1119-6127 and HESS J1356-645 for PSR J1357-6429.

\subsection{PSR J1420-6048 and PSR J1418-6058 in the Kookaburra}
Two other potential soft \gr-ray pulsars are located in the ``Kookaburra'' radio complex \citep{roberts1999}, coindicent with \cgro\ EGRET source 3EG 1420-6038 (and GeV J1417-6100).
Within this complex, encompassing an area of about $1^{\circ} \times 1^{\circ}$, two young and energetic pulsars have been detected at an angular separation of only $16\farcm5$. One, 
PSR J1420-6048 ($\tau\simeq 13.5$ kyr; $L_{\hbox{\scriptsize sd}}\simeq 1.0\times10^{37}$ erg/s) is located in the North-Eastern wing as radio pulsar \citep{damico2001} and is later also detected as 
high-energy \gr-ray pulsar \citep{weltevrede2010}. The other, PSR J1418-6058 ($\tau\simeq 10.2$ kyr; $L_{\hbox{\scriptsize sd}}\simeq 5.0\times10^{36}$ erg/s) is in the South-Western 
wing, coincident with a bright radio nebula called the ``rabbit", and is detected in a blind search at high-energy \gr-rays \citep{abdo2009d}. 

Moreover, \citet{aharonian2006b} detected TeV counterparts for both pulsars, HESS J1420-607 and HESS J1418-609, while \citet{hoffmann2007} reported on a soft \gr-ray source positionally
consistent with PSR J1420-6048 \citep[see also][]{fiocchi2010}. 

A (very) hard X-ray point source, embedded in a weak diffuse nebula, has been identified for PSR J1420-6048 within the (radio) K3 nebula \citep{roberts1999} by \citet{ng2005} who found 
for a 10 ks \cxo\ ACIS-S observation only 26 counts within an aperture of $1\arcsec$ around PSR J1420-6048. Further spectral analysis yielded a photon index $\Gamma$ of $\sim -1$ 
with a very large uncertainty due to the poor source statistics. We analysed a 90.7 ks on-axis \cxo\ ACIS-I observation performed on December 8, 2010 (obs.id. 12545; unpublished 
so far) to determine the (total) spectrum of the pulsar.
Adopting an absorbed power-law model and using a Maximum Likelihood (energy resolved) PSF fit method to extract the source counts (now 233 counts for the 1.5--10 keV range), we derived
a photon index $\Gamma$ of $-0.46\pm0.07$, an unabsorbed/absorbed 2--10 keV (total) flux of $(1.33\pm 0.10)\cdot 10^{-13}$ / $(1.14\pm 0.30)\cdot 10^{-13}$ erg/cm$^2$s and a Hydrogen 
column density N$_{\hbox{\scriptsize H}}$ of $(3.35_{-0.51}^{+0.74}) \times 10^{22}$ cm$^{-2}$, which is slightly larger than the Galactic column of $(1.58-2.13)\times 10^{22}$ cm$^{-2}$.
This confirms the hardness and weakness of the source with greatly improved accuracy. Our derived column density is also consistent with earlier X-ray derived values, using different
instruments aboard \xmm\ and \Suzaku\ and adopting much larger extraction radii, and thus including also the softer diffuse nebula emission \citep[see e.g.][]{vanetten2010,marelli2011,kishishita2012}. 

However, so far in the canonical X-ray band (0.3-10 keV) pulsed emission has {\it not} been securely detected yet from PSR J1420-6048 \citep{ng2005} in spite of some weak indications in 
\asca\ GIS data \citep{roberts2001}. Given the hardness of the source spectrum and encouraging results from W3PIMMS concerning its detectability in \rxte\ PCA data\footnote{Assuming a pulsed fraction of 
100\% for PSR J1420-6048 the measured \cxo\ spectrum predicts for the \rxte\ PCA with 3 PCU's operating a $5\sigma\ (3\sigma)$ detection for a total observing time of 125 (45) ks in 
the PCA 2-22 keV band.} we searched for the non-thermal pulsed component of PSR J1420-6048 above $\sim 2$ keV analysing a 66.4 ks on-axis 
\rxte\ PCA observation (obs.id. 80088) performed in the period August 4--6, 2003. The pulsar was observed in a post-glitch episode with valid (radio) ephemerides before and after the glitch \citep{yu2013}, thereby considerably narrowing the period search window. However, also now for the 2--30 keV PCA band the non-thermal pulsed emission component could not be detected 
within the predicted period search window. 


The region around the other pulsar, PSR J1418-6058, within the ``Rabbit" nebula has been resolved by \cxo\ \citep[obs.id. 2794; 10.1 ks; September 22, 2002;][]{ng2005} into two
unresolved point sources , denoted R1 and R2, and an extended source North of these sources. In a later \cxo\ observation (obs.id. 7640; 71.1 ks; June 14, 2007) 
source R2 was undetectable and is likely associated with an AGN, while the other source R1 and the extended source behaved as steady sources. Updated positional information for 
PSR J1418-6058 from Fermi LAT timing analysis \citep[see the TEMPO parameter file in the supplementary material provided in][]{abdo2013} yielded a position consistent with R1 \citep[see also Section on PSR J1418-6058 in][]{ray2011b}. Searches for pulsed X-ray emission from PSR J1418-6058 so far focussed on source R2 \citep[ACIS in Continuous Clocking mode and XMM-EPIC pn 
in Small Window Mode;][]{ng2005}, while the extended source North to R1 and R2 was favoured as counterpart in \citet{marelli2012} (see Chapter 4 page 154--157), who performed a 
spectral analysis on this proposed counterpart subsequently.

We performed a period search using EPIC-pn Small Window Mode data from a long 125.2 ks \xmm\ observation performed on Februari 21--23, 2009 (obs.id. 0555700101) selecting events from a 
$15\arcsec$ aperture centered on R1. We could {\it not} detect pulsed emission in the expected (a Fermi LAT based ephemeris overlapping the XMM observation period is available) period range, 
neither in the 0.3-2 keV nor in the 2--10 keV band.

\subsection{The Fermi blind-search pulsars PSR J1023-5746, PSR J1838-0537 and PSR J1826-1256}
PSR J1023-5746, detected in a blind search in (\fermi) LAT data, is located near star forming region RCW 49 and its OB association Westerlund 2. It has a characteristic age of only 4.6 kyr and spin-down luminosity of
$1.1\times 10^{37}$ erg/s \citep{sazparkinson2010}. Archival off-axis \cxo\ ACIS observations revealed a hard X-ray source, CXOU J102302.84-574606.9, coincident with the Fermi LAT timing 
position with a typical (hard) pulsar-like power-law spectrum with a photon index $\Gamma$ of $-1.2\pm0.2$ and unabsorbed flux of $(1.2\pm0.3)\times 10^{-13}$ erg/cm$^2$s \citep{sazparkinson2010}. 
The detection was later confirmed in an on-axis 9.9 ks \cxo\ ACIS observation within the \cxo\ Pulsar Survey (ChaPS) project \citep{kargaltsev2012b}.
The pulsar location is also positionally consistent with TeV source HESS J1023-575 \citep{hesscol2011}. Unfortunately, the current (hard) X-ray archive does not contain suitable observations to perform period searches in order to reveal the possible/expected pulsed fingerprint of the pulsar.

PSR J1838-0537 ($\tau \simeq 5$ kyr; $L_{\hbox{\scriptsize sd}}\simeq 5.9\times10^{36}$ erg/s ) was discovered by \citet{pletsch2012} as \fermi\ LAT blind-search pulsar applying 
a novel data-analysis search technique. Its sky position coincides with the extended TeV source HESS J1841-055, which is likely composed of multiple sources. Recently, a 43.3 ks on-axis 
\xmm\ observation (obs.id. 0720750201; Oct. 14, 2013) has been performed. The EPIC-pn exposure was split in two chunks of
24.7 and 12.5 ks, while it was operating in Full Frame mode (time resolution 73.4 ms, which is just too low to catch the 145.77 ms timing signal). The second chunk was heavily contaminated 
by soft proton flares (only 3.7 ks of exposure left after screening) and subsequently ignored in our further analysis. The first exposure part, however, was unpolluted, and has an effective 
exposure of 22.1 ks. In this exposure we detected at a $6\sigma$ level in the 2--10 keV band a (very) weak source positionally consistent with PSR J1838-0537. 
We found $61 \pm 13$ counts for the source in the 2--10 keV band, while below 2 keV no detection could be claimed. The count rate, however, is too low to determine and constrain the X-ray 
spectral characteristics of the pulsar. Unfortunately, there are no further suitable X-ray observations in the (HEASARC) archive providing
sufficient time resolution to search for a possible timing signal of this pulsar.

Another \fermi\ LAT blind-search pulsar is PSR J1826-1256 \citep[$\tau \simeq 14.4$ kyr; $L_{\hbox{\scriptsize sd}}\simeq 3.6\times10^{36}$ erg/s;][]{abdo2009d}, not to be confused 
with the nearby energetic pulsar PSR J1826-1334 (PSR B1823-13; $\tau \simeq 21$ kyr; $L_{\hbox{\scriptsize sd}}\simeq 2.8\times10^{36}$ erg/s) about 38\arcmin\ away, which has not been detected 
(yet) as high-energy \gr-ray pulsar contrary to its PWN \citep{grondin2011}. 
PSR J1826-1256 coincides with the Northern extension of the extended TeV source HESS J1825-137 \citep{aharonian2006c}.
This TeV structure is positionally consistent with \cgro\ EGRET source ($>1$ GeV) GeV J1825-1310 \citep[also 3EG J1826-1302; see][]{roberts2001b}. A plausible soft X-ray counterpart is \asca\ GIS 
source AX J1826.1-1257 located within the error contours of GeV J1825-1310. The X-ray source region was later resolved by \cxo\ (obs.id. 3851; 15.2 ks; ACIS-I; Feb. 17, 2003) in a 
point-like component, positionally consistent with the pulsar, with a faint, $\sim 4\arcmin$ long trail of hard ($> 2$ keV) X-ray emission, resembling an ``Eel'' \citep{roberts2007}. 
Using data from a much deeper more recent \cxo\ ACIS-I observation (obs.id. 7641; 74.8 ks; July 26, 2007) the spectral characteristics of the counterpart could be determined yielding 
a (very hard) photon index $\Gamma$ of $-0.79\pm0.39$ and an unabsorbed 0.3-10 keV flux of $(1.12\pm0.25)\times 10^{-13}$ erg/cm$^2$s absorbed through a Hydrogen column N$_{\hbox{\scriptsize H}}$ of 
$(1.26_{-0.46}^{+0.53}) \times 10^{22}$ cm$^{-2}$ \citep[see Chapter 4, page 202--204 of][]{marelli2012}.
In the X-ray archive we found a 92.9 ks \rxte\ observation (obs.id. 90069; Nov. 24--29, 2004) with PSR J1826-1256 5\farcm7 off-axis.  A subsequent search for a possible timing signal in the 2--30 keV 
band yielded a negative result, not surprising given the source flux and its prediction for the PCA instrument requiring observation times in the range 450--650 ks for a $5\sigma$ detection. 
A decisive answer about the X-ray timing properties will be given soon from the data analysis of a 140 ks \xmm\ observation performed on Oct. 11, 2014.

\begin{table*}
\renewcommand{\tabcolsep}{1.8mm}
\begin{center}
\caption{Rotation powered (non-recycled) pulsars with (securely detected) pulsed emission in the hard X-ray band ($\ga 20$ keV). 
 \label{tabrpp}}
\begin{tabular}{lccccclcc}
\hline\noalign{\smallskip}
name                          & period & age    & $^{10}\log(L_{\hbox{\scriptsize sd}})$  & $S_{1400}$    & pulse  & photon & \fermi\ & TeV \\
                              &  (ms)  &(kyr)   &                      &  (mJy)        & shape  & index  & LAT/Pulsed  & PWN/Pulsed \\
\hline\noalign{\smallskip}
\noalign{\smallskip}
\psrf\ (3C58)                 & 65.7   & 5.4    & 37.43  & $0.045^{1}$  & two sharp pulses     & -1.1(1)  & yes   & yes\\
\crabj/B0531+21 (Crab)        & 33.5   & 1.23   & 38.66  & $14$         & two pulses           & curved   & yes   & yes/yes \\
\psrg\ (N157B in LMC)         & 16.1   & 4.9    & 38.69  & $<0.06^{2}$  & single, sharp        & -1.57(1) & no    &yes \\
\lmcpsra/B0540-69             & 50.5   & 1.7    & 38.18  & $0.106^{3}$  & structured broad     & curved   & yes   & ... \\
\hspace{0.5cm}(N158A in LMC)                &        &        &        &              &                      &          &       &   \\
\velaj/B0833-45 (Vela)        & 89     & 11     & 36.84  & $1100$       & multiple sharp       & -1.1     & yes   & yes/yes \\
\psrigra\, \igra              & 31.2   & 12.7   & 37.71  & $0.242$      & single, broad        & -1.95(4) & no    & no \\
\mshj/B1509-58                & 150    & 1.6    & 37.26  & $0.94$       & single broad         & curved   & yes   & yes \\
\hspace{0.5cm}(MSH 15-52)                   &        &        &        &              &                      &          &       &   \\
\psrc\                        & 69.0   & 8.0    & 37.20  & $0.5^{4}$    & single, broad        & -1.42(2) & no    & ?  \\
\psrcanda\ (G338.3-0.0)       & 206    & 3.4    & 36.64  & $<1.0^{5}$   & single               & -1.3     & no    & yes\\
\psrb\ (G11.2-0.3)            & 65.0   & 24.0   & 36.81  & $<0.07^{2}$  & single, broad        & -1.11(1) & no    & no \\
\psrcandb                     & 48.1   & 43.0   & 36.80  & $<0.017^{6}$ & two pulses           & -0.85(3) & yes   & ...\\
\psri\ (G12.82-0.02)          & 44.7   &  5.6   & 37.75  & $<0.01^{7}$  & single, broad        & -1.30(3) & no    & yes\\
\psraxj\, \axj                & 70.5   & 23.0   & 36.75  & $<2$         & structured broad     & -1.12(1) & no    & yes\\
\psra\ (Kes 75)               & 324    & 0.72   & 36.91  & $<0.027^{8}$ & single, broad        & -1.20(1) & no    & yes\\
\psrigrb\, \igrb              & 38.5   & 42.8   & 36.99  & $<0.853^{9}$ & single, broad        & -1.37(1) & no    & yes\\
\psrd\ (G54.1+0.3)            & 136    & 2.9    & 37.08  & $0.045^{1}$  & single, broad        & -1.21(1) & no    & yes\\
\psrh\ (G76.9+1.0)            & 48.6   &  8.9   & 37.47  & $0.067^{10}$ & two sharp pulses     & -1.20(2) & no    & no \\
\psre\ (G106.6+2.9)           & 51.6   & 10.5   & 37.34  & $0.25$       & two pulses           & -1.11(3) & yes   & yes\\
\noalign{\smallskip}
\hline
\multicolumn{9}{l}{Notes.}\\
\multicolumn{9}{l}{Column 1 gives source name(s) and associated SNR or PWN (between brackets), when applicable.}\\
\multicolumn{9}{l}{Column 3 gives the characteristic age ($\tau=-0.5\nu/\dot{\nu}$)}\\
\multicolumn{9}{l}{Column 4 the spin-down power ($L_{\hbox{\scriptsize sd}}=4\pi^2I\nu\dot{\nu}$), in erg/s.}\\
\multicolumn{9}{l}{Column 5 gives the radio flux density (or upper limit) at 1400 MHz ($S_{1400}$; see Section 6), taken from the ATNF database except for the noted entries, where: }\\
\multicolumn{9}{l}{\hspace{0.5cm}(1) \citet{camilo2002}, (2) \citet{crawford1998}, (3) \citet{manchester1993},(4) \citet{kaspi1998}, (5) \citet{castelletti2011}, (6) \citet{ray2011b},}\\
\multicolumn{9}{l}{\hspace{0.5cm}(7) \citet{halpern2012}, (8) \citet{archibald2008}, (9) \citet{pandey2006}, (10) \citet{arzoumanian2011}.}\\ 
\multicolumn{9}{l}{Columns 6 and 7  give for the 2--150 keV band a description of the pulse shape and the energy spectrum (power-law photon index or curved spectrum).}\\
\multicolumn{9}{l}{Column 8 indicates whether the \fermi\ LAT detected a pulsed signal.}\\
\multicolumn{9}{l}{Finally, column 9 indicates whether a PWN and pulsed emission have been detected at TeV energies.}\\
\end{tabular}
\end{center}
\end{table*}

\subsection{Young and energetic pulsars not detected by {\it Fermi } LAT: PSR J1301-6305, PSR J1341-6220, PSR J1614-5048, PSR J1803-2137 and PSR J1856+0245}
Table 13 of \citet{abdo2013} lists 28 pulsars not detected by the LAT with spin-down powers exceeding $10^{36}$ erg/s. Apart from PSR J1747-2809 (in G0.9+0.1) all top 12 pulsars 
listed in this table, 
those with the highest spin-down power, {\it are} detected as soft \gr-ray pulsars \citep[note: \lmcpsr\ has now been detected also as \fermi\ pulsar; ][]{martin2014}.
Of the additional 15 pulsars we consider those pulsars which are young and/or associated with TeV counterparts the best candidates to detect as soft \gr-ray pulsars.  

In particular, PSR J1301-6305 ($\tau \simeq 11$ kyr; $L_{\hbox{\scriptsize sd}}\simeq 1.7\times10^{36}$ erg/s), associated with HESS J1303-631 \citep{aharonian2005c,hesscol2012}, and PSR J1341-6220 
(PSR B1338-62; $\tau \simeq 12.1$ kyr; $L_{\hbox{\scriptsize sd}}\simeq 1.4\times10^{36}$ erg/s), associated with SNR G308.7+0.0/G308.8-0.1 \citep{caswell1992,kaspi1992} are good candidates to harbour 
a soft \gr-ray pulsar. For the latter pulsar, a frequent glitcher, now also a weak X-ray counterpart is visible (this work) in archival XMM Newton data (obs.id. 0301740101; July 26, 2005, 37.8 ks).

Other interesting candidates are the 7.4 kyr old pulsar PSR J1614-5048 (PSR B1610-50; $L_{\hbox{\scriptsize sd}}\simeq 1.6\times10^{36}$ erg/s) showing weak $>2$ keV X-ray emission in \asca\ GIS images \citep{kawai1996,brinkmann1999}, but doubted by \citet{pivovaroff2000}, and PSR J1803-2137 (PSR B1800-21; $\tau \simeq 15.8$ kyr; $L_{\hbox{\scriptsize sd}}\simeq 2.2\times10^{36}$ erg/s) possibly associated with HESS J1804-216 \citep{kargaltsev2007b,lind2013}. For the latter pulsar \citet{kargaltsev2007a} detected weak, but hard pulsar emission in a 30.2 ks \cxo\ ACIS observation.

Noteworthy is also the 81 ms pulsar PSR J1856+0245 ($\tau \simeq 21$ kyr; $L_{\hbox{\scriptsize sd}}\simeq 4.6\times10^{36}$ erg/s) discovered in the Arecibo PALFA survey \citep{hessels2008} coincident with TeV source HESS J1857+026. At X-rays the pulsar was clearly detected in a 39 ks \cxo\ ACIS-I observation \citep[obs.id. 12557; Feb. 28, 2011][]{rousseau2012}, but the low number of source counts prevented an accurate spectral characterisation of the pulsar at X-rays. 
Data from an archival 55.3 ks \xmm\ observation (obs. id. 0505920101; March 27, 2008) have been analysed in this work applying our Maximum Likelihood source fitting method per energy band to disentangle the source contributions
from the pulsar and an unrelated source at an angular distance of about $20\arcsec$, rendering about 300 pulsar counts above $2$ keV for further spectral analysis. 
Adopting an absorbed power-law model we obtained a very hard photon index $\Gamma$ of $-0.34\pm 0.08$ absorbed through a Hydrogen column N$_{\hbox{\scriptsize H}}$ of 
$(1.8_{-0.7}^{+1.2}) \times 10^{22}$ cm$^{-2}$, consistent with the Galactic value of (1.5--1.8)$\times 10^{22}$ cm$^{-2}$. The unabsorbed 2--10 keV flux was $(9.52_{-0.80}^{+0.91})\times 10^{-14}$ erg/cm$^2$s, consistent with the value given in \citet{rousseau2012}. 
Our results are, however, somewhat different from those reported by \citet{nice2013}, who found a softer photon index and a somewhat lower unabsorbed 2--10 keV flux, mainly 
due to the higher absorbing Hydrogen column density N$_{\hbox{\scriptsize H}}$ of $(4.9_{-2.4}^{+3.2}) \times 10^{22}$ cm$^{-2}$ that they derived.

\subsection{Another soft \gr-ray pulsar candidate: IGR J11014-6103}

A puzzling soft \gr-ray pulsar candidate is \integral\ source IGR J11014-6103 \citep{bird2010}, studied in detail by \citet{pavan2011} using all available archival X-ray data of the region.
These authors concluded that the source is probably a PWN surrounding an energetic pulsar given the morphology of the source composed of two distinct sources separated by $\sim 22\arcsec$ and a dimmer
elongated structure (tail) of $\sim 4\arcmin$ size. 

A subsequent \cxo\ observation shed light on the spatial structure of the emission components, and showed that the putative pulsar moves away 
from SNR MSH 11-61A, located 11\arcmin away from IGR J11014-6103 \citep{pavan2014}. The modulated main jet and counter jet structures are perpendicularly oriented with respect to the propagation direction
away from the SNR. Spectral analysis yielded for the pulsar counterpart, adopting an absorbed power-law model, a hard photon index $\Gamma$ of $-1.1\pm0.2$ with a 2--10 keV flux of $(6.1\pm0.6)\times 10^{-13}$ erg/cm$^2$s.

The putative pulsar has been detected recently by \citet{halpern2014}, who found 62.8 ms pulsations in two different \xmm\ observations performed on July 21, 2013 and June 8, 2014. 
From the timing characteristics they deduced a spin-down luminosity $L_{\hbox{\scriptsize sd}}$ of $\simeq 1.36\times10^{36}$ erg/s and a characteristic age $\tau$ of $116$ kyr, rather old
for such a system. These values should be taken with some care, because it had been assumed that no glitch or glitches had occured between the two observations. 

In this work we have also determined the pulsed spectrum of IGR J11014-6103 (analoguous to the method outlined for \psri\ in Sect. \ref{sectpsri} ), employing EPIC-pn data from the two \xmm\ observations. We found for spectrum of the pulsed emission a photon index $\Gamma$ of $-1.13\pm 0.11$ and an unabsorbed 2--10 keV flux of $(3.93\pm 0.35)\times 10^{-13}$ erg/cm$^2$s, absorbed through an Hydrogen column N$_{\hbox{\scriptsize H}}$ of $(0.95_{-0.25}^{+0.38}) \times 10^{22}$ cm$^{-2}$. The latter value is consistent with the column density derived by \citet{pavan2014}, who used \cxo\ ACIS-I data, for the total pulsar emission. From the pulsed and total (unabsorbed) 2--10 keV fluxes we derived a pulsed fraction of about 64\%.
This demonstrates that IGR J11014-6103 is an excellent candidate to show up as soft \gr-ray pulsar in the future, given its spectral hardness and the already established soft \gr-ray nature for its total source emission.

\section{Summary \& Discussion}
In section 5 we presented in detail the multi-wavelengths characteristics of the currently 18 members of the soft \gr-ray pulsar population, and provided for 14 of them new high-energy results from 
this work. For \crab\ (\crabj), \lmcpsr\ (\lmcpsra), \psrf\ and \psra\ we reviewed the current status at high energies referring to up-to-date literature.

For all of them the pulse profiles have been detected significantly above 20 keV and the spectral shapes of the pulsed-emission spectra have been determined (curved or power-law shapes). However, these pulsars are weak emitters in the hard X-ray window, with \crabj\ (the Crab pulsar) and \mshj\ being the exceptions.
By exploiting the extensive \rxte\ and \integral\ archives for accumulating sufficient exposures on these pulsars at hard X-rays, we could achieve this major progress.  Phase-coherent ephemerides were required to find the hard X-ray pulsed signals for all observations used. An extreme example is the case of \psrg\ for which we derived and used 30 ephemerides spread over 7.7 years to arrive at a {$\sim 10\sigma$} detection of the pulse profile between 15 and 50 keV, in this case in \rxte\ HEXTE data. In several cases we also obtained new or statistically improved results at lower X-ray energies using archival \xmm\ data.

An interesting example is  \velaj\ (the Vela pulsar), for which we present in Fig. \ref{vela_prof} pulse profiles from the radio to the high-energy \gr-ray band. High statistics \integral\ ISGRI, \rxte\ PCA and \xmm\ EPIC-pn data reveal a plethora of sharp pulses, which have counterparts at lower energies in the radio, optical and/or UV bands as well as at the high-energy \gr-rays. One can identify five, possibly more sharp pulses, exhibiting very different variations in shape and intensity over the electromagnetic spectrum. This complexity has not yet been addressed in theoretical works, and it is beyond this paper to address this unique behaviour of the Vela pulsar.

A pulsar, for which we present an extensive study is \psrigrb, discovered as \integral\ source \igrb\, and associated with the TeV source HESS J1849-000. The latter association suggested the pulsar to be located in a PWN. Using again archival \rxte\ PCA and \xmm\ data, we derived a new high-statistics spectrum for the pulsed emission (2-10 keV) that can be well described with a power-law shape with index of $-1.37\pm 0.01$ and pulsed flux of $(4.21 \pm 0.08) \times 10^{-12}$ erg/cm$^{2}$s. The latter flux is a factor about four higher than reported by \citet{gotthelf2011}. This pulsed spectrum could be extended up to 60 keV by analysing \rxte\ HEXTE  data.
Using \integral\ ISGRI data, we obtained a new spectrum for the total emission in the energy band 20-300 keV with a softer spectral index of $-1.82\pm 0.14$. Taking the latter spectrum into account, the spectrum of the pulsed emission is better described with a curved spectral shape from 2 keV up to $\sim 300$ keV (see Fig. \ref{igrj18490acisspc}). We applied for CXO observations (with the HRC-S for timing and the ACIS-S for spatially resolved spectral analysis) to resolve the contributions from the point source and the PWN. 
These HRC-S and the \rxte\ PCA observations both give a pulsed fraction of $\sim 80 \%$, about three times the value published by \citet{gotthelf2011}. The ACIS-S observation reveals for energies 2-10 keV structured diffuse emission in a rectangular region ($75\arcsec \times 37\farcs5$ box in Fig.\ref{igrj18490cxoacis}) around the point source with a photon spectral index of $-1.13 \pm 0.06$, to be compared with an index of $-1.75 \pm 0.05$ that we measured with XMM for a more extended region. This difference we explain as Synchrotron cooling towards the outskirts of the PWN.

\begin{figure*}
  \begin{center}
     \includegraphics[width=16.5cm,height=20.5cm,angle=0,bb=100 105 515 680]{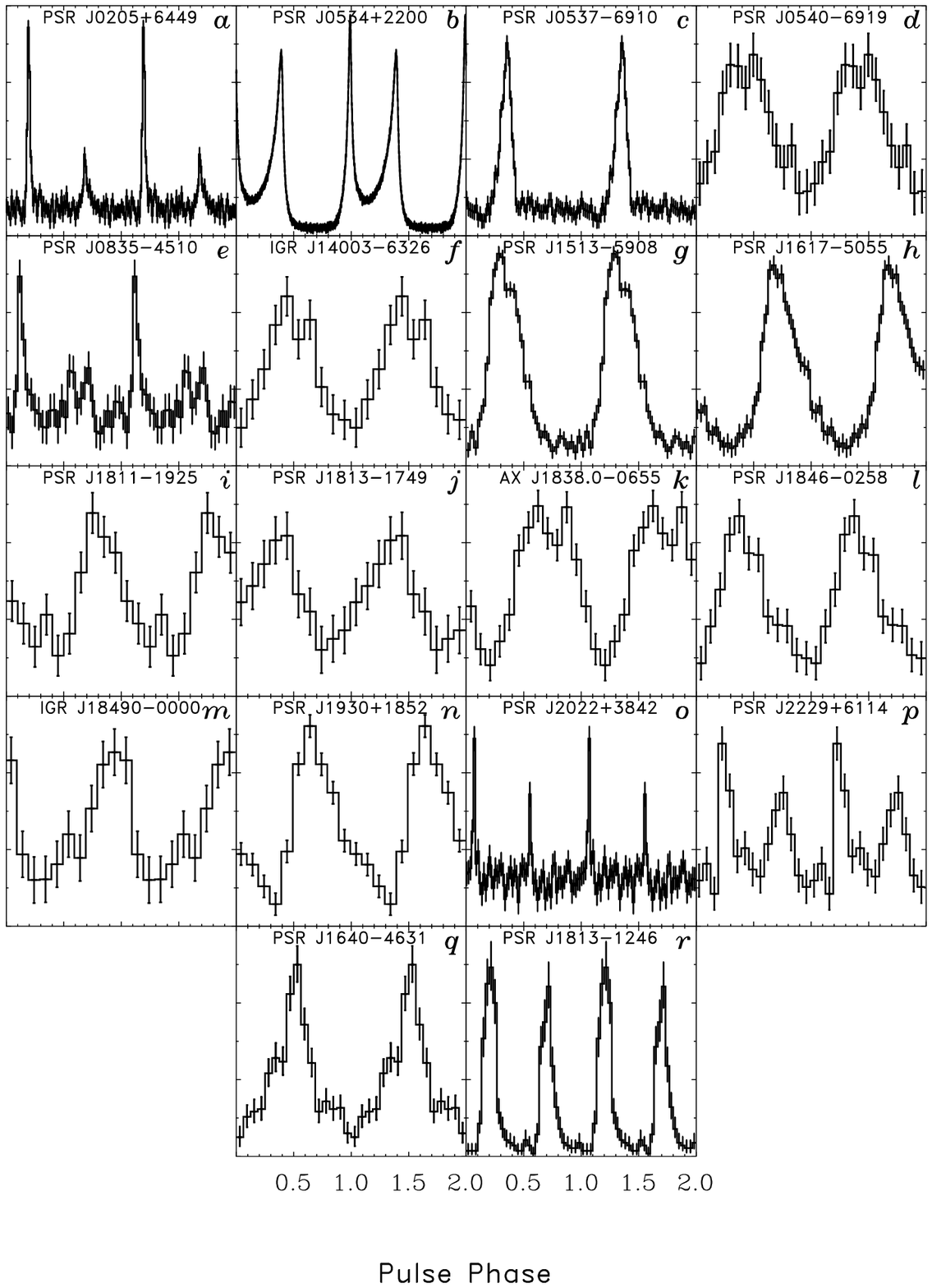}
     \caption{\label{he_psr_morph}The pulse shapes of the soft \gr-ray pulsar population members: {\it a)} \psrf, PCA 8.2--27.8 keV, {\it b)} PSR J0534+2200/B0531+21, ISGRI 20--300 keV, 
                                  {\it c)} \psrg, PCA $\ga 10$ keV, {\it d)} PSR J0540-6919/B0540-69, HEXTE 12--48 keV, {\it e)} PSR J0835-4510/B0833-45, ISGRI 20--300 keV, {\it f)} \igra, PCA 7.6--32.2 keV,
                                  {\it g)} PSR J1513-5908/B1509-58, ISGRI 20--300 keV, {\it h)} \psrc, PCA 8.2--32.0 keV, {\it i)} \psrb, ISGRI 15--135 keV, {\it j)} \psri, PCA 8.1--27.5 keV,
                                  {\it k)} \axj, ISGRI 20--150 keV, {\it l)} \psra, ISGRI 20--150 keV, {\it m)} \igrb/PSR J1849-0001, HEXTE 14--150 keV, {\it n)} \psrd, PCA 8.2--32.1 keV,
                                  {\it o)} \psrh, PCA 11.2--27.3 keV, {\it p)} \psre, PCA 9.4--22.4 keV, and the recently reported hard X-ray pulsars {\it q)} \psrcanda, \nustar\ 3--25 keV,
                                  {\it r)} \psrcandb, XMM EPIC-pn 8--12 keV (significantly smeared due to the rather coarse time resolution of $\sim 5.7$ ms given its period 
                                  of 48.1 ms; see also Fig. \ref{psr1813pca}). 
                                  }
  \end{center}
\end{figure*}

\begin{figure*}
  \begin{center}
     \includegraphics[width=8.75cm]{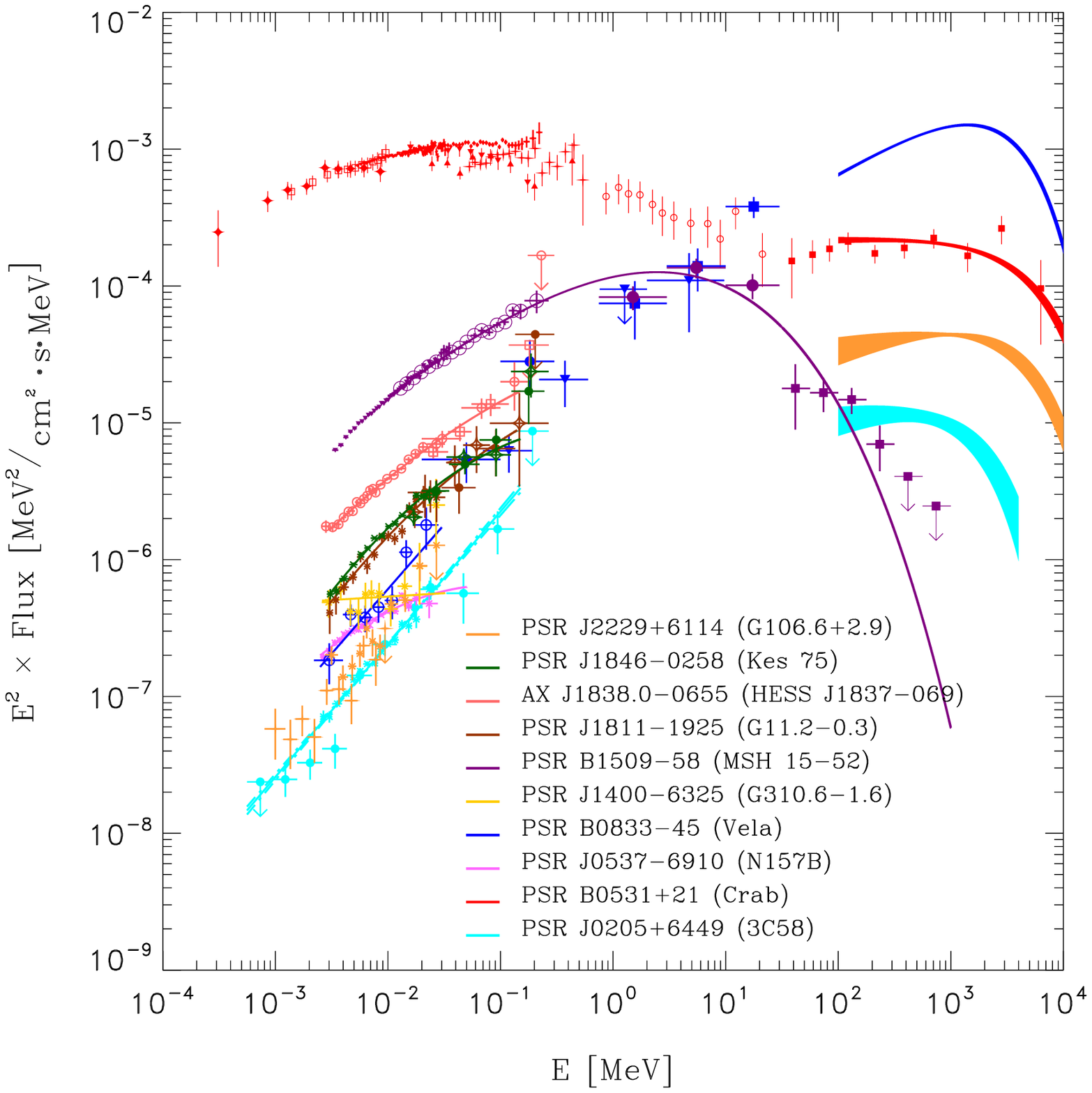}
     \includegraphics[width=8.75cm]{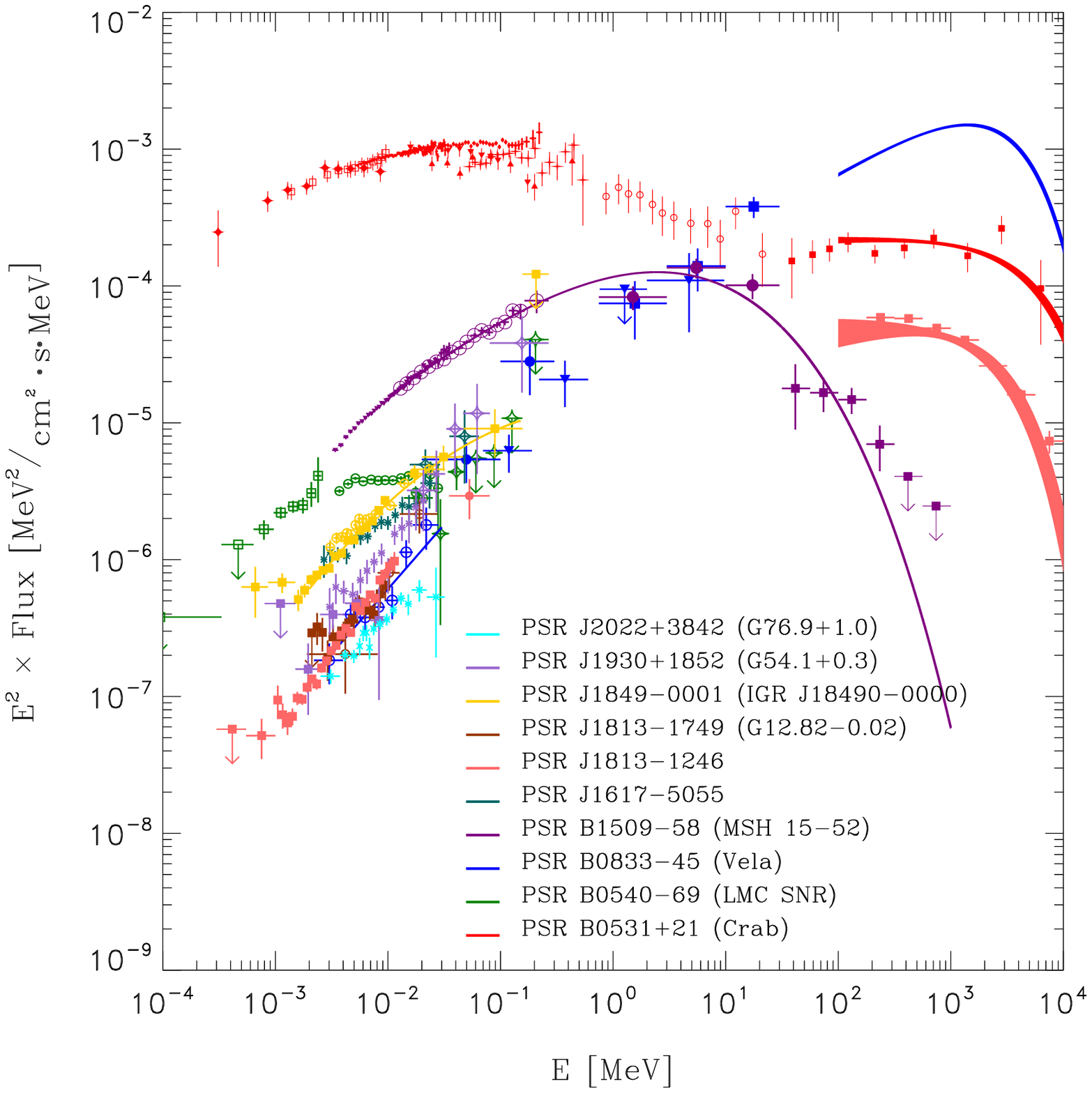}
     \caption{\label{spectralcompilation} The high-energy {\it pulsed} emission spectra for 17 of the 18 soft \gr-ray pulsar population members from 0.1 keV -- 10 GeV spread across 
     two panels. The spectra of \crab\ (Crab), \vela\ (Vela) and \msh\ are shown in both panels for reference purposes. The names of the soft \gr-ray pulsar members, ordered according to their right ascension increasing from bottom to top, are indicated in the plots along with color coding. Symbols/data points are for flux measurements and solid lines (filled regions for the Fermi passband) for the model fits. The pulsed spectrum of \psrcanda\ is not shown in either of the panels, because only its total spectrum has been reported on.}
  \end{center}
\end{figure*}

\subsection{Characteristics of the soft \gr-ray pulsar population}

In the following we will summarize and discuss the characteristics of the population of soft \gr-ray pulsars in comparison with those of the sample of non-recycled pulsars in the \fermi\ LAT second pulsar catalogue. 
Table \ref{tabrpp} summarizes the main characteristics of the 18 soft \gr-ray pulsars. They are all fast rotators with 
spin-periods ($P=1/\nu$) between 16.1 ms and 324 ms, most of them (14) even spinning well below 100 ms (see column 2 of Table \ref{tabrpp}). 
The characteristic ages ($\tau=-0.5\nu/\dot{\nu}$) of this pulsar sample range from 1.23 kyr (\crab) up to 43 kyr (\psrcandb), thus representing a very young
population, well below 50 kyr. Moreover, the population is very energetic with spin-down powers ($L_{\hbox{\scriptsize sd}}=4\pi^2I\nu\dot{\nu}; I$ is the moment of inertia of the neutron star assumed to be $10^{45}$ g cm$^2$) above $4.4\times 10^{36}$ erg/s (the lowest value measured for \psrcanda). 

The radio flux densities (or upper limits) at 1400 MHz ($S_{1400}$) are shown in column 5 of Table \ref{tabrpp}, taken from the ATNF Pulsar Catalog, except for the noted entries. When there are no published values at 1400 MHz, we extrapolated to $S_{1400}$ from measurements/upper limits at other frequencies, assuming $S_{\nu} \propto \nu^{\alpha}$ with $\alpha = -1.7$ or using a published $\alpha$ value for a particular pulsar.
For 8 members no radio pulsations have yet been detected, however, two (\psrcanda\ and \psrigrb) have upper limits at 1400 MHz as high as $\sim$ 1 mJy. \citet{abdo2013} defined for the \fermi\ LAT catalogue a pulsar as ``radio-loud''  if $S_{1400} > 30 \mu Jy$, and ``radio-quiet'' if the measured radio flux density is lower. In their sample of 77 young non-recycled pulsars 41 (53\%) are radio-loud, and the fraction of radio-loud pulsars increases with increasing $L_{\hbox{\scriptsize sd}}$  (consistent with 100\% for $L_{\hbox{\scriptsize sd}}> 10^{37}$ erg/s). Adopting that definition, in Table \ref{tabrpp} there are three radio-quiet pulsars \psrcandb, \psri\ and \psra\ with upper limits below $30 \mu Jy$. There are also three weak radio-loud pulsars listed with flux densities 
$30 \mu Jy < S_{1400} < 100 \mu Jy$ and two with upper limits in that range. Only the Vela and Crab pulsars have flux densities above 1 mJy (11\%). 
The sample of LAT radio-loud non-recycled pulsars has only three entries below $100 \mu Jy$ and 18 (44\%) above 1 mJy (see Fig. 3 of \citet{abdo2013}). This indicates that the radio-loud soft \gr-ray pulsars appear to be on average weaker radio pulsars than their LAT counterparts. 
However, the fraction of radio-loud pulsars of the energetic ($L_{\hbox{\scriptsize sd}}>4.4\times 10^{36}$ erg/s) soft \gr-ray sample is similar to that fraction of the LAT pulsars, namely in the range 56 \% to 83\%, depending on how many of the so far not detected pulsars are radio-quiet, and seems to agree 
with the trent reported above for the \fermi\ LAT pulsars.

Another noteworthy finding is that only 7 of the 18 members of the soft \gr-ray pulsar population are detected as \gr-ray pulsars in the hard \gr-ray band above 100 MeV \citep[see column 8, with as latest addition the Fermi detection of \lmcpsra, the twin of the Crab pulsar in the LMC;][]{martin2014}. Furthermore, \citet{abdo2013} list in their Table 13 28  rotation-powered pulsars which are not (yet) detected by \fermi\ despite of their high spin-down powers (above $10^{36}$ erg/s). 
Of the top 12 in that table, those with the highest spin-down power, 11 are listed in Table \ref{tabrpp} of this paper. Interestingly, those most energetic pulsars missed by the LAT, are reported by us to be soft \gr-ray pulsars.

In the last column of Table \ref{tabrpp} one can see that at least 12 members do have, often bright, 
TeV counterparts, associated with the TeV emission from their PWNe.  For three of these pulsars (\psri\ , \psraxj\ and \psrigrb) we present in this paper the 
total (PWN+pulsar) source spectra as measured with the imaging instrument \integral\ ISGRI from 20 keV up to $\sim$ 150 keV, together with the pulsed-emission spectra. The difference between these two spectra gives an estimate of the PWN spectrum at hard X-rays. This PWN spectrum is shown for \psraxj\ in 
Fig.\ref{psraxj_spc}  for energies between 3 and $\sim 60$ keV, and suggests a break/bend around 50 keV. For \psrc\ and \psrb\, the latter located in SNR G11.2-0.3, we also present the pulsed and total-emission spectra, but these pulsars are not detected at TeV energies, so far.
In only two cases, \crabj\ (Crab) and \velaj\ (Vela), pulsed TeV emission has been detected. 

The variation in morphology of the pulse profiles (described in column 6 of Table \ref{tabrpp}) in the hard X-ray band can be seen in Fig. \ref{he_psr_morph}.
This figure shows that the large majority (11 members) has broad/structured  single pulses, and  one (\psrg) has a sharper single pulse, making the fraction 
with single pulses 67\%. These are all, except the canonical soft \gr-ray pulsar \mshj\ and \lmcpsra, {\it not} detected at high-energy \gr-rays ($>100$ MeV) 
by \fermi\ LAT.
The remaining 6 (33\%) have double or even multiple (\velaj; Vela pulsar) pulses, of which 5 {\it are} detected at high-energy \gr-rays (except so far \psrh).
It is interesting to note that of the 77 non-recycled pulsars in the second \fermi\ pulsar catalogue 58 (or 75\%) show two strong, caustic peaks 
significantly separated, thus a very much larger fraction than for the soft \gr-sample. In fact, 91\% (10/11) of the soft 
\gr-ray pulsars which are {\it not} detected by \fermi\ LAT show a single, mostly broad/structured pulse.

\begin{figure*}
  \begin{center}
     \includegraphics[width=5.75cm]{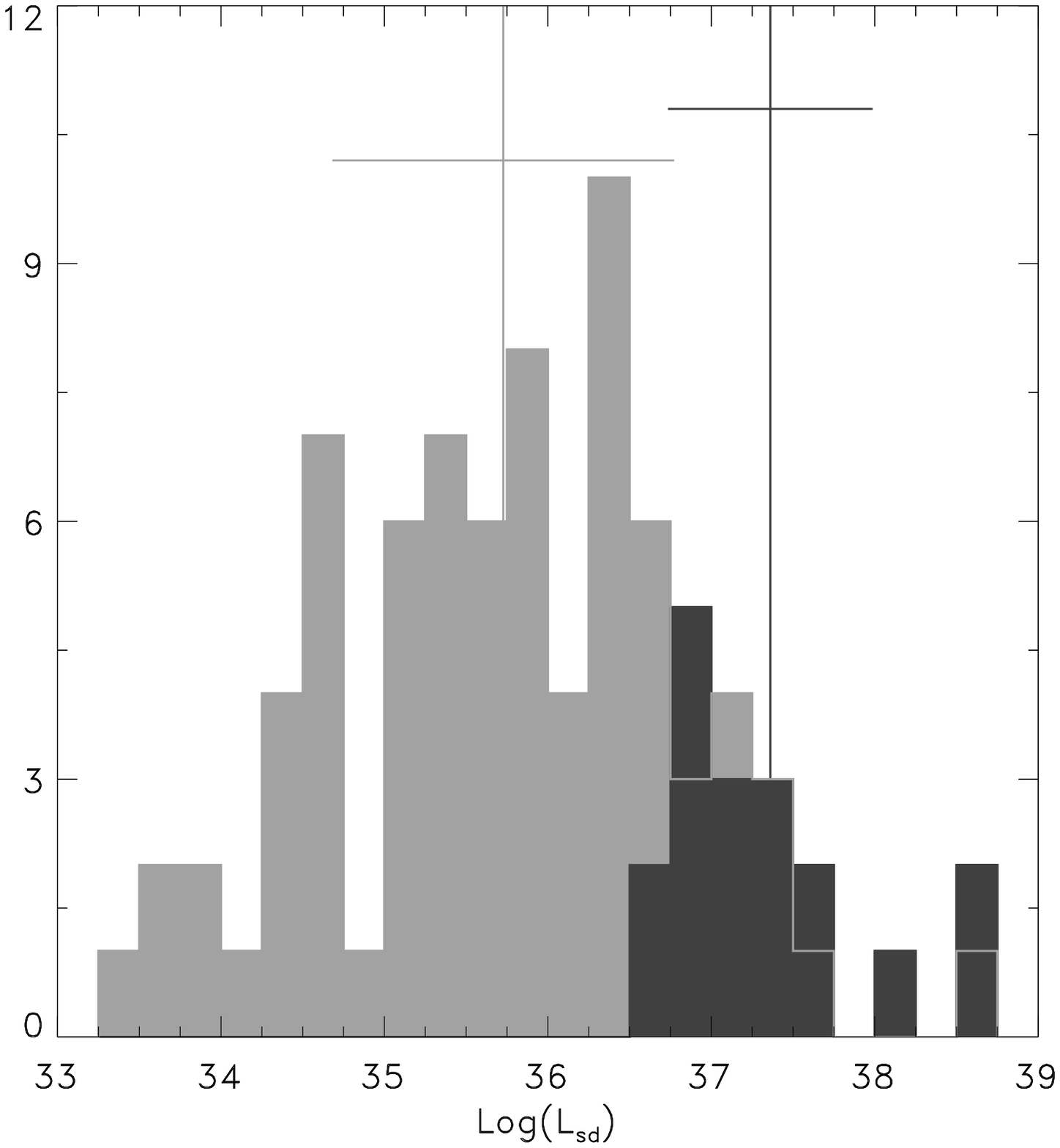}
     \includegraphics[width=5.75cm]{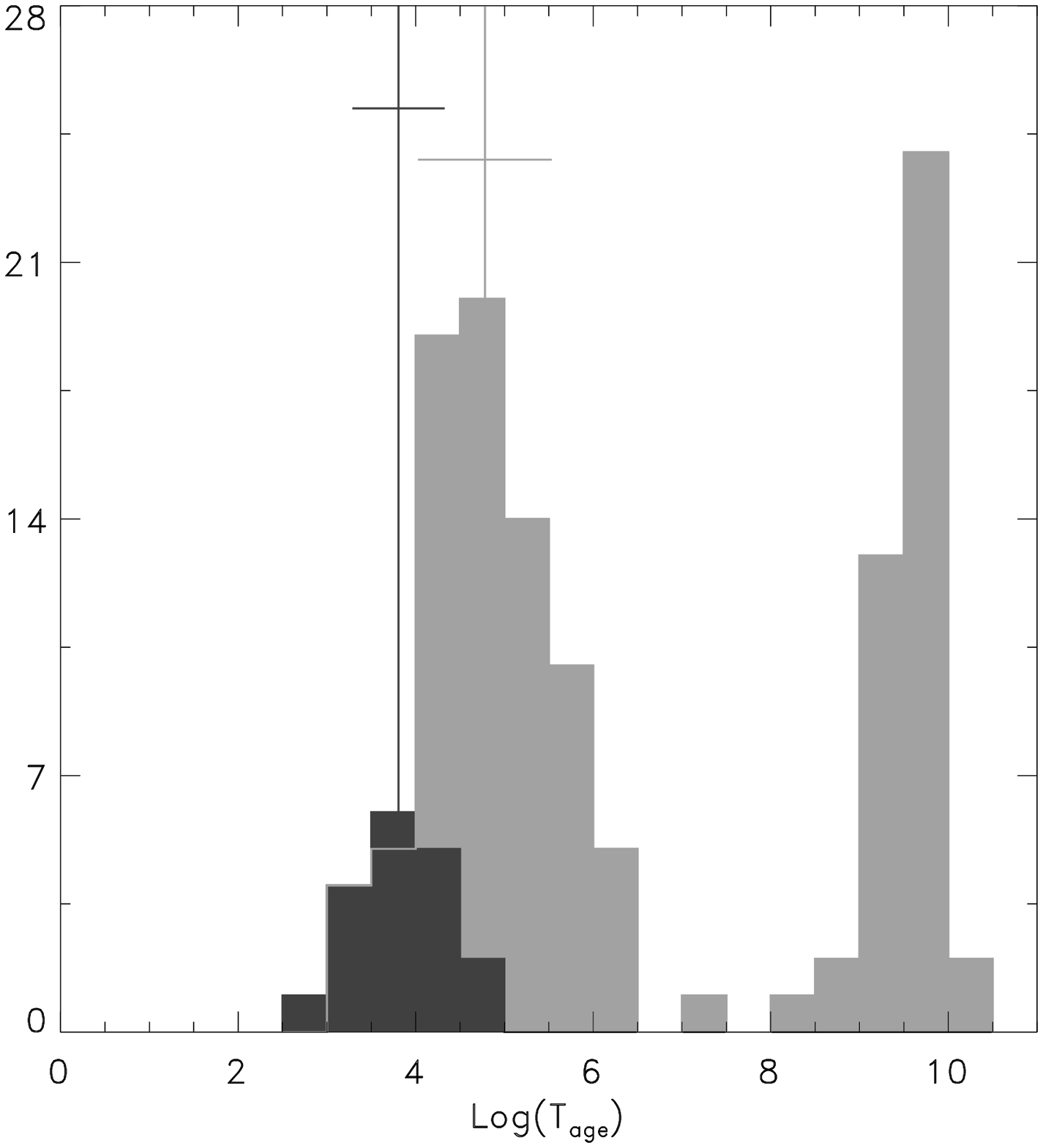}
     \includegraphics[width=5.75cm]{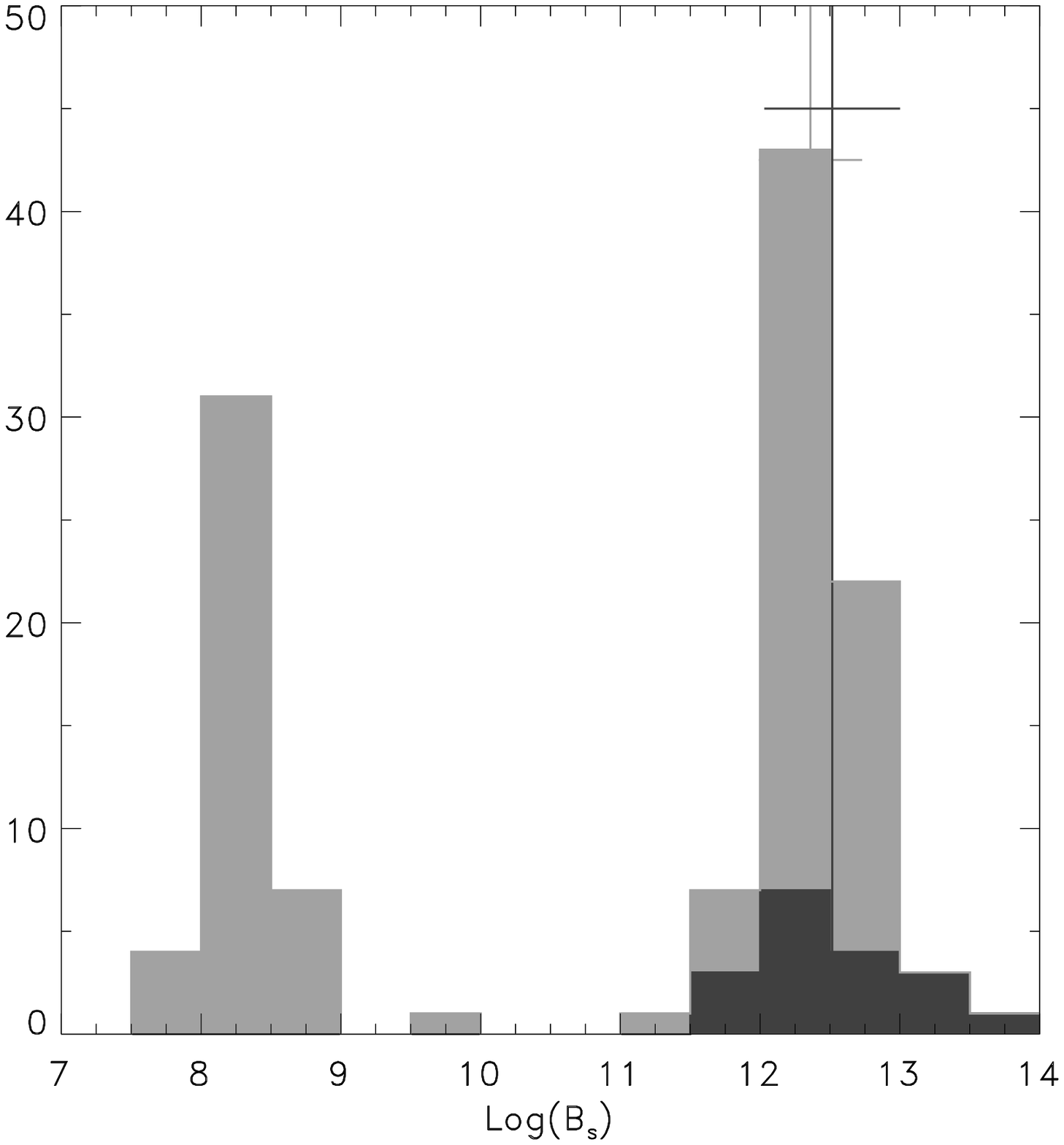}
     \caption{\label{populationdst} The distributions of the soft \gr-ray (dark grey; 18 members) and hard \gr-ray (light grey, \fermi\ LAT) pulsar 
     populations as a function of spin-down luminosity (left panel; including only non-recycled \fermi\ LAT pulsars; 77 members), characteristic age (middle panel) and surface (dipole) magnetic field strength (right panel). The means (vertical) and standard deviations (horizontal) for both populations are indicated by the solid lines. The recycled pulsar subset (43 members) of the \fermi\ LAT population has been ignored in the estimation of these numbers. The soft \gr-ray pulsar population is on average $\sim 50$ times more 
luminous (left panel) and $\sim 10$ times younger than the \fermi\ LAT pulsar population, while the $B_S$ distributions are comparable.}
  \end{center}
\end{figure*}

Finally, column 7 of Table \ref{tabrpp} gives information about the spectral shape at hard X-rays. All members, except \crabj, 
\lmcpsra\ and \psrigra, show very hard spectra in the soft \gr-ray band with photon indices in the range -0.85 -- -1.57. Fig. \ref{spectralcompilation} shows 
in two panels the spectral energy distributions over the high-energy band from $\sim 2$ keV up to 10 GeV of the pulsed emissions of the soft 
\gr-ray pulsars. The preliminary \fermi\ spectrum of \lmcpsra\ \citep{martin2014} is not included, but the pulsed spectrum over the total 1 keV - 10 GeV 
band mimics the shape of that of \crabj. Also \psrcanda\ is not included, because spectral information on the pulsed emission has not yet been reported 
for energies above 10 keV.
For reference, the high-energy pulsed spectra of \crabj, \velaj\ and \mshj\ are shown in both panels. \velaj\ exhibits a pulsed spectrum that is 'typical' for 
the \fermi\ LAT-detected pulsars with maximal luminosity at GeV energies. \mshj\ has also been (weakly) detected by \fermi\ above 100 MeV, but reaches 
its maximum luminosity at MeV energies. 
It is clear from this figure that the pulsed emission spectra of the soft \gr-ray pulsar population, given the hardness of the spectra at hard X-rays/soft
\gr-rays and the non-detections above 100 MeV (11 members) or the detections as soft \fermi\ LAT pulsars (5 members), also reach maximum luminosities in 
the MeV range. 
This is thus remarkably different from the typical spectral characteristics of the \fermi\ LAT-detected pulsars. 
Note, that \velaj\ has been detected in the soft \gr-ray band, only, because its distance is $\sim 290$ pc.

We also compared the distributions of the spin-down power, characteristic age and surface magnetic field strength at the pole 
($B_{\hbox{\scriptsize s}}\simeq 3.2\times 10^{19}\sqrt{-\dot{\nu}/\nu^3}$ G) of the soft- and hard \gr-ray pulsar populations. 
The results are shown in the three panels of Fig. \ref{populationdst} for spin-down power (left), characteristic age (middle) and surface magnetic field 
strength (right). The averages (vertical lines) and standard deviations (horizontal lines) of the $^{10}\log$ of 
the spin-down powers for the \fermi\ LAT non-recycled (77 members; light grey histogram left panel Fig. \ref{populationdst}) and soft \gr-ray 
(18 members; dark grey histogram left panel Fig. \ref{populationdst}) pulsar population are $35.73\pm 1.04$ and $37.36\pm 0.63$, respectively. 
Thus the latter is on average $\sim 43\times$ more energetic than the \fermi\ LAT population. 
The averages and standard deviations of the $^{10}\log$ of the characteristic ages are $4.78\pm 0.75$ and $3.81\pm 0.52$ for the non-recycled 
\fermi\ LAT pulsar- and soft \gr-ray pulsar population, respectively. This indicates that the soft \gr-ray pulsar population is on average 
$\sim 9.3\times$ younger than the non-recycled \fermi\ LAT pulsar population.
Finally, the right panel of Fig. \ref{populationdst} shows that the distributions of the surface magnetic field strengths are comparable with 
averages $^{10}\log B_{\hbox{\scriptsize s}}$ of $12.36\pm0.37$ and $12.52\pm0.49$ for the non-recycled \fermi\ LAT pulsar- and soft \gr-ray 
pulsar population, respectively.

\subsection{ Too many young luminous \gr-ray pulsars?}

The advent of the \fermi\ launch and the milestone later achieved with the release of the second \fermi\ pulsar catalogue triggered studies of the collective properties of the pulsar population for comparisons with predictions from competing \gr-ray pulsar models, constraining the birth properties, beaming and evolution \citep{gonthier2007, watters2011, takata2011, pierbattista2012}. One of the conclusions from these studies is that (one-pole) outer-gap type models and slot-gap type models can reproduce many of the average radio and high-energy \gr-ray characteristics. However, significant differences in properties between the simulated and the observed \fermi\ LAT populations are also noted. The sample of soft \gr-ray pulsars is too small for a similar study, and the peculiar characteristics of the young and energetic pulsars {\it not} detected
at soft \gr-ray energies does not make it a fully coherent picture. Nevertheless, it is interesting to shortly address how this sample fits in the general picture emerging from the theoretical studies of the \fermi\ population of young non-recycled pulsars. 

In Table 2, we are dealing with very young energetic pulsars, which are mostly {\it not} detected by \fermi\ LAT because they reach their maximum luminosities in the MeV band and become too weak for detection above 100 MeV. All high-energy emission models propose contributions to the non-thermal pulsar spectrum of synchrotron radiation produced deeper in the magnetosphere, reaching its maximum luminosity at MeV energies, and of higher-energy curvature radiation and inverse Compton radiation produced higher-up  in the magnetosphere, closer to the light cylinder.
Apparently, the contribution from synchrotron emission is in our sample dominating in the shape of the measured average spectrum over the  components produced closer to the light cylinder. These curvature and inverse Compton components appear to be too weak for detection by \fermi\ LAT. Some of these pulsars will likely be detected by \fermi\ after the accumulated exposures have been sufficiently increased over the coming years (a first example is the recently reported detection of \lmcpsr). This is particularly interesting because \citet{pierbattista2012} conclude that their simulations for all models (Polar Cap, Slot Gap, Outer Gap and One Pole Caustic) under predict the detected number of LAT pulsars with high $L_{\hbox{\scriptsize sd}}$. Our sample of soft \gr-ray pulsars increases even further the number of pulsars with the highest $L_{\hbox{\scriptsize sd}}$ that emit non-thermal gamma-ray radiation, thus increases the found discrepancy.

In section \ref{sect_cand_pul} we discussed a sample of young luminous pulsars which are not yet detected in the soft \gr-ray  band, but are excellent candidates for detection in dedicated observations with X-ray instruments with large sensitive areas and operating in modes with sufficient time resolution. \nustar\ (in the 3--80 keV band) and \xmm\ (2--12 keV band) are currently the only observatories, which can provide the required capabilities. An increase of the size of the detected soft \gr-ray pulsar population and, in particular, a more detailed measurement of the timing and spectral characteristics of individual pulsars and the sample in the soft \gr-band up to MeV energies will give important constraints for the above mentioned comparisons between the different models aiming at explaining the production of high-energy radiation in the pulsar magnetospheres. It is therefore unfortunate that no future space missions are currently planned to bridge the observational gap between 100 keV 
and 100 MeV.

\section*{Acknowledgments}
This research has made extensive use of data and/or software provided by the High Energy Astrophysics Science Archive Research Center (HEASARC), which is a 
service of the Astrophysics Science Division at NASA/GSFC and the High Energy Astrophysics Division of the Smithsonian Astrophysical Observatory. 
We also acknowledge the use of NASA's Astrophysics Data System (ADS) and of the SIMBAD database, operated at CDS, Strasbourg, France.
In this work we analysed observations performed with \integral, an ESA project with instruments and science data centre funded by ESA member states (especially the PI countries: Denmark, France, Germany, Italy, Switzerland, Spain) and with the participation of Russia and the USA, and we also analysed observations obtained with \xmm, an ESA science mission with instruments and contributions directly funded by ESA Member States and the USA (NASA).
Moreover, this research has made use of data obtained from the \cxo\ Data Archive.

%
%

\bsp

\label{lastpage}

\end{document}